\documentclass[12pt,leqno]{article}

 \usepackage{amsmath,amssymb,theorem,latexsym,graphics,epsfig,eufrak}

 \usepackage[all]{xy}
 \usepackage{color}
 \usepackage{psfrag}

 \newtheorem{theo}{Theorem}[section]

 {\theorembodyfont{\rmfamily} \newtheorem{rem}[theo]{Remark}}
 {\theorembodyfont{\rmfamily} }
 {\theorembodyfont{\rmfamily} }
 {\theorembodyfont{\rmfamily} }

 \DeclareFontFamily{U}{rsf}{}
 \DeclareFontShape{U}{rsf}{m}{n}{
   <5> <6> rsfs5 <7> <8> <9> rsfs7 <10->  rsfs10}{}
 \DeclareMathAlphabet{\mathscr}{U}{rsf}{m}{n}
 \newcommand{\mycal}[1]{\mathscr{#1}}
 
 \newcommand{\oX}{{\overline X}}
 \newcommand{\opi}{{\overline \pi}}
 \newcommand{\oH}{{\overline H}}
 \newcommand{\wX}{{\widetilde{X}}}

 \newcommand{\CF}{{\cal{F}}}
 \newcommand{\CS}{{\mathcal S}}
 \newcommand{\half}{\frac{1}{2}}
 \newcommand{\bS}{\boldsymbol{S}}
 \newcommand{\bM}{{\boldsymbol{M}}}
 \newcommand{\bL}{{\boldsymbol{L}}}
 \newcommand{\CO}{{\cal{O}}}
 \newcommand{\CE}{{\cal{E}}}
 \newcommand{\CH}{{\cal A}}
 \newcommand{\ra}{{\longrightarrow}}
 \newcommand{\dbar}{{\overline \partial}}
 \newcommand{\tr}{{\rm{Tr}}}
 \newcommand{\be}{\begin{equation}}
 \newcommand{\ee}{\end{equation}}
 \newcommand{\IP}{{\mathbb P}}
 \newcommand{\IC}{{\mathbb C}} 
 \newcommand{\IZ}{{\mathbb Z}} 
 \newcommand{\CG}{{\mathcal G}} 
 \newcommand{\CM}{\mycal{M}} 
 \newcommand{\CQ}{{\mathcal Q}}
 \newcommand{\CC}{{\mathcal C}} 
 \newcommand{\pt}{{\hbox{Pre-Tr}}}
 \newcommand{\wCC}{{\widetilde {\mathcal C}}}
 \newcommand{\CK}{{{\mathcal K}}}
 \newcommand{\CB}{{{\mathscr B}}}
 \newcommand{\CD}{{\mathcal D}}
 \newcommand{\wCD}{{\widetilde {\mathcal D}}}
 \newcommand{\CT}{{\mathcal T}}
 \newcommand{\II}{{\mathbb I}}
 \newcommand{\CL}{{\mathcal L}}
 
 \newcommand{\lb}{{\lambda}}

 \newcommand{\ckk}{{C({\mathcal K}, {\mathcal K}')}}
 \newcommand{\tp}{{\widetilde p}}
 \newcommand{\tq}{{\widetilde q}}

 \newcommand{\CX}{{\mathscr X}}
 \newcommand{\IF}{{\mathbb F}}

 \newcommand{\oW}{{\overline W}}

 \newcommand{\wm}{{\widetilde m}}
 \newcommand{\wbM}{{\widetilde {\boldsymbol{M}}}}
 \newcommand{\locext}{{\underline{\operatorname{Ext}}}}
 \newcommand{\lochom}{{\underline{\operatorname{Hom}}}}

 \newcommand{\CJ}{\mycal{J}}
 \newcommand{\ws}{{\widetilde s}}
 \newcommand{\wY}{{\widetilde Y}}
 \newcommand{\CW}{{\mathcal W}}  
 \newcommand{\IR}{{\mathbb R}} 
 \newcommand{\wbS}{{\widetilde {\boldsymbol S}}}
 \newcommand{\ext}{{\hbox{Ext}}}
 \newcommand{\Hom}{{\hbox{Hom}}}
 \newcommand{\CN}{\mycal{N}}
 \newcommand{\wsigma}{{\widetilde \Sigma}} 
 \newcommand{\CV}{\mycal{V}}

 \newcommand{\op}[1]{\operatorname{#1}}
 \newcommand{\bh}{\boldsymbol{h}}
 
 \newcommand{\bB}{\boldsymbol{B}}
 \newcommand{\bR}{\boldsymbol{R}}
 \newcommand{\bbad}{\boldsymbol{a}\boldsymbol{d}}
 \newcommand{\str}{{\sf str}}

 \newcommand{\les}[6]{\xymatrix{       &      & ...  \ar[r]  &  {#1}
 \ar@{->}`r/10pt[d] `[l] `^dl[dlll]  `^dr/10pt[dll]    [dll] \\
 &  {#2} \ar[r] & {#3} \ar[r] & {#4}
 \ar `r/10pt[d] `[l]  `^dl[dlll] `^dr/10pt[dll]   [dll] \\
 & {#5} \ar[r]  & {#6} \ar[r] & \ldots }}

 \newcommand{\Appendix}[1]{%
   \refstepcounter{section}%
   \addcontentsline{toc}{section}%
     {\bfseries\appendixname~\thesection\ #1}%
     {\medskip\noindent \Large\bfseries\appendixname\ \thesection\ #1}%
 \sectionmark{#1}\smallskip\noindent
 \renewcommand{\theequation}{\thesection.\arabic{equation}}
 }

\begin{document}
 \title{Geometric transitions and integrable systems}
 \author{D.-E. Diaconescu$^\flat$, R. Dijkgraaf$^{\natural}$, R. Donagi$^{\dagger}$, 
C. Hofman$^{\sharp}$ and T. Pantev$^{\dagger}$\\
$^\flat$ {\small New High Energy Theory Center, Rutgers University}\\
{\small 126 Frelinghuysen Road, Piscataway, NJ 08854}\\
$^\natural$ {\small Institute for Theoretical Physics and KdV Institute for Mathematics}\\
{\small University of Amsterdam, Valckenierstraat 65, 1018 XE Amsterdam, The Netherlands}\\
$^{\sharp}$ {\small The Weizmann Institute for Science, Department of Particle Physics}\\
{\small Herzl Street 2, 76100 Rehovot, Israel}\\
$^{\dagger}$ {\small Department of Mathematics, University of Pennsylvania} \\
{\small David Rittenhouse Lab., 209 South 33rd Street, Philadelphia, PA  19104-6395}}

 \date{June 2005}
 \maketitle

 \begin{abstract}
 We consider {\bf B}-model large $N$ duality for a new class of
 noncompact Calabi-Yau spaces modeled on the neighborhood of a ruled
 surface in a Calabi-Yau threefold. The closed string side of the
 transition is governed at genus zero by an $A_1$ Hitchin integrable
 system on a genus $g$ Riemann surface $\Sigma$. The open string side
 is described by a holomorphic Chern-Simons theory which reduces to a
 generalized matrix model in which the eigenvalues lie on the compact
 Riemann surface $\Sigma$. We show that the large $N$ planar limit of
 the generalized matrix model is governed by the same $A_1$ Hitchin
 system therefore proving genus zero large $N$ duality for this class
 of transitions.

 \end{abstract}

 \tableofcontents

 \section{Introduction} \label{sec:introduction}

 Large $N$ duality \cite{GV:corresp} has been at the center of many
 recent developments in topological string theory.  In particular {\bf
 B}-model transitions \cite{DV:matrix,DV:geometry} have revealed a
 fascinating interplay of random matrix models, integrable systems and
 Calabi-Yau geometry.

 In this paper we generalize the results of
 \cite{DV:matrix,DV:geometry} reviewed in section two to a new class of
 conifold transitions among noncompact Calabi-Yau threefolds. As
 explained in section three, the starting point of our construction is
 a configuration
 \[ \begin{array}{ccccc}
 \widetilde{\bS} & \subset & \widetilde{\bM} & &\\
 \downarrow & & \downarrow && \\
 \bS & \subset & \bM & \subset & \bL.
 \end{array}
 \] 
 of moduli spaces which generalizes the essential geometric features of
 the local transitions studied in \cite{DV:matrix}. Here $\bL$ is a
 component of the moduli space of projective or quasi-projective
 Calabi-Yau threefolds and $\bM$ is a subspace of $\bL$ parameterizing
 Calabi-Yau manifolds with isolated conifold singularities which admit
 a (quasi-)projective small resolution.  The deepest stratum $\bS$,
 which is a key element of the construction, parameterizes Calabi-Yau
 spaces with a genus $g$ curve $\Sigma$ of $A_1$ singularities. The
 spaces $\wbS,\wbM$ are moduli spaces of the resolution of
 Calabi-Yau spaces in $\bS$ and respectively $\bM$. Such geometric
 structures have been considered before in the physics literature
 \cite{KMP,KKLM:super,KKLM:mirror} in relation to $N=2$ gauge theories
 and open string superpotentials. Here we will show that they play a
 key role in {\bf B}-model geometric transitions.

 Our main construction is carried out in section four. We consider
 noncompact Calabi-Yau spaces fibered by affine quadrics over a fixed
 genus $g$ curve $\Sigma$.  A special feature of this model is that the
 moduli spaces $\wbM$, $\bL$ are isomorphic to the total spaces of
 vector bundles over $\bS$.

 Large $N$ duality is an equivalence between {\bf B}-type open-closed
 topological strings on a resolved threefold corresponding to a point in
 $\wbM$ and closed topological strings on a generic threefold in
 $\bL$. In the present paper we establish this result for genus zero
 topological amplitudes in the geometric framework described above. The
 proof involves two parts.

 The genus zero dynamics for closed {\bf B}-topological strings on
 Calabi-Yau spaces is usually encoded in the intermediate Jacobian
 fibration over the moduli space, which supports an integrable system
 structure \cite{DM:spectral,DM:cubics}. In our case we show that the
 relevant integrable system for a family of threefolds parameterized
 by a normal slice to $\bS$ in $\bL$ is the $A_1$ Hitchin integrable
 system. In particular, the normal slice $\bL_s$ at a point $s\in \bS$
 is isomorphic to the space of quadratic differentials on $\Sigma$,
 which is the base of the Hitchin systems.  This follows from a
 structure result for the intermediate Jacobians of the Calabi-Yau
 threefolds in $\bL$ proved in section five.

 The second part of the proof is more physical in nature and involves
 {\bf B}-topological open string dynamics on a small resolution
 parameterized by a generic point in $\wbM$. In section six we
 construct the holomorphic Chern-Simons theory which captures open
 string target space dynamics using the formalism of D-brane
 categories.  Then we argue that the holomorphic Chern-Simons
 functional integral reduces to a finite dimensional integral on a real
 cycle in the product $\hbox{Sym}^N(\Sigma)
 \times\hbox{Sym}^N(\Sigma)$.  This can be regarded as a generalized
 matrix model in which the eigenvalues are parameterized by a compact
 Riemann surface. The final result of this section is that the large
 $N$ planar limit of this generalized matrix model is captured by the
 same $A_1$ Hitchin system that was found in section five.  This concludes the
 physical proof of genus zero large $N$ duality.

{\it Acknowledgments.} We are very grateful to Bogdan Florea and Antonella 
Grassi for collaboration
at an early stage of the project and many useful discussions. We would also 
like to thank Jacques Distler, Sheldon Katz, Bal\'azs Szendr\"oi and Cumrun Vafa for 
helpful discussions. 
D.-E. D. would also like to acknowledge the partial 
support of the Alfred P. Sloan 
foundation and the hospitality of KITP Santa Barbara and The Aspen Center 
for Physics where part of this work was performed. 
The research of R.D. was supported by a NWO Spinoza grant and the FOM program 
{\it String Theory and Quantum Gravity}. 
R.D. was partially supported by NSF grant DMS 0104354 and FRG grant 0139799 
for ``The Geometry of Superstrings''. T.P. was partially supported by 
NSF grants  FRG 0139799 and DMS 0403884. 
The work of C.M.H. was supported in part by a Marie Curie Fellowship under
contract MEIF-CT-2003-500687, the Israel-US Binational Science Foundation,
the ISF Centers of Excellence Program and Minerva.

 \section{Review of Dijkgraaf-Vafa transitions} \label{sec:DV}

 In this section we will review large $N$ duality for a class of
 geometric transitions among noncompact Calabi-Yau threefolds first
 studied in \cite{DV:matrix}. Adopting the common terminology in the
 physics literature, for us a geometric transition will be an extremal
 transition connecting two different components of a moduli space of
 Calabi-Yau threefolds through a degeneration. The degenerations
 usually considered in this context are nodal Calabi-Yau threefolds
 with isolated ODP singularities. More complicated singularities, such
 as rational double points, can also appear as junctions of
 geometric transitions and support very interesting large $N$
 physics \cite{F:planar}. We will not look at these more complicated
 geometries here but they certainly deserve a thorough investigation
 from the point of view of Dijkgraaf-Vafa quantization. 

 In the situation considered in \cite{DV:matrix}, we have a moduli
 space $\bL$ of noncompact Calabi-Yau hypersurfaces in $X_l\subset \IC^4$
 defined by equations of the form 
\be\label{eq:geomtransB} uv + y^2 - l(x) =0 
\ee 
where $l(x)$ is an arbitrary polynomial of degree
 $2n$. The moduli space $\bL$ is the complex vector space
 of dimension $2n+1$ parameterizing the coefficients of $l(x)$.  The
 degeneration takes place along a subvariety $\bM\subset \bL$
 characterized by the property that 
 \be\label{eq:geomtransC} m(x) =
 (W_m'(x))^2 
 \ee 
 for $m\in \bM$, where $W_m(x)$ is an arbitrary
 polynomial of degree $n+1$. Therefore $\bM$ is a $(n+1)$-dimensional
 subvariety in $\bL$. In the following we will call $W_m(x)$ the classical
 superpotential for reasons that will shortly become clear. It is easy
 to check that if the roots of $W'_m(x)$ are distinct, $X_m$ has $n$
 isolated ODPs given by solutions of the equations
 \[ 
 u=v=y=0,\qquad W'_m(x)=0
 \]
 We will refer to such points in $\bM$ as generic points. If $W'_m(x)$
 has coincident roots, $X_m$ develops more complicated singularities.
 A special role in the theory will be played by the singular point
 $s$ of $\bM$ for which the polynomial $s(x)$ (and hence $W_m'(x)$) is
 identically zero.

 For a generic point $m \in \bM$ we can easily construct a quasi-projective
 crepant resolution of $X_m$ by blowing-up $\IC^4$ along the subvariety
 \[ 
 u=0,\qquad y-W'_m(x)=0
 \] 
 This resolution is not unique since we can obtain a different one for
 example by blowing up $\IC^4$ along
 \[ 
 v=0,\qquad y+W'_m(x) =0, 
 \] 
 and we can also consider obvious variations. However all these
 resolutions are related by flops.  Therefore we will have a moduli
 space $\wbM$ of smooth Calabi-Yau threefolds $\wX_\wm$ and a finite-to
 one (in fact in this case two to one) surjective map $\rho:\wbM\to
 \bM$ so that $\wX_\wm$ is a quasi-projective crepant resolution of
 $X_{\rho(\wm)}$. We will denote by $C_1, \ldots, C_n$ the exceptional
 curves on $\wX_\wm$.

Resolving the singular threefolds $X_m$ corresponding to the special
 points in $\bM$ where $W'_m(x)$ acquires multiple roots is more
 involved. Here we will only discuss the extreme case of the threefold
 $X_s$ corresponding to the singular point $s\in \bM$. Note that $s$ is a
 branch point for the cover $\rho:\wbM\to \bM$. The inverse image
 $\rho^{-1}(s)$ consists of a single point $\ws\in \wbM$.  Since
 $W_s(x)\equiv 0$, the singular threefold $X_s$ is determined by the
 equation
 \[ 
 uv +y^2=0. 
 \] 
 Therefore $X_s$ is isomorphic to a direct product of the form
 $\IC\times Y$ where $Y$ is the singular quadric surface described by
 the same equation in $\IC^3$ with coordinates $(u,v,y)$. In particular
 $X_s$ has a line $u=v=y=0$ of $A_1$ singularities. The resolution
 $\wX_\ws$ is isomorphic to the direct product $\IC\times {\widetilde
 Y}$, where ${\widetilde Y}$ is the minimal resolution of $Y$;
 ${\widetilde Y}$ is isomorphic to the total space of the line bundle
 $\CO(-2)$ over $\IP^1$, and the map ${\widetilde Y}\to Y$ is the
 contraction of the zero section.

 Note that a resolution $\wX_\wm$ corresponding to a generic point is
 related to $\wX_\ws$ by a complex structure deformation and has the
 structure of a fibration in quadrics over the complex line.  Consider
 the projection map ${\widetilde \pi}:\wX_\wm \to \IC$ defined in terms of local
 coordinates by forgetting $(y,u,v)$. Using the equation
 \eqref{eq:geomtransB}, the fibers of this projection are easily seen
 to be affine quadrics in $\IC^3$. The fibers over points in $\IC$
 different from the zeroes of $W_m(x)$ are smooth affine quadrics which
 can be described as smoothings of the A$_1$ singularity.  The fiber
 over a zero of $W_m(x)$ is isomorphic to the resolution $\wY$ of the
 $A_1$ singularity.  In particular these fibers contain the exceptional
 curves $C_1, \ldots, C_n$ which may be identified with the zero
 section of $\wY$.

 In the framework of topological {\bf B}-strings, large $N$ duality
 relates topological open strings on a resolution $\wX_\wm$ to
 topological closed strings on a smoothing $X_l$. The topological open
 string {\bf B}-model is constructed by wrapping $N_i$ topological {\bf
 B}-branes on the $i$-th exceptional curve $C_i$, $i=1,\ldots, n$ in
 $\wX_\wm$. The precise statement for extremal transitions among
 general projective or quasi-projective Calabi-Yau varieties is not
 known\footnote{To first order however, large N duality for general
 transitions can be made precise, and is formulated and proved in
 \cite{DDGP:mixed}.}. Here we will explain how large $N$ duality works
 for the special class of transitions introduced above.  Our discussion
 follows closely \cite{DV:matrix}.

 First we should explain what we mean by a topological open {\bf
 B}-model defined by wrapping branes on the exceptional curves. From a
 rigorous mathematical point of view, topological ${\bf B}$-branes
 should be described in terms of derived objects -- or, more concretely
 complexes of vector bundles -- on the threefold $\wX_\wm$. Although
 this formalism will be very useful for generalizations, in the present
 case it suffices to think informally of a D-brane with multiplicity
 $N_i$ wrapping an exceptional curve $C_i$ as a rank $N_i$ vector
 bundle $E_i$ on $C_i$. More specifically for the purpose of large $N$
 duality we will consider trivial bundles of the form $E_i=C_i \times
 \IC^{N_i}$.  In the following we fix a
 threefold $\wX_\wm$ with  $m=\rho(\wm)$ generic.

 The dynamics of such a brane should be described in terms of a set of
 off-shell fields which in this case are $C^\infty$ bundle valued
 differential forms on $C_i$ of the form
 \[
 A^{0,p}\left(\hbox{End}(E_i)\otimes \Lambda^q N_{C_i/\wX_\wm} \right)
 \] 
 subject to the constraint $p+q=1$. In principle, one would like to
 write down a holomorphic Chern-Simons action functional for such
 fields using first principles \cite{EW:CS} and work out the rules
 for quantization.  However, this is quite difficult to do in 
 practice because the quantization of holomorphic
 Chern-Simons theories is typically untractable.

 One of the main insights of \cite{DV:matrix} is that this program can
 actually be carried out in the present geometric situation as
 follows. First construct the holomorphic Chern-Simons action for {\bf
 B}-branes on the resolution of the singular threefold $\wX_\ws$. Then
 construct the holomorphic Chern-Simons action for branes on a generic
 threefold $\wX_\wm$ by adding a superpotential deformation. The end
 result is that the dynamics of {\bf B}-branes at a generic point
 $\wm\in \wbM$ is captured by a holomorphic matrix model.

 The first step is easy. The resolved threefold $\wX_\ws$ is isomorphic
 to the product $\IC\times \wY$, where $\wY$ is isomorphic to the total
 space of $\CO(-2)$ over $\IP^1$.  Therefore $\wX_\ws$ contains a ruled
 surface $S=\IC\times \IP^1$ where the rational fibers of the ruling
 are exceptional curves on $\wX_\ws$ obtained by resolving the line of
 $A_1$ singularities. In particular all fibers of $S \to \IP^{1}$ are
 $(0,-2)$ curves on $\wX_\ws$.  We consider a system of $N$ topological
 {\bf B}-branes wrapping a given fiber $C$ of the ruling, where
 \[ 
 N = \sum_{i=1}^n N_i
 \]  
 Informally this means that we pick up a trivial bundle $E$ of the form
 $C\times \IC^N$.  We want to write down a holomorphic Chern-Simons
 action functional on the set of off-shell fields
 \[ 
 A^{0,p}\left(\hbox{End}(E)\otimes \Lambda^q N_{C/\wX_\ws} \right)= 
 A^{0,p}\left(\Lambda^q N_{C/\wX_\ws} \right)\otimes M_N(\IC) 
 \] 
 with $p+q=1$. Since $N_{C/\wX_\ws} \simeq \CO_C\oplus \Omega^1_C$ we
 are left with three off-shell fields
 \[ 
 \phi^{0,1} \in A^{0,1}\otimes M_N(\IC),\qquad \phi^{0,0} \in
 A^{0,0}\otimes M_N(\IC),\qquad \phi^{1,0}\in A^{0,0}\otimes
 \Omega^1_C\otimes M_N(\IC).
 \] 
 In order to write down the holomorphic Chern-Simons action, we have to
 regard the holomorphic bundle $E$ as a $C^\infty$ bundle equipped with
 a $(0,1)$ connection which in this case can be taken to be the trivial
 Dolbeault operator $\dbar$ on $C$.  The field $\phi^{0,1}$ represents
 an off-shell deformation of the background connection which can be
 eliminated by performing a gauge transformation. Therefore it suffices
 to write down an action for the remaining fields $\phi^{0,0},
 \phi^{1,0}$. This is simply given by \be\label{eq:dvactA} S_\ws =
 \int_C \tr\left(\phi^{1,0} \wedge \dbar \phi^{0,0}\right).  \ee This
 is the holomorphic Chern-Simons action for {\bf B}-branes on
 $\wX_\ws$. The Chern-Simons actions for the branes on an arbitrary
 threefold $\wX_\wm$ is constructed in \cite{DV:matrix} by adding a
 superpotential term to the functional \eqref{eq:dvactA}.

 From our perspective, this construction can be best summarized as
 follows. Our final goal is to construct the partition function for
 {\bf B}-branes on $\wX_\wm$, at least as a formal perturbative
 expansion.  The partition function of any gauge theory is obtained by
 formally integrating over all fields in the action and dividing by
 the volume of the gauge group. Usually, if the fluctuations of the
 theory are described by some complex fields $\psi$, the path integral
 can be formally written as an integral over the space of fields $\int
 D\psi D{\overline \phi}e^{-S(\phi,{\overline \phi})}$.  However, the
 holomorphic Chern-Simons action depends only on the holomorphic part
 of the fields; the antiholomorphic part is absent. Then the quantum
 theory should be formally defined by integrating the holomorphic
 measure $D\phi e^{-S(\phi)}$ over a suitable middle dimensional real
 cycle $\Gamma$ in the space of fields
 \cite{EW:CS,DV:matrix,EW:twistors}.  Therefore the formal expression
 of the functional integral is \be\label{eq:fctintA} Z =
 \frac{1}{\hbox{vol}(\CG)}  \int_\Gamma D\phi e^{-S(\phi)}.  \ee The cycle
 $\Gamma$ should be regarded as part of the data specifying the
 quantum theory.  The dependence of the physical quantities of the
 choice of $\Gamma$ has been thoroughly investigated for holomorphic
 matrix models in \cite{CL:matrix}. Since we are interested only in a
 semiclassical expansion around the critical points, the choice of the
 contour
 is irrelevant. We will only assume that one can find such a cycle so
 that the functional integral is (at least formally) well defined and
 the usual perturbative techniques are valid.

 As always, the functional integral \eqref{eq:fctintA} reduces
 to an integral over (a middle dimensional real cycle in) the moduli
 space $\CM$ of critical points of the action modulo gauge
 transformations. The holomorphic measure of the moduli space integral
 should be determined in principle by integrating out the massive
 modes, provided that the original measure $D\phi$ is at least formally
 well defined.

 For the action written down in equation \eqref{eq:dvactA}, the
 critical point equations are
 \[ 
 \dbar\phi^{0,0}=0, \qquad \dbar \phi^{1,0}=0.
 \]
 The set of solutions is parameterized by the complex vector space of
 $N\times N$ complex matrices.  As explained above in order to
 formulate the quantum theory we have to specify a real middle
 dimensional cycle in the space of solutions $M_N(\IC)$. A convenient
 choice is to restrict the functional integral to the subspace of
 hermitian matrices. Therefore the partition function of {\bf B}-branes
 on $\wX_\ws$ is given by the matrix integral \be\label{eq:partfctA}
 Z_\ws = \frac{1}{\hbox{vol}(U(N))} \int d\phi \ee where $d\phi$ is the
 linear measure on the space of hermitian $N\times N$ matrices. The
 coefficient $1/ \hbox{vol}(U(N))$ is due to a residual $U(N)$ gauge
 symmetry which acts on $\phi$ by conjugation.

 Now, recall that our goal is to find an effective description for the
 dynamics of off-shell fluctuations around a {\bf B}-brane
 configuration on $\wX_\wm$ specified by the multiplicities $N_1$,
 $i=1,\ldots, n$. According to \cite{DV:matrix}, the quantum
 fluctuations about such a background are governed by the perturbative
 expansion of a deformed matrix integral of the form
 \be\label{eq:partfctB} Z_\wm = \frac{1}{\hbox{vol}(U(N))} \int d\phi
 e^{-\frac{1}{g_s}W_m(\phi)} \ee where $m=\rho(\wm)$, and $W_m(x)$ is
 the polynomial function introduced in equation \eqref{eq:geomtransC}.
Although this deformation has been recently derived by topological open 
string computation in \cite{AK}, a more geometric treatment is better
 suited for our  purposes. 

 The main observation is that on a generic threefold $\wX_\wm$ there
 exists a family of transverse holomorphic deformations of the
 exceptional curves $C_i$, $i=1,\ldots, n$ parameterized by the complex
 line. Transverse holomorphic deformations of a holomorphic cycle are
 $C^\infty$-deformations of the cycle which depend holomorphically on
 the deformation parameters. To construct a transverse holomorphic
 family which includes the $C_{i}$'s, we will use the special geometry
 of the fibration  $\wX_{\wm} \to \IC$. We constructed $\wX_{\wm}$ as a
 quasi-projective small resolution of the singular threefold $X_{m}$
 defined by $uv + y^{2} = (W'_{m}(x))^{2}$, given by a generic
 polynomial $W_{m}$ of degree $n+1$. Viewing the polynomial $W'_{m}(x)$
 as a degree $n$ map from ${\mathbb C}$ to ${\mathbb C}$ we can 
 identify $X_{m}$ with the fiber product 
 \[
 \xymatrix{
 X_{m} \ar[r] \ar[d] & Z \ar[d] \\
 \IC \ar[r]_-{W'_{m}} & \IC
 }
 \]
 where $Z$ is the conifold hypersurface $uv + y^{2} = w^{2}$ in
 $\IC^{4}$ with coordinates $(w,y,u,v)$. Hence we can obtain the
 quasi-projective small resolution $\wX_{\wm}$ of $X_{m}$ as the fiber
 product
 \[
 \xymatrix{
 \wX_{\wm} \ar[r]^-{p_{\widetilde{Z}}} \ar[d] & \widetilde{Z} \ar[d] \\
 \IC \ar[r]_-{W'_{m}} & \IC
 }
 \] 
 with a resolved conifold. Explicitly $\widetilde{Z}$ is the total
 space of the vector bundle ${\mathcal O}(-1)\oplus {\mathcal O}(-1)$
 on ${\mathbb P}^{1}$ and the map $\widetilde{Z} \to {\mathbb C}$
 corresponds to the natural epimorphism of vector bundles ${\mathcal
   O}(-1)\oplus {\mathcal O}(-1) \to {\mathcal O}$. In this picture for
 $\wX_{\wm}$, the exceptional curves $C_{i}$, $i = 1, \ldots, n$ are
 simply the preimages of the zero section $C \subset \widetilde{Z}$ of 
 \[
 \widetilde{Z} = \operatorname{{\mathcal O}(-1)\oplus {\mathcal O}(-1)} \to
 {\mathbb P}^{1}.
 \]
 Now, the fibration $\widetilde{Z} \to \IC$ is well known
 \cite{hitchin-construction} to be the
 complement of the fiber at infinity of the twistor family for the
 Taub-NUT hyperkahler metric on the surface $\wY =
 \operatorname{tot}({\mathcal O}_{{\mathbb P}^{1}}(-2))$. In particular,
 the family of twistor lines on $\widetilde{Z}$ gives a $C^{\infty}$
 trivialization $\tau :  \widetilde{Y}\times {\mathbb C} \widetilde{\to}
  \widetilde{Z}$ of the fibration $\widetilde{Z} \to
 \IC$, which is transverse holomorphic by construction. Note that the
 twistor-line trivialization identifies the surface
 $\widetilde{Y}\times \{0\} \subset \widetilde{Y}\times {\mathbb C}$
 holomorphically with the zero fiber of $\widetilde{Z} \to {\mathbb
 C}$. and $C \subset \widetilde{Y} \cong \widetilde{Z}_{0}$ can be
 viewed either as the zero section of ${\mathcal O}(-2)$ or as the zero
 section of ${\mathcal O}(-1)\oplus {\mathcal O}(-1)$. The $\tau$-image
 of  the holomorphic
 family
 \[
 \xymatrix@C-1pc@M+4pt{
 C\times {\mathbb C} \ar@{^{(}->}[r] \ar[rd] &
  \widetilde{Y}\times
 {\mathbb C}  \ar[d] \\ 
 & {\mathbb C}
 }
 \]
 gives a transverse holomorphic family of two spheres
 \[
 \xymatrix@C-1pc@M+4pt{
 \tau(C\times {\mathbb C}) \ar@{^{(}->}[r] \ar[rd] &
  \widetilde{Z} \ar[d] \\ 
 & {\mathbb C}
 }
 \]
 which over $0 \in {\mathbb C}$ specializes to the holomorphic
 $(-1,-1)$ curve $C \subset \widetilde{Z}$. Finally, we can pull back
 this family to the fiber product $\wX_{\wm}$ to obtain a transverse
 holomorphic family
 \[
 \xymatrix@C-1pc@M+4pt{
 \mycal{C} := p_{\widetilde{Z}}^{-1}(\tau(C\times {\mathbb C}))
 \ar@{^{(}->}[r] \ar[rd] & 
  \wX_{\wm} \ar[d] \\ 
 & {\mathbb C}
 }
 \]
 of 2-spheres, which is parameterized by ${\mathbb C}$ and includes the
 exceptional curves $C_{i}$.

 Given such a family, there is a pure geometric construction for a
 holomorphic function $\CW$ on the parameter space whose critical
 points coincide with the locations of the holomorphic fibers. Such a
 function is called a superpotential and is determined by the
 Abel-Jacobi map associated to the transverse holomorphic family. To
 construct this function for the present model, pick an arbitrary
 reference point $p_0$ in $\IC$ and for any point $p\in
 \IC\setminus\{p_0\}$ pick an arbitrary path $\gamma$ joining $p_0$ and
 $p$. For each path $\gamma$ we can construct a canonical three-chain
 $\Gamma$, $i=1,\ldots, n$ with boundary
 \[
 \partial \Gamma = \mycal{C}_{p}-\mycal{C}_{p_0}
 \]
 which is swept by the cycles $\mycal{C}_q$ in the family with $q\in
 \gamma$.  Then we have \be\label{eq:clsupA} \CW = \int_{\gamma}
 \Omega_{\wX_\wm} \ee where $\Omega_{\wX_\wm}$ is a global holomorphic
 three-form on $\wX_\wm$. Usually such a function would be multivalued
 because of the choices involved, but this complication does not appear
 in the present example because $H_3(\wX_\wm,\IZ)=0$.  Finally, one can check
 by an explicit computation that $\CW$ agrees with $W_m$ up to an
 irrelevant additive constant. This is the geometric interpretation of
 \eqref{eq:partfctB}.

 Using standard matrix model technology, the matrix integral
 \eqref{eq:partfctB} can be rewritten as an integral over the
 eigenvalues $\{\lambda_a\}$, $a=1,\ldots, N$ of $\phi$
 \be\label{eq:partfctC} Z_{\wm} = {\frac{1}{N!}} \int_{\IR^N}\prod_{a=1}^N
 d\lambda_a \Delta(\lambda_a) e^{-{\frac{1}{g_s}} \sum_{a=1}^N
 W_m(\lambda_a)}.  \ee where
 \[ 
 \Delta(\lambda_a) = \prod_{a\neq b} (\lambda_a-\lambda_b).
 \] 
 The critical points of the classical superpotential are in one to one correspondence with 
 partitions of the form 
 \[ 
 N= N_1+N_2 + \cdots + N_n.
 \]
 Each such partition corresponds to a distribution of eigenvalues among the $n$ zeros 
 of $W_m$. 
 For the purpose of large $N$ duality we are interested in the perturbative expansion of 
 the matrix integral around such a critical point in the large $N$ limit. The duality 
 predicts a precise relation between this expansion and the perturbative expansion of the 
 closed topological string on a smooth threefold $X_l$.

 Here we will only explain how duality works for genus zero amplitudes. The genus zero amplitudes of the 
 topological open string are captured by the large $N$ planar limit of the matrix integral. This 
 means that we take the limits 
 \[ 
 N\to \infty, \qquad g_s\to 0 \qquad \text{ while keeping $\mu = Ng_s$ fixed}. 
 \] 
 In this limit, the perturbative expansion
 of the matrix model free energy about a classical vacuum is encoded in
 a geometric structure associated to the semiclassical equations of
 motion.  The semiclassical equations of motion can be obtained by
 applying the variational principle to the effective superpotential
 \[ 
 W^{scl} = \sum_{a=1}^N W_m(\lambda_a) -g_s \hbox{log} \Delta(\lambda_a).
 \]
 The distribution of eigenvalues is 
 characterized by the resolvent 
 \[ 
 \omega(x)= \frac{1}{N} \tr\left({\frac{1}{x-\phi}}\right)
 \] 
 which is a rational function on the complex plane with poles at the
 locations of the eigenvalues.  In the $N\to \infty$ limit, the
 eigenvalues behave as a continuous one dimensional fluid with density
 $\rho(\lambda)$ normalized so that
 \[ 
 \int_\IR \rho(\lambda)d\lambda = 1. 
 \]
 The large $N$ limit of the resolvent is 
 \[ 
 \omega_\infty (x) = \int_\IR \frac{\rho(\lambda)d\lambda}{x-\lambda}.
 \]
 Then one can show that the semiclassical vacua at large $N$ are in one
 to one correspondence to solutions to the algebraic equation
 \be\label{eq:largeNA} 
 \omega_\infty^2 - \frac{1}{\mu}W'_m(x)
 \omega_\infty = g(x) 
 \ee 
 where $g(x)$ is a polynomial of degree
 smaller or equal to $n-1$.  More precisely, for each choice of $f(x)$,
 one can find a large $N$ semiclassical vacuum by solving equation
 \eqref{eq:largeNA}. The distribution of eigenvalues in such a
 semiclassical vacuum is supported on a disjoint union of line segments
 $\Gamma_1, \ldots, \Gamma_n$ centered around the roots of $W_m(x)$.
 Note that equation \eqref{eq:largeNA} determines a hyperelliptic
 curve, which is a double cover of the complex plane with branch points
 situated at the endpoints of the segments $\Gamma_i$, $i=1,\ldots,
 n$. The line segments $\Gamma_i$, $i=1,\ldots, n$ are branch cuts for
 the double cover.

 Next we claim that the genus zero free energy can be
 expressed in terms of period integrals of the meromorphic one-form
 $\eta$ on the hyperelliptic curve.  Given the branch cuts $\Gamma_i$,
 $i=1,\ldots, n$ we can choose $2(n-1)$ contours $(A_1,\ldots, A_{n-1},
 B_1,\ldots, B_{n-1})$ in the complex plane ${\mathbb C}$,  which give rise to a
 symplectic basis of cycles on the hyperelliptic curve.  The periods of
 $\eta$ on the $A$-cycles
 \[
 s^i = \frac{1}{2\pi i} \int_{A_i} \omega_\infty(x) dx 
 \]
 determines the filling fractions associated to the cuts $\Gamma_1,
 \ldots, \Gamma_{n-1}$.  Note that there are only $n-1$ independent
 filling fractions since their sum should be 1.  According to
 \cite{DV:matrix}, the integrals of the same differential
 $\omega_\infty(x)dx$ compute the amount of energy necessary for moving
 an eigenvalue from the $i$-th branch cut $\Gamma_i$, $i=1,\ldots, n-1$
 to $\Gamma_n$.  Therefore, if we denote by $\CF^{\text{op}}$ the free energy
 of the matrix model, we have
 \[ 
 \frac{\partial \CF^{\text{op}}}{\partial s^i} = \int_{B_i} \omega_\infty(x)
 dx,  \qquad i=1,\ldots, n-1.  
 \]

 The main claim of large $N$ duality is that the large $N$ limit of
 the open string theory on $\wX_\wm$ is equivalent to a closed string
 theory on a threefold $X_l$ of the form \eqref{eq:geomtransB} for
 some $l(x)$ of the form $l(x) = f(x) + (W_{m}'(x))^{2}$. The above
 geometric interpretation of the matrix model free energy makes this
 correspondence very transparent. To the spectral curve of the matrix
 model \eqref{eq:largeNA} we can associate a noncompact Calabi-Yau
 threefold $X_{m,f}$ determined by the equation \be\label{eq:largeNB}
 uv + y^2 - (W_m'(x))^2 =f(x) \ee where
 \[ 
 y = 2\mu\omega_\infty - W_m'(x), \qquad f(x) = 4\mu g(x). 
 \]
 This threefold is a fibration in quadrics over the complex plane with
 coordinate $x$.  The generic fiber is a smooth affine quadric in
 $\IC^3$, while the fibers over the roots of the polynomial
 $(W_m'(x))^2+f(x)$ are singular quadrics with isolated $A_1$
 singularities. As explained above, one can construct a smooth two
 cycle homeomorphic to a two-sphere on each smooth fiber. This cycle
 shrinks to a point on the singular fibers. Therefore $X_{m,f}$ can be
 viewed as the total space of a family of affine quadrics which
 degenerate to singular $A_1$ quadrics at finitely many points.  The
 smooth cycles in the smooth fibers are vanishing cycles with respect
 to the degenerations.

 Each contour $A_i$ in the complex plane gives rise to closed
 three-cycle $S_i$ which is swept by the two-cycles in the fibers over
 the points of $A_i$. We can perform the same construction for the
 contours $B_i$. In this case the fiber two-cycles shrink at the
 endpoints of the contour resulting again in a closed three-cycle $T_i$
 on $X_{m,f}$.  With a suitable normalization of the global holomorphic
 three-form on $X_{m,f}$, one now gets that
\be\label{eq:largeND} s^i
 = \int_{S_i} \Omega_{X_{m,f}}, \qquad \frac{\partial
 \CF^{\text{op}}}{\partial s^i} = \int_{T_i} \Omega_{X_{m,f}} \ee 
for
 $i=1,\ldots, n-1$.  However these are precisely the defining relations
 for the special geometry holomorphic prepotential associated to the
 family of smooth Calabi-Yau threefolds $\bL$. Therefore we can
 conclude that the open string free energy $\CF^{\text{op}}$ can be
 interpreted in the large $N$ planar limit as the genus zero closed
 string free energy for a ${\bf B}$-model on the threefold $X_{m,f}$.

 \section{Dijkgraaf-Vafa limits of compact Calabi-Yau spaces}
 \label{sec:DVcompact} 

 In the previous section we have studied large N duality for a special
 class of extremal transitions among noncompact Calabi-Yau
 threefolds. The geometric features of those models enabled us to
 formulate and give a physical proof of the duality conjecture. In this
 section we will develop a general geometric framework for geometric
 transitions by generalizing the essential features of the local
 models. The first important aspect of Dijkgraaf-Vafa transitions which
 can be taken as a guiding principle for more general geometric
 situations is a stratified structure of the moduli space which will be
 described in detail below.

 \subsection{Stratification of moduli} \label{ss:stratification}

 The general setup for geometric transitions consists of the following
 data.  We take $\bL$ to be a fixed component of the moduli space of
 Calabi-Yau threefolds. Let $\bM$ be the subvariety of $\bL$
 parameterizing the singular threefolds with a fixed number $n$ of
 isolated ODPs which admit a crepant projective resolution. We will
 denote by $l$, $m$ the points of $\bL$ and respectively $\bM$ and by
 $X_l$, $X_m$ the corresponding Calabi-Yau threefolds.  For a fixed
 generic point $m\in \bM$, $X_m$ may have several distinct projective
 crepant resolutions related by flops. Therefore the moduli space
 $\wbM$ of the resolution is a finite cover of $\bM$. We denote by
 $\rho:\wbM\to \bM$ the finite to one map which associates to any
 smooth crepant resolution the singular threefold obtained by
 contracting the exceptional locus. We will also denote by $\wm$ a
 point in $\wbM$ so that $\rho(\wm)=m$.  Then the extremal transition
 can be represented by a diagram of the form \be\label{eq:geomtransA}
 \xymatrix{ & {\wX}_{\wm} \ar[d] \\ X_{l} \ar@{~>}[r] & {X}_{m}.  } \ee
 where $X_{l}$ is a smooth Calabi-Yau space corresponding to a point
 $l\in \bL$, $\xymatrix@1{{X}_{l} \ar @{~>}[r] & {X}_{m}}$ is a
 degeneration of $X_{l}$ to a Calabi-Yau variety $X_{m}$, $m\in \bM$
 having $n$ ordinary double points, and $\wX_{\wm} \to X_{m}$, $\wm\in
 \wbM$ is a crepant quasi-projective resolution of $X_{m}$.

 In the previous section, the connection between holomorphic
 Chern-Simons theory and matrix models was based on the existence of a
 maximally degenerate point $s\in \bM$ where the Calabi-Yau threefold
 develops a curve of $A_1$ singularities. In a general situation, we
 should be looking for a singular subspace $\bS\subset \bM$
 characterized by the property that the Calabi-Yau threefolds
 parameterized by points $s\in \bS$ have curves of singularities.  It is
 not known if such a boundary locus of $\bL$ exists in general, but
 from now on we will restrict our considerations to components with
 this property. We will discuss several examples later in this section.
 Note that the threefolds $X_s$, $s\in \bS$ have a unique projective
 crepant resolution, as opposed to Calabi-Yau varieties $X_m$
 parameterized by generic points $m\in \bM$. Therefore the moduli space
 $\wbS$ of the resolution is a subspace of the ramification locus of
 the map $\rho:\wbM\to \bM$ isomorphic to $\bS$.  Therefore we obtain a
 stratified structure of the moduli spaces $\bL$, $\wbM$ described by
 the following diagram \be\label{eq:strata}
 \begin{array}{ccccc}
 \widetilde{\bS} & \subset & \widetilde{\bM} & &\\
 \downarrow & & \downarrow && \\
 \bS & \subset & \bM & \subset & \bL.
 \end{array}
 \ee From now on we will refer to extremal transitions with this
 structure as stratified extremal transitions.  Let us now discuss some
 examples.

 \subsection{Examples of nested moduli} \label{ss:nested}

 Several examples of stratified extremal transitions among compact
 Calabi-Yau threefolds have been constructed in the literature
 \cite{KMP,KKLM:super,KKLM:mirror,BS:curves}.  In all these examples,
 the Calabi-Yau threefolds are hypersurfaces or complete intersections
 in weighted projective spaces. Other examples can be obtained using
 the Borcea-Voisin construction.

 Here we will discuss in detail one of the complete intersection
 models. We consider the moduli space $\bL$ of complete intersections
 in ${\mathbb P}^{5}$ of the form 
 \be\label{eq:complintA} 
 Q_2(z_i)=0,
 \qquad Q_4(z_i)=0 
 \ee where 
 $Q_2(z_i), Q_4(z_i)$ are homogeneous
 polynomials of degree $2,4$ in the projective coordinates
 $[z_1:z_2:\ldots : z_6]$. The stratification we are interested is
 induced by a stratification of the space of quadric polynomials
 $Q_2(z_i)$. Let us denote by $Z_2$, $Z_4$ the four dimensional quadric
 and quartic in $\IP^5$ cut by the equations \eqref{eq:complintA}. If
 $Q_2$ is a quadric of rank three, then $Z_{2}$ will be singular along a linear
 subspace $\IP^2\subset \IP^5$ which intersects a generic $Z_{4}$ along
 curve $\Sigma$ of genus $g=3$. The resulting Calabi-Yau threefold has
 $A_1$ singularities at the points of $\Sigma$. If $Q_2$ is a quadric of rank
 four, $Z_{2}$ will be singular along a line $\IP^1\subset
 \IP^4$, which intersects a generic quartic $Z_{4}$ in four points. Therefore
 the complete intersection has four isolated ODPs.  If $Q_2$ is a
 quadric of rank greater or equal to $5$, the generic complete
 intersection is smooth.  Therefore we obtain a stratified moduli space
 \[ 
 \bS \subset \bM\subset \bL
 \] 
 where $\bS$, $\bM$, $\bL$ have dimensions $83, 86$ and respectively
 $89$. The general point in $\bS$, $\bM$, $\bL$ is a complete
 intersection of a generic $Z_{4}$ and a $Z_{2}$ defined by a quadric
 $Q_{2}$ of rank $3$, $4$, and $\geq 5$ respectively. For future
 reference let us write down equations for the generic
 threefolds in each stratum.  Up to automorphisms of $\IP^5$, we can
 write the equation of a rank three quadric in the form
 \[ 
 z_1^2-z_2z_3=0.
 \] 
 The singular locus is cut by the equations $z_1=z_2=z_3=0$. 
 The quadrics corresponding to points in $\bM$ can be written as 
 \[ 
 z_1^2-z_2z_3+Q_1(z_4,z_5,z_6)^2=0
 \]
 where $Q_1(z_4,z_5,z_6)$ is a homogeneous polynomial of degree one in
 $z_4,z_5,z_6$.  The singular locus is given in this case by
 $z_1=z_2=z_3=Q_1(z_4,z_5,z_6)=0$.  The quadrics corresponding to
 points in $\bL$ can be written as
 \[ 
 z_1^2-z_2z_3+Q_2(z_4,z_5,z_6)=0
 \] 
 where $Q_2(z_4,z_5,z_6)$ is a homogeneous polynomial of degree two in
 $z_4,z_5,z_6$ which is not necessarily a perfect square.

 Given a stratified extremal transition among compact Calabi-Yau
 threefolds, one can obtain a similar transition among noncompact
 spaces by linearization. This is an algebraic-geometric process which
 embodies the notion of local limit of a Calabi-Yau space often used in
 the physics literature.  For a stratified extremal transition as
 above, we would like to perform linearization of the singular
 threefolds $X_s$ along the curves of singularities $\Sigma$. Taking
 into account the complete intersection presentation of these models,
 we can achieve this goal by linearizing the ambient projective space
 $\IP^5$ along the subspace $\IP^2\subset \IP^5$ cut by the equations
 $z_1=z_2=z_3=0$.

 Consider the direct product $\IC\times \IP^5$ regarded as a trivial
 fibration over the complex line, and perform a blow-up along the
 subspace $\{0\}\times \IP^2$. The resulting space has the structure of
 a fibration over $\IC$ in which the central fiber is reducible and
 consists of two components.  One component ${\overline P}$ is
 isomorphic to the total space of the projective bundle
 \be\label{eq:projbundle} \IP(\CO(1)^{\oplus 3}\oplus \CO) \ee over
 $\IP^2$. The second component is isomorphic to the blow-up of $\IP^5$
 along $\IP^2$.  The two components intersect transversely along the
 section of ${\overline P}$ over $\IP^2$ defined by the trivial summand
 in \eqref{eq:projbundle}. The linearization of $\IP^5$ along $\IP^2$
 is given by the complement $P$ of this section in ${\overline P}$. By
 construction, $P$ is isomorphic to the total space of the bundle
 $\CO(1)^{\oplus 3}$ over $\IP^2$.

 Now let us consider a complete intersection $X_s$ of the form
 \eqref{eq:complintA} with a quadric polynomial of the form
 $Q_2(z_i)=z_1^2-z_2z_3$. The curve $\Sigma$ of $A_1$ singularities on
 $X_s$ is cut by the equations
 \[ 
 z_1=z_2=z_3=0,\qquad Q_4(z_i)=0.
 \]
 Therefore $\Sigma$ is a quartic in the projective plane $\IP^2\subset
 \IP^5$ defined by  
 \[P_4(z_4,z_4,z_6)\equiv Q_4(0,0,0,z_4,z_5,z_6)=0.\]
 The linearization of $X_s$ along $\Sigma$ can be accordingly described
 as a complete intersection in $P$. Let $x_1, x_2, x_3$ denote the
 canonical coordinates along the fibers of $P$, that is tautological
 sections of $\CO(1)^{\oplus 3}$ pulled back to $P$.  Then the
 linearization of $X_s$ is given by the following equations in $P$
 \be\label{eq:linA} x_1^2-x_2x_3=0,\qquad P_4(z_4,z_5,z_6)=0.  \ee
 Therefore we obtain a moduli space $\bS$ of singular threefolds
 isomorphic to the space of quartics in $\IP^2$. The higher strata
 $\bM, \bL$ can be described as moduli spaces of smoothings of
 \eqref{eq:linA} by deforming the quadric equation as explained above.

 Note that in this case the linearized threefolds $X_s$, $s\in \bS$ are
 isomorphic to a direct product of the form $\Sigma \times Y$ where $Y$
 is the canonical $A_1$ surface singularity. The resolution $\wX_s$
 will therefore be isomorphic to a direct product $\Sigma \times
 {\widetilde Y}$, just as in the previous section. Moreover, the
 deformations $X_m, X_l$, as well as the resolution $\wX_\wm$ have the
 structure of affine quadric bundles over $\Sigma$. Therefore the local
 transition obtained by linearization is a direct generalization of the
 Dijkgraaf-Vafa transitions. This motivates calling such models
 Dijkgraaf-Vafa limits of compact Calabi-Yau spaces.  We will construct
 below a more general class of extremal transitions among noncompact
 Calabi-Yau threefolds which exhibit the same properties.

 \section{Linear transitions} \label{sec:linear}

 The local extremal transition obtained in the previous section by
 linearization is a special case of a general abstract construction
 which will be the main focus of this section. These transitions will
 be called linear transitions because the moduli spaces
 $\widetilde{\bM}$ and $\bL$ are vector
 bundles over the bottom stratum $\bS$ as will become clear
 below. First we present the geometric construction, and then explain
 the connection between our noncompact Calabi-Yau threefolds and
 Hitchin's integrable system.

 Let $\Sigma$ denote a smooth projective curve of genus $g\geq
 2$. Take $\bS$ to be the moduli space of pairs $(\Sigma,V)$, where $V
 \to \Sigma$ is a semi-stable rank two bundle
 equipped with a fixed isomorphism $\Lambda^2V \simeq K_\Sigma$.  For each
 such pair  $s = (\Sigma, V)$, we can construct a singular Calabi-Yau
 threefold $X_s$ as follows.  Let $T_s$ be the total space of $V$ and
 let $\xi_s:T_s\to T_s$ be the holomorphic involution acting by
 multiplication by $(-1)$ on each fiber. Then we take $X_s$ to be the
 quotient $T_s/(\xi_s)$, where $(\xi_s)$ denotes the finite group with
 two elements generated by $\{1,\xi_s\}$.  Note that $\xi_s$ fixes the
 zero section of $V\to \Sigma$, therefore $X_s$ has a curve of $A_1$
 singularities isomorphic to $\Sigma$. Blowing up $X_{s}$ along its
 singular locus $\Sigma \subset X_{s}$ yields a canonical
 quasi-projective crepant resolution $\wX_{s} \to X_{s}$ of the
 singularities of $X_{s}$. The exceptional locus $S_{s}$ of the map
 $\wX_{s} \to X_{s}$ is the projectivization of the normal cone of
 $\Sigma \subset X_{s}$ and so can be identified with the geometrically
 ruled surface ${\mathbb P}(V)$. For future reference, note that
 $\wX_{s}$ can be naturally identified with the total space of a line
 bundle over $S_{s}$, so that the exceptional divisor $S_{s} \subset \wX_{s}$
 becomes the zero section of this line bundle. Indeed, this follows by
 noting 
 the blow-up of the vertex of a two dimensional affine quadric is the
 total space of ${\mathcal O}_{{\mathbb P}^{1}}(-2)$. Explicitly we
 get $\wX_{s} \cong \operatorname{Tot}({\mathcal O}_{S_{s}}(-2))$. 

 In order to describe $\bM$ and $\bL$ we have first to understand the
 deformation theory of a singular threefold $X_s$ as well as the
 deformation theory of its resolution $\wX_s$.

 \subsection{Deformations of singular threefolds} \label{sec:singdefos}

 For simplicity we will drop the subscript $s$ in this subsection, denoting 
 a singular threefold by $X$. The small resolution $\wX_s$
 will be denoted  by $\wX$. 
 The space of infinitesimal deformations of $X$ is 
 $\ext^1_X(\Omega^1_X, \CO_X)$. The local to global spectral sequence
 yields the following  
 exact sequence
 \be\label{eq:locglobA}
 \begin{aligned}
 0& \to H^1(\locext^0_X(\Omega^1_X,\CO_X)) \to \ext^1(\Omega^1_X,\CO_X) \to 
 H^0(\locext^1_X(\Omega^1_X, \CO_X))\cr & \to
 H^2(\locext^0(\Omega^1_X,\CO_X)) 
 \end{aligned}
 \ee 
 The first term of this sequence
 $H^1(\locext^0_X(\Omega^1_X,\CO_X))$ parameterizes equisingular
 deformations of $X$, therefore it is isomorphic to the tangent space
 $T_s\bS$ to the bottom stratum $\bS$. Our main interest in this
 section is in the quotient space \be\label{eq:defquotA}
 \frac{\ext^1(\Omega^1_X,\CO_X)}{H^1(\locext^0_X(\Omega^1_X,\CO_X))}\simeq
 \hbox{ker}\left(H^0(\locext_X^1(\Omega^1_X, \CO_X))\to
 H^2(\locext^0(\Omega^1_X,\CO_X)) \right) \ee which parameterizes
 general deformations of $X_s$  modulo
 equisingular deformations.

 We claim that for the singular threefolds $X=T/(\xi)$ constructed
 above, this space is isomorphic to the space of holomorphic quadratic
 differentials on $\Sigma$
 \[ 
 H^0(\Sigma,K_{\Sigma}^{\otimes 2}).
 \]
 In order to prove this claim, it will be useful to find a presentation
 of $X$ as a singular hypersurface in the total space of a holomorphic
 bundle over $\Sigma$. We will first explain this construction in a
 simplified situation when $V$ is complex vector space of dimension two
 and $\xi$ is a holomorphic involution acting on $V$ by multiplication
 by $-1$.  In this case $V/(\xi)$ can be realized as a
 hypersurface in the complex vector space $W=S^2(V)$. The map $V/(\xi)
 \hookrightarrow W$ is given by the $\xi$-invariant polynomials on
 $V$. Explicitly we can describe this map as follows.

 The second symmetric tensor power $S^2(V)$ is a subspace of $\Hom(V^\vee,V)$. 
 Given a linear map $\phi: V^\vee \to V$, there is an induced determinant map 
 \[ 
 \hbox{det}(\phi):= \wedge^{2}\phi \Lambda^2(V^\vee) \to \Lambda^2(V).
 \]
 Let us write an arbitrary element of $S^2(V)$ in the form 
 \[ 
 \phi = u e_1\otimes e_1 + v (e_1\otimes e_2+e_2\otimes e_1) + w
 e_2\otimes e_2 
 \] 
 where $\{e_1, e_2\}$ is a basis of $V$. Let $\{f^1,f^2\}$ be the dual
 basis of $V^\vee$. Then a  
 straightforward linear algebra computation shows that 
 \[ 
 \hbox{det}(\phi)(f^1\wedge f^2) = (uw-v^2) e_1\wedge e_2. 
 \] 
 Therefore the hypersurface 
 \[ 
 \hbox{det}(\phi)=0
 \] 
 is isomorphic to the canonical $A_1$ singularity $V/(\xi)$. 

 If $V$ is a rank two bundle over $\Sigma$, we can perform this
 construction fiberwise, obtaining a  
 hypersurface in the total space $W$ of the second symmetric power
 $S^2(V)$. Let us denote by $\pi_W:W 
 \to \Sigma$ the projection map. Then the above computation shows that
 there is a tautological  
 section $det_W$ of the pull-back bundle 
 \[
 \pi_W^*\lochom_\Sigma (\Lambda^2V^\vee, \Lambda^2 V)= \pi_W^*
 (\Lambda^2V)^{\otimes 2} \simeq \pi_W^*  
 K_{\Sigma}^{\otimes 2}
 \]
 to the total space $W$. The zero locus of this section is a singular
 hypersurface $X$  
 in $W$ isomorphic to $V/(\xi)$. The singular locus of $X$ is
 isomorphic to $\Sigma$.  

 Using this presentation of $X$, we can compute the space of normal
 deformations \eqref{eq:defquotA}.  
 We have an exact sequence 
 \[
 \CJ_X / \CJ_X^2 \to \Omega^1_W\otimes_{\CO_W} \CO_X \to \Omega^1_X \to
 0. 
 \] 
 Since $X$ is a Cartier divisor in $W$, it follows that the first 
 map in this sequence is injective. Therefore we obtain an exact sequence 
 \be\label{eq:normdefB}
 0\to \CO_{W}(-X)_{|X}\to \Omega^1_{W|X}\to \Omega^1_X \to 0.
 \ee
 Thus the complex $[\CO_{W}(-X)_{|X}\to \Omega^1_{W|X}]$ is a 
 resolution of the sheaf of K\"ahler  
 differentials  
 $\Omega^1_X$ by locally free ${\mathcal O}_{X}$-modules. The Ext
 sheaves $\locext_X^{i}(\Omega^1_X, \CO_X)$ are 
 the sheaf cohomology groups of the dual complex 
 \be\label{eq:normdefC} 
 \Theta_{W|X} \stackrel{f}{\ra}\CO_{W}(X)_{|X} 
 \ee
 obtained from $[\CO_{W}(-X)_{|X}\to \Omega^1_{W|X}]$ by applying the
 functor $\lochom_X(\ 
 \ ,\CO_X)$.  Here 
 $\Theta_W := \lochom(\Omega^1_W,\CO_W)$ denotes the sheaf of
 holomorphic tangent vectors  
 to $W$. 
 Computing locally we see that the first local Ext sheaf ($=$  the
 sheaf cokernel   
 of $f$)  is given by 
 \be\label{eq:normdefD} 
 \locext_X^{1}(\Omega^1_X, \CO_X)\simeq \CO_{X}(X)_{|\Sigma}. 
 \ee
 By construction, 
 \[
 \CO_{X}(X)_{|\Sigma} \simeq K_{\Sigma}^{\otimes 2} 
 \]
 therefore we find that 
 \be\label{eq:normdefE} 
 H^0(X,\locext_X^{1}(\Omega^1_X, \CO_X))\simeq
 H^0(\Sigma,K_{\Sigma}^{\otimes 2}) 
 \ee
 is the space of holomorphic quadratic differentials on $\Sigma$. 

 In order to finish the computation, we have to determine the kernel of
 the map  
 \be\label{eq:obsmap} 
 H^0(X,\locext_X^{1}(\Omega^1_X, \CO_X))\to
 H^2(X,\locext_X^{0}(\Omega^1_X, \CO_X)). 
 \ee
 Note that the projection $\pi:X \to \Sigma$ is an affine map,
 therefore we have  
 \[ 
 H^2(X, \CF) = H^2(\Sigma, \pi_*\CF)=0 
 \]
 for any coherent sheaf $\CF$ on $X$. Therefore the map
 \eqref{eq:obsmap} is trivial, and we  
 obtain the desired result. 

 \subsection{Deformations of the resolution}  \label{ss:deformations}

 We now turn to the deformation theory of the resolution $\wX_s$.  By
 analogy with the previous subsection, our aim is to understand all
 deformations of $\wX_s$ modulo the
 deformations of the pair $(\wX_{s},S)$.  We will keep using the notation
 of the previous paragraph omitting the subscript $s$.

 The infinitesimal deformations of the pair $(\wX,S)$ are parameterized
 by the first  
 hypercohomology group of the two term complex 
 \[ 
 \CT:\quad \Theta_\wX \to i_{S*}N_{S/\wX}
 \] 
 where the first term is placed in degree zero and $i_{S} : S
 \hookrightarrow \wX$ is the natural inclusion.  
 The hypercohomology spectral sequence reduces to an exact sequence
 which reads in part   
 \[ 
 \les{H^0(\wX,N_{S/\wX})}{{\mathbb H}^1(\CT)}{H^1(\wX,\Theta_\wX)}  
 {H^1(\wX,N_{S/\wX})}{{\mathbb H}^{2}(\CT)}{H^2(\wX,\Theta_\wX)} 
 \]
 By construction, the normal bundle $N_{S/\wX}$ is isomorphic to the
 canonical bundle $K_S$ 
 of $S$. Since $S=\IP(V) = \operatorname{Proj}(S^{\bullet}V^{\vee})$ is
 a projective bundle over $\Sigma$, we have  
 \be\label{eq:mixedefA}
 \begin{aligned}
 K_S & = q^*K_\Sigma \otimes \omega_{S/\Sigma} \cr
 & \simeq q^*\left(K_\Sigma\otimes
 \wedge^2V^{\vee}\right)\otimes \CO_{S}(-2)\cr 
 & = \CO_S(-2)\cr
 \end{aligned} 
 \ee
 where $\CO_S(1)$ denotes the relative hyperplane bundle of $S$ over
 $\Sigma$, and  
 we have used the relation \cite[Chapter~III.8, exercise 8.4]{RH:algeom} 
 \[ 
 \omega_{S/\Sigma}\simeq q^*\Lambda^2V^{\vee} \otimes \CO_S(-2). 
 \]
 This shows that
 \[ 
 H^0(\wX,N_{S/\wX}) \simeq  H^0(S,K_S) = 0. 
 \] 
 In addition, it is not hard to check that the map 
 \be \label{eq:rightend}
 {\mathbb H}^{2}(\wX,\CT) \to H^{2}(\wX,\Theta_{\wX})
 \ee
 is injective. Indeed, by definition $\Theta_{\wX} \to i_{S*}N_{S/\wX}$
 is surjective. The kernel sheaf $\Theta_{X,S} = \ker[\Theta_{\wX} \to
   i_{S*}N_{S/\wX}]$ is locally free and is the sheaf of germs of vector
 fields on $\wX$ that at the points of $S$ are tangent to $S$. Write $a :
 X \to S$ for the natural projection. As explained at the beginning of
 section \ref{sec:linear}, the  resolution $\wX$ of $X$ is the total space of
 the line bundle ${\mathcal O}_{S}(-2) \cong K_{S} \cong N_{S/\wX}$ on $S$.
 In particular, the vertical tangent bundle $\Theta_{\wX/S}$ can be
 identified with the line bundle $a^{*}K_{S}$ and so the tangent
 sequence for the map $a : \wX \to S$ reads:
 \be \label{eq:tangentseq}
 0 \to a^{*}K_{S} \to \Theta_{\wX} \stackrel{da}{\to} a^{*}T_{S} \to 0.
 \ee
 Combining \eqref{eq:tangentseq} with the normal-to-$S$  sequence
 \[
 0 \to \Theta_{\wX,S} \to \Theta_{X} \to i_{S*}N_{S/\wX} \to 0
 \]
 we obtain a commutative diagram with exact rows and columns:
 \be \label{eq:defosofpair}
 \xymatrix{
 & 0 \ar[d] & 0 \ar[d] &  & \\
 0 \ar[r] & a^{*}K_{S}(-S) \ar[d] \ar[r] & \Theta_{\wX,S} \ar[d] \ar[r]^-{da}
 & a^{*}T_{S} \ar@{=}[d]  \ar[r] & 0 \\
 0 \ar[r] & a^{*}K_{S} \ar[d] \ar[r] & \Theta_{\wX} \ar[d] \ar[r]^-{da}
 & a^{*}T_{S} \ar[r] & 0 \\
 & i_{S*}N_{S/\wX} \ar@{=}[r] \ar[d] & i_{S*}N_{S/\wX} \ar[d]
 & & & \\
 & 0 & 0 & & &
 }
 \ee
 Since $S \subset \wX = \operatorname{Tot}(K_{S})$ is identified with
 the zero section of $K_{S}$, it follows that the tautological section
 of $a_{*}K_{S}$ vanishes precisely at $S$, and so $a^{*}K_{S} \cong
 \CO_{\wX}(S)$. Consequently, the diagram \eqref{eq:defosofpair}
 becomes 
 \[
 \xymatrix{
 & 0 \ar[d] & 0 \ar[d] &  & \\
 0 \ar[r] & \CO_{\wX} \ar[d] \ar[r] & \Theta_{\wX,S} \ar[d] \ar[r]^-{da}
 & a^{*}T_{S} \ar@{=}[d]  \ar[r] & 0 \\
 0 \ar[r] & a^{*}K_{S} \ar[d] \ar[r] & \Theta_{\wX} \ar[d] \ar[r]^-{da}
 & a^{*}T_{S} \ar[r] & 0 \\
 & i_{S*}N_{S/\wX} \ar@{=}[r] \ar[d] & i_{S*}N_{S/\wX} \ar[d]
 & & & \\
 & 0 & 0 & & &
 }
 \]
 Looking at the long exact sequences in cohomology for the first two
 rows we get 
 \[
 \xymatrix{
 H^{1}(a^{*}T_{S}) \ar[r] \ar@{=}[d] & H^{2}(\CO_{\wX}) \ar[r] \ar[d] &
 H^{2}(\Theta_{\wX,S})  \ar[r] \ar[d] & H^{2}(a^{*}T_{S}) \ar@{=}[d] \\
 H^{1}(a^{*}T_{S}) \ar[r] & H^{2}(a^{*}K_{S}) \ar[r]  &
 H^{2}(\Theta_{\wX}) \ar[r] & H^{2}(a^{*}T_{S})
 }
 \]
 and so the map ${\mathbb H}^{2}(\CT) = H^{2}(\Theta_{\wX,S}) \to
 H^{2}(\Theta_{\wX})$ will be injective if the map \linebreak $H^{2}(\CO_{\wX})
 \to H^{2}(a^{*}K_{S})$ is injective. Since $a : \wX \to S$ is an
 affine map we have 
 \[
 \begin{split}
 H^{2}(\wX,\CO_{\wX}) & = H^{2}(S,a_{*}\CO_{\wX}) = H^{2}(S, \oplus_{i
   \geq 0} K_{S}^{-i}) \\
 H^{2}(\wX,a^{*}K_{S}) & = H^{2}(S,a_{*}a^{*}K_{S}) = H^{2}(S, \oplus_{i
   \geq -1} K_{S}^{-i}).
 \end{split}
 \] 
 The map $H^{2}(\CO_{\wX})
 \to H^{2}(a^{*}K_{S})$ is induced from the obvious inclusion of
 sheaves \linebreak 
 $\oplus_{i
   \geq 0} K_{S}^{-i} \to \oplus_{i
   \geq -1} K_{S}^{-i}$ and so we have an exact sequence 
 \[
 H^{1}(S,\oplus_{i
   \geq -1} K_{S}^{-i}) \to H^{1}(S,K_{S}) \to H^{2}(\CO_{\wX})
 \to H^{2}(a^{*}K_{S}).
 \]
 Since the map $H^{1}(S,\oplus_{i \geq -1} K_{S}^{-i}) \to
   H^{1}(S,K_{S})$ is induced from the projection \linebreak $K_{S}\oplus
   (\oplus_{i \geq 0} K_{S}^{-i}) \to K_{S}$, it is clearly surjective
   and hence $H^{2}(\CO_{\wX}) \to H^{2}(a^{*}K_{S})$ must be
   injective. This implies the injectivity of ${\mathbb H}^{2}(\CT) \to
   H^{2}(\Theta_{\wX})$ and so we get a short exact sequence of the form 
 \be\label{eq:mixedefB} 
 0 \to {\mathbb H}^1(\CT)\to H^1(\wX,\Theta_\wX) \to 
 H^1(\wX,N_{S/\wX})\to 0.
 \ee
 The first term in \eqref{eq:mixedefB} parameterizes infinitesimal
   deformations of the  
 pair $(\wX,S)$, hence it is isomorphic to the tangent space $T_s\wbS$
   to $\wbS$ at $s$.  
 The middle term parameterizes infinitesimal deformations of $\wX$
   regardless of the behavior of  
 $S$. Therefore the quotient 
 \be\label{eq:defquotB}
 \frac{H^1(\wX,\Theta_\wX)}{{\mathbb H}^1(\CT)}\simeq H^1(\wX,i_{S*}N_{S/\wX})
 \ee  
 parameterizes infinitesimal deformations of 
 $\wX$ not preserving $S$ modulo deformations of the pair$(\wX,S)$. 
 Using the fact that $K_{S} = \CO_{S}(-2)$ and the Leray spectral
   sequence, we can easily compute  
 \be\label{eq:mixedefC} 
 H^1(\wX,N_{S/\wX})\simeq H^1(S, K_S) \simeq
   H^0(\Sigma,K_{\Sigma}).
 \ee

 \subsection{Higher strata} \label{ss:higher}

 Let us now construct the higher strata $\wbM, \bL$. Note that assuming
 these spaces exist, the quotient spaces \eqref{eq:defquotA},
 \eqref{eq:defquotB} are isomorphic to the fibers of the normal bundle
 $N_{\bS/\bL}$ and respectively $N_{\wbS/\wbM}$ at $s$.  In this
 section we will show that these infinitesimal normal deformations can be
 integrated to finite linear deformations. More precisely, we will
 construct families of noncompact Calabi-Yau manifolds parameterized by
 \[ 
 \bL=\bS \times H^0(\Sigma, K_{\Sigma}^{\otimes 2}),\qquad 
 \wbM = \wbS \times H^0(\Sigma, K_{\Sigma})
 \]
 together with a quadratic map $\Pi:\wbM \to \bL$ which form a diagram
 of the form  
 \eqref{eq:strata}. 

 We start with the construction of $\wbM$. The main observation here is
 that the space $H^1(S,K_S) = H^{0}(\Sigma,K_{\Sigma})$ parameterizes
 deformations of the 
 canonical bundle of $S$ as an affine bundle over $S$. For a fixed $S$
 there is a linear family of such deformations which can be
 constructed synthetically  as follows.

 An element $\alpha \in H^{0}(\Sigma,K_{\Sigma}) = H^1(S,K_S)\simeq
 \ext^1(\CO_S,K_S)$ 
 determines (up to isomorphism) an extension 
 \be\label{eq:extensionalpha}
 0\to K_S \to E_\alpha \to \CO_S\to 0,
 \ee
 where $E_\alpha$ is a locally free sheaf of rank two on $S$.
 Let $\oX_\alpha$ be the total space of the $\IP^{1}$ bundle
 $\IP(E_\alpha)$ over $S$,  and let 
 $\wX_\alpha$ be the complement of the infinity section $H_{\alpha}:= \IP(K_S)
 \subset \IP(E_\alpha)$. Then $\wX_\alpha$ is an 
 affine bundle over $S$, or more formally an $K_S$-torsor over $S$. 
 The threefolds $\wX_\alpha$ form a linear family 
 \[
 {\widetilde \CX} \to \wbM_{s}, \quad \wbM_{s} := H^0(\Sigma, K_\Sigma).
 \] 
 of noncompact Calabi-Yau manifolds. 

 For future reference let us summarize some elementary geometric
 properties of the generic fiber 
 $\wX_\alpha$. Assume that $\alpha$ 
 has distinct simple zeroes.  
 Let $\pi_\alpha : E_\alpha \to \Sigma$ denote the projection map to
 $\Sigma$.  
 By construction, the restriction of $E_\alpha$ to a generic fiber
 $S_p$, $p\notin \hbox{div}(\alpha)$  
 is the unique (up to isomorphism) nontrivial extension 
 \[
 0\to \CO(-2) \to \CO(-1)\oplus \CO(-1) \to \CO\to 0 
 \] 
 of $\CO$ by $\CO(-2)$ over $\IP^1$. The restriction of $E_\alpha$ to a
 special fiber $S_p$ of the  
 ruling with $p\in\hbox{div}(\alpha)$ is the trivial extension 
 \[ 
 0\to \CO(-2) \to \CO(-2)\oplus \CO \to \CO\to 0 
 \] 
 Therefore the generic fibers of $\pi_\alpha:E_\alpha \to \Sigma$
  are isomorphic to the total space of the rank two bundle
 \[
 \CO(-1)\oplus \CO(-1) \to \IP^{1}, 
 \] 
 whereas the special fibers
  are isomorphic to the total space of the rank two bundle  
 \[ 
 \CO(-2)\oplus \CO \to \IP^1.
 \]
 It follows that the projective bundle $\IP(E_\alpha)$ is a projective
 quadric fibration  
 $\opi_\alpha :\IP(E_\alpha)\to \Sigma$ with generic fibers isomorphic
 to $\IF_0=\IP^1\times \IP^1$ and special fibers isomorphic to the
 Hirzebruch surface $\IF_2$.  

 Recall that the noncompact threefold $\wX_\alpha$ is the complement in
 $\oX_\alpha$ of the section at infinity
 $H_\alpha=\IP(K_S)$. From the point of view of the
 fibration structure over $\Sigma$, $H_\alpha$ intersects the generic
 $\IF_0$ fiber of $\opi_\alpha :\IP(E_\alpha)\to \Sigma$ along a
 $(1,1)$ curve and the special $\IF^2$ fibers along a section of
 $\IF^{2} \to \IP^{1}$ of self-intersection $+2$. Therefore the
 noncompact threefold $\wX_\alpha$ contains 
 $2g-2$ projective rational curves $C_1,\ldots, C_{2g-2}$ which can be
 identified with the negative sections of the special $\IF_2$ fibers. A
 straightforward local computation confirms that each of these curves is a
 $(-1,-1)$ curve on $\wX_\alpha$. Therefore for generic $\alpha$,
 $\wX_\alpha$ contains exactly $2g-2$ isolated $(-1,-1)$ curves as
 expected.

 If $\alpha$ is non generic, that is it has multiple zeroes, a similar
 analysis shows that $\wX_\alpha$ contains a projective rational curve
 for each zero of $\alpha$. However the curve corresponding to a
 double zero is a rigid $(0,-2)$ curve as opposed to a $(-1,-1)$
 curve as in the generic case.

 Next we will show that one can contract the exceptional curves
 constructed above on each $\wX_\alpha$ obtaining a singular threefold
 $X_{\alpha^2}$ which depends only on $\alpha^2\in
 H^0(\Sigma,K_{\Sigma}^{\otimes 2})$.  We claim that there exists a
 $\IP^{2}$-bundle $\opi_W:\oW\to \Sigma$, so that for any $\alpha \in
 H^0(\Sigma,K_\Sigma)$ there exists a canonical map
 $\phi_\alpha :\oX_\alpha \to \oW$ which contracts the exceptional
 curves. The image $\phi_\alpha(\oX_\alpha)$ is a singular hypersurface
 $\oX_{\alpha^2}$ in $\oW$ depending only on $\alpha^2$ sitting in a
 fixed (independent of $\alpha$) linear system on $\oW$.

 Moreover, there is a preferred hyperplane at infinity $h_\infty$ in
 $\oW$ so that  
 the restriction 
 \[ 
 \phi_\alpha\big|_{\wX_\alpha}: \wX_\alpha \to X_{\alpha^2}
 \] 
 is a small contraction of  $\wX_\alpha$ onto the noncompact nodal 
 Calabi-Yau threefold \[X_{\alpha^2} = \oX_{\alpha^2} \setminus 
 \left(\oX_{\alpha^2}\cap h_\infty\right).\]

 To prove this claim, take $\oW$ to be the $\IP^{3}$-bundle $\IP(S^2V
 \oplus \CO_\Sigma)$  
 over $\Sigma$, and take $h_\infty$ to be the hyperplane
 $\IP(S^2V)$. In order to construct the  
 map $\phi_\alpha: \oX_\alpha \to \oW$ it suffices to
 exhibit a line bundle  
 $\xi_\alpha$ on $\oX_\alpha$ so that 
 \[ 
 \opi_{\alpha*} \xi_\alpha \simeq  \left(S^2V\oplus \CO_\Sigma\right)^\vee = S^2(V^\vee)
 \oplus \CO_\Sigma.
 \]
 Let $\xi_\alpha$ be the relative hyperplane bundle $\xi_\alpha =
 \CO_{\oX_{\alpha}}(H_\alpha)$ for  
 $r_\alpha:\oX_\alpha\to S$. 
 The restriction of $\xi_\alpha$ to each fiber of 
 $r_\alpha: \oX_\alpha \to S$ is isomorphic to $\CO_{\IP^1}(1)$ and 
 $r_{\alpha*} \xi_\alpha \simeq E_\alpha^\vee$. 
 Moreover, the restriction of $\xi_\alpha$ to a generic $\IF_0$ fiber
 of the quadric fibration  
 $\opi_\alpha : \oX_\alpha \to \Sigma$ is isomorphic to
 $\CO_{\IF_0}(1,1)$ whereas the  
 restriction to a special $\IF_2$ fiber is isomorphic to
 $\CO_{\IF_2}(\Delta_{0})$ where  
 $\Delta_{0}$ denotes the zero section on $\IF_2$, $\Delta_{0}^2=2$.

 Let us compute 
 \[ 
 \opi_{\alpha*} \xi_\alpha = q_*(r_{\alpha*} \xi_\alpha) \simeq
 q_*E_\alpha^\vee. 
 \]
 By construction, $E_\alpha^\vee$ is an extension of the form 
 \[ 
 0\to \CO_S \to E_\alpha^\vee \to K_S^{-1} \to 0
 \] 
 on $S$. Taking direct images, we find the following extension on $\Sigma$ 
 \[ 
 0\to \CO_\Sigma \to q_*E_\alpha^\vee \to q_*(K_S^{-1})\to 0
 \] 
 where we have used $R^1q_*\CO_S =0$.
 Now $K_S\simeq \CO_S(-2)$, hence $q_*K_S^{-1}\simeq S^{2}V^\vee$. 
 Therefore we have an extension of the form 
 \be\label{eq:bundlextA} 
 0\to  \CO_\Sigma \to q_*E_\alpha^\vee \to S^{2}V^\vee \to 0
 \ee
 on $\Sigma$. In order to construct a contraction map $\phi_\alpha$ we
 have to prove that this  
 extension splits and we also have to choose a splitting. 

 First note that if $V$ is a stable rank two bundle on $\Sigma$, 
 \[
 \ext^1(\CO_\Sigma, S^2V) \simeq H^0(\Sigma, S^{2}V^\vee\otimes
 K_{\Sigma})^\vee =0  
 \] 
 since $S^2V^{\vee}\otimes K_\Sigma$ is a stable bundle with trivial
 determinant.   
 Therefore in that case, the extension \eqref{eq:bundlextA} splits. 
 If $V$ is semistable, but not stable, the extension group
 $\ext^1(\CO_\Sigma, S^2V)$ is not necessarily 
 trivial. However, we claim that the extension \eqref{eq:bundlextA} is
 still split for an arbitrary
 semistable bundle $V$. This claim is equivalent to the statement that
 the pushforward map  
 \[ 
 q_*:H^1(S,K_S) \to H^1(\Sigma, (q_*K_{S}^{-1})^\vee) 
 \] 
 is trivial. 
 Using Serre duality and respectively relative Serre duality for the
 map $q:S\to \Sigma$,  
 we obtain a dual map 
 \[ 
 q_*^\vee :H^0(\Sigma, q_*(K_{S/\Sigma}^{-1}))\to H^1(S,\CO_S)
 \simeq H^1(\Sigma, \CO_\Sigma) 
 \] 
 where $K_{S/\Sigma}$ is the relative dualizing sheaf for $q:S\to
 \Sigma$.  
 This map is the connecting homomorphism for the short exact sequence
 of sheaves 
 \be\label{eq:dualeulerA}
 0\to \CO_\Sigma \to q_*\left(q^*V\otimes \CO_S(1)\right)\to
 q_*(K_{S/\Sigma}^{-1})\to 0. 
 \ee
 obtained by pushing forward the dual of the relative Euler sequence on
 $S$. 
 We can easily compute the terms in the above exact sequence obtaining  
 \[ 
 0\to \CO_\Sigma \to V^\vee\otimes V \to \hbox{End}_0(V) \to 0
 \] 
 where $\hbox{End}_0(V)$ is the bundle of traceless endomorphisms of
 $V$. However this  
 sequence is canonically split, hence the connecting homomorphism
 vanishes. Therefore we  
 can conclude that the extension \eqref{eq:bundlextA} is
 split. Choosing a splitting we obtain  
 a (non-canonical) isomorphism 
 \[ 
 q_*(E_\alpha^\vee) \simeq S^2V \oplus \CO_\Sigma
 \]
 which defines a map $\phi_\alpha: \oX_\alpha \to \oW$ as claimed above. 
 Since the divisor $\oH_\alpha$ does not intersect the exceptional
 curves on $\oX_\alpha$,  
 it follows that these curves are contracted by
 $\phi_\alpha$. Therefore $\phi_\alpha$ maps  
 $\oX_\alpha$ onto a nodal hypersurface in $\oW$. 

 Next, we will show that the image of $\phi_\alpha$ moves in a fixed
 linear system on $\oW$.  
 Let us fix an $\alpha \in H^0(\Sigma,K_\Sigma)$. By construction, the
 image of $\phi_\alpha$  
 must be the zero locus of a section of a line bundle $\CL$ on $\oW$. 
 Since $\phi_\alpha: \oX_\alpha \to \oW$ is a small contraction, we can
 apply the adjunction formula  
 obtaining 
 \be\label{eq:adjunct}
 K_{\oX_\alpha} = \phi_\alpha^*(K_\oW\otimes \CL).
 \ee
 A routine computation using the relative Euler sequence yields 
 \be\label{eq:canonicalB}
 K_{\oX_\alpha}\simeq \xi_\alpha^{\otimes -2},\qquad 
 K_\oW\simeq \CO_\oW(-4)\otimes \opi_W^*K_{\Sigma}^{\otimes 2}.
 \ee
 By direct substitution in \eqref{eq:adjunct}, we obtain 
 \[ 
 \xi_\alpha^{\otimes -2} \simeq \xi_\alpha^{\otimes -4} \otimes
 \opi_\alpha^*K_\Sigma^{\otimes -2}\otimes  
 \phi_\alpha^*\CL. 
 \] 
 Since this equation is valid for any value of $\alpha$ it follows that 
 \be\label{eq:linsystA} 
 \CL \simeq \CO_\oW(2) \otimes \opi^*_W K_\Sigma^{\otimes 2}
 \ee
 Therefore we can conclude that the hypersurfaces
 $\phi_\alpha(\oX_\alpha)$ belong to  
 the linear system \linebreak 
 $|2h_\infty + \opi^*_W K_\Sigma^{\otimes 2}|$
 for any $\alpha$.  

 Let us compute the space of global sections of $\CL$. We have 
 \[ 
 \begin{aligned} 
 \opi_{\oW*}\CO_\oW(2) & = S^2(\CO_\Sigma\oplus S^2V^\vee) \cr
 & = \CO_\Sigma \oplus S^2 V^\vee \oplus S^2(S^2 V^\vee)\cr
 & = \CO_\Sigma \oplus S^2 V^\vee \oplus S^2(\Lambda^2 V^\vee )\oplus
 S^4 V^{\vee} \cr
 \end{aligned} 
 \] 
 hence, using the isomorphisms $\Lambda^2V \simeq K_\Sigma$, and
 $S^{4}V^{\vee} \otimes K_{\Sigma}^{\otimes 2} \cong S^{4}V\otimes
 K_{\Sigma}^{\otimes -2}$, 
 we find  
 \be\label{eq:linsystB}  
 \begin{aligned} 
 H^0(\oW,\CO_\oW(2) \otimes & \opi^*_W K_\Sigma^{\otimes 2})
 = \cr
 &  H^0(\Sigma, \CO_\Sigma)\oplus H^0(\Sigma,K_\Sigma^{\otimes 2})\oplus 
 H^0(\Sigma, S^2V) \oplus H^0(\Sigma, S^4V \otimes K_{\Sigma}^{\otimes -2}).\cr
 \end{aligned} 
 \ee

 Let us interpret the terms in the right hand side of equation
 \eqref{eq:linsystB}.  
 By construction, a non-zero element in $H^0(\Sigma, \CO_\Sigma)$
 viewed as a section in $\CO_\oW(2) \otimes \opi^*_W
 K_\Sigma^{\otimes 2}$ will vanish exactly along the image $\phi_0(\oX)$
 of the undeformed  
 threefold $\oX$ in $\oW$. This is the projective completion of the
 singular affine hypersurface  
 $X\subset W$ constructed in  section \ref{sec:singdefos} as the zero
 locus of the 
 determinant $\det_W \in H^0(W, \pi_W^* K_{\Sigma}^{\otimes 2})$. We
 will denote by $\det_{\oW}$ the section of $\CL$ whose zero locus is  
 $\oX$. The second term parameterizes hypersurface deformations of
 $\oX$ of the form  
 \[ 
 {\det}_{\oW} - \beta=0 
 \] 
 where $\beta\in H^0(\Sigma,\CO_\Sigma(2K_\Sigma))$ is a quadratic
 differential on $\Sigma$.  
 By construction, the defining equation of the hypersurface
 $\phi_\alpha(\oX_\alpha)$ is in the affine  
 subspace 
 \be\label{eq:affine}
 H^0(\Sigma, \CO_\Sigma) \oplus\{-\alpha^2\} \oplus H^0(\Sigma, S^2V).
 \ee
 All divisors in this affine subspace are isomorphic since any two
 divisors are related by a global automorphism of $W$ which is a
 translation by a section
 in $H^0(\Sigma, S^2V)$. In fact the affine space \eqref{eq:affine}
 also parameterizes the choice of a 
 splitting of the exact sequence \eqref{eq:bundlextA}. Therefore each
 point $a$ in this affine space  
 represents the image of $\oX_\alpha$ through a map $\phi_{\alpha, a}$
 which depends on the choice  
 of the splitting. In particular for some point $a$ we will obtain a
 hypersurface in $\oW$ with defining  
 equation 
 \[ 
  {\det}_{\oW}-\alpha^2=0 
 \] 
 We will denote this hypersurface by $\oX_{\alpha^2}$. 
 Finally, the last term in the right hand side of the decomposition
 \eqref{eq:linsystB} corresponds  
 other deformations of $\oX$ which will not be considered in this paper
 (note that for stable $V$ these extra deformations vanish). 

 This gives rise to the following picture. For a fixed point $s\in
 \bS\simeq\wbS$ we obtain a  
 linear deformation space $\bL_s=H^0(\Sigma, K_{\Sigma}^{\otimes 2})$ 
 parameterizing noncompact Calabi-Yau threefolds $X_\beta$ determined by
 equations of the form  
 \be\label{eq:bigeq} 
 {\det}_W - \pi_W^* \beta =0 
 \ee
 We also have a linear deformation space $\wbM_s=H^0(\Sigma,
 K_{\Sigma})$ parameterizing the  
 threefolds $\wX_\alpha$. Moreover we have a quadratic map 
 \be\label{eq:quadrmap}
 \Pi_s : \wbM_s \to \bL, \qquad \alpha \to \alpha^2 
 \ee 
 which corresponds to a small contraction of $\wX_\alpha$. The image of
 this quadratic map  
 is a singular subvariety in $H^0(\Sigma, K_\Sigma^{\otimes 2})$ isomorphic to 
 $H^0(\Sigma, K_\Sigma)/(\pm 1)$. 
 Note that $\bM_s, \bL_s, \wbM_s$ do not depend on the point $s\in \bS$. 

 Then we can construct the higher strata $\bM, \bL, \wbM$ as direct products 
 \be\label{eq:strataB} 
 \begin{aligned} 
 & \bM= \bS\times H^0(\Sigma, K_\Sigma)/(\pm 1)\cr
 & \wbM = \wbS \times  H^0(\Sigma, K_\Sigma)\cr
 & \bL= \bS \times H^0(\Sigma, K_\Sigma^{\otimes 2}).\cr
 \end{aligned} 
 \ee 
\

\bigskip

\begin{rem} \label{rem:balazs} The explicit geometric description of
  the non-compact Calabi-Yau spaces $X_{s}$, $X_{l}$ and
  $\widetilde{X}_{s}$ above can be extended to ``linear'' Calabi-Yau
  varieties with an arbitrary $ADE$ singularity along a curve
  $\Sigma$. More precisely suppose $R \subset SL(2,\mathbb{C})$ is a
  fixed finite subgroup. We now can look at the moduli space $\bS$ of
  pairs $(\Sigma,V)$, where $\Sigma$ is a smooth curve of genus $g
  \geq 2$ and $V \to \Sigma$ is a rank two holomorphic vector bundle
  which has canonical determinant and is equipped with a fiberwise
  $R$-action. Again for each $s = (\Sigma,V)$ we can form the
  non-compact Calabi-Yau variety $X_{s} = \op{tot}(V)/R$. The variety
  $X_{s}$ has a curve of singularities if type $R$ and canonical
  minimal crepant resolution $\widetilde{X}_{s}$. We can again look at
  the moduli spaces $\widetilde{\bM}$ and $\bL$ of $\widetilde{X}_{s}$
  and $X_{s}$ and try to describe them explicitly. A uniform
  description of these spaces for all groups $R$
  was given in \cite{S:artin}. Again, it turns out that $\bL$ and
  $\widetilde{\bM}$ are total spaces of vector bundles on
  $\bS$. In \cite{S:artin} Szendr\"{o}i identifies the fibers $\bL_{s}$
  and $\widetilde{\bM}_{s}$ over a point $s = (\Sigma,V)$ with the
  vector spaces
\[
\begin{split}
\widetilde{\bM}_{s} & = H^{0}(\Sigma,K_{\Sigma}\otimes \mathfrak{t})
\\
\bL_{s} & = H^{0}(\Sigma,(K_{\Sigma}\otimes \mathfrak{t})/W),
\end{split}
\]
where $\mathfrak{t}$ and $W$ denote the Cartan algebra and the Weyl
group of the complex $ADE$ group corresponding to $R$ under the McKay
correspondence.

Furthermore \cite{S:artin} describes explicitly the universal families
of deformations of $\widetilde{X}_{s}$ and $X_{s}$ over
$\widetilde{\bM}_{s}$ and $\bL_{s}$ and shows that $\bM_{s} \subset \bL_{s}$ is
naturally isomorphic to the cone 
\[
H^{0}(\Sigma,(K_{\Sigma}\otimes
\mathfrak{t}))/W \subset H^{0}(\Sigma,(K_{\Sigma}\otimes
\mathfrak{t})/W). 
\]

We will analyze the large $N$ physics of these more general
transversal geometries  in the forthcoming paper \cite{DDP} but for
now we concentrate on the case $R = \mathbb{Z}/2$.
\end{rem}

 \section{Intermediate Jacobians and Hitchin Pryms} \label{sec:jacobians}

 As we saw in the previous section the (normal to $\bS$) 
 loci $\bL_s\subset \bL$ are isomorphic to the base $H^0(\Sigma,
 K_{\Sigma}^{\otimes 2})$ of the $A_1$-Hitchin system on $\Sigma$.  This
 raises the question whether there is a more intrinsic geometric
 connection between our noncompact Calabi-Yau threefolds and Hitchin
 systems.  In this section we will give a positive answer to this
 question developing an intrinsic geometric relation between Hitchin
 Pryms and the intermediate Jacobians of the Calabi-Yau threefolds
 $X_\beta$, $\beta \in H^0(\Sigma,
 K_{\Sigma}^{\otimes 2})$. We begin with a brief review of the
 Hitchin system.  

 \subsection{Hitchin integrable systems and Pryms} \label{ss:pryms}

 For simplicity we will consider here only the $SL(2,\IC)$ Hitchin
 system which is relevant for our problem. Recall that an algebraically
 completely  integrable  Hamiltonian system (ACIHS)
 \cite{DM:spectral,DM:cubics} is defined by the following data \label{p:acihs}

 \begin{description}
 \item[(i)] a nonsingular complex algebraic variety $\CN$ equipped with
 a non-degenerate global holomorphic $(2,0)$ form $\sigma$
 \item[(ii)] a projection $\bh:\CN\to \CB$ where $\CB$ is a
   nonsingular complex algebraic variety so  
 that the fibers $\CN_\beta$ of $\bh$ are abelian varieties satisfying
 \item[(iii)] $\CN_\beta$ is a Lagrangian subvariety of $\CN$ for any
   point $\beta\in \CB$.  
 \end{description}

 \

 Let us now recall the construction of the Hitchin integrable system
 \cite{NH:stable} following \cite{DM:spectral, DM:cubics}.  A
 $SL(2,\IC)$ Higgs bundle $(E,\phi)$ on $\Sigma$ consists of a rank
 two holomorphic bundle on $\Sigma$ with trivial determinant and a
 global section $\phi \in H^0(\Sigma, \op{End}_0(E) \otimes
 K_{\Sigma})$. Here $\op{End}_0(E)$ denotes the bundle of traceless
 endomorphisms of $E$.  Such a pair is called stable (semistable) if
 there are no $\phi$-invariant subbundles of $E$ that violate the
 usual slope inequality \cite{NH:stable}.  In this case, there exists
 a quasi-projective moduli variety $\CM := \CM_{SL(2,\IC)}$ of
 semistable Hitchin pairs of complex dimension $6g - 6$. More
 generally \cite{NH:stable}, we can consider $G$ Higgs bundles for a
 general complex reductive group $G$. By definition these are
 semistable pairs $(P,\phi)$ with $P$ a principal $G$-bundle on
 $\Sigma$ and $\phi \in H^{0}(\Sigma, \op{ad}(P)\otimes
 K_{\Sigma})$. Again there is a quasi-projective moduli space of such
 Higgs bundles and much of the discussion below generalizes to these
 moduli spaces. For the purposes of this paper we will ignore these
 more general moduli spaces with one exception, namely the moduli
 space $\CM_{{\mathbb P}GL(2,\IC)}$ of topologically trivial ${\mathbb
 P}GL(2,\IC)$ Higgs bundles on $\Sigma$. As we will see below, the
 spaces $\CM_{SL(2,\IC)}$ and $\CM_{{\mathbb P}GL(2,\IC)}$ are closely
 related. They  are the only moduli spaces of Hitchin pairs corresponding
 to a structure group of type $A_{1}$ and will not reappear in the
 next section as families of intermediate Jacobians for Calabi-Yaus in
 the moduli space $\bL$.

 The key element in the construction of the ACIHS is the notion of a
 spectral cover introduced in \cite{NH:stable}. The spectral cover
 $p_{\beta} : \wsigma = \wsigma_{\beta} \to \Sigma$ of a
 pair $(E,\phi)$ is a curve in the total space of the cotangent bundle
 $T^*\Sigma$ of $\Sigma$ defined by the eigenvalue equation
 \be\label{eq:spectralA} 
 \det(y\cdot \op{id} -p^{*}\phi)=0.  
 \ee 
 Here $p : T^{*}\Sigma \to \Sigma$ is the natural projection and $y \in
 H^{0}(T^{*}\Sigma, p^{*}T^{*}\Sigma)$ is the tautological section. 
 Since $\phi$ is a
 traceless endomorphism of $E$, this equation can be rewritten as
 \be\label{eq:specialB} y^2-\beta=0 \ee 
 where
 $\beta=\det(\phi)\in H^0(\Sigma, K_{\Sigma}^{\otimes 2})$ is the
 determinant of $\phi$. This shows that the spectral cover $\wsigma$ is
 smooth reduced and irreducible if and only if $\beta$ has distinct
 simple zeroes. Moreover, $\widetilde{\Sigma}$ is invariant under the
 holomorphic 
 involution $\iota: T^*\Sigma \to T^*\Sigma$ which acts by
 multiplication by $(-1)$ on each fiber.

 According to \cite{NH:stable}, the map 
 \be\label{eq:projmapA}
 \bh:\CM \to H^0(\Sigma, K_{\Sigma}^{\otimes 2}),\qquad 
 (E,\phi)\to \det(\phi)
 \ee
 is proper and surjective. We will denote by $B=H^0(\Sigma,
 K_{\Sigma}^{\otimes 2})$ 
 the space of quadratic differentials on $\Sigma$ and by 
 $\CB \subset B$ the open subset 
 consisting of quadratic differentials $\beta$ with simple zeroes. Let
 $\CN = \bh^{-1}(\CB)$  
 denote the inverse image of $\CB$ in $\CM$. 

 In order to determine the fiber of $\bh$ at a point $\beta \in \CB$,
 note that a pair $(E,\phi)\in \CN$ gives rise to a pair $(\wsigma,
 L)$ where $L$ is a complex holomorphic line bundle on $\wsigma$ of
 degree $2g-2$. $L$ is defined by the property that the fiber
 $L_\lambda$ over a point $\lambda\in \wsigma$ is the eigenspace of
 $\phi$ corresponding to the eigenvalue $\lambda$.  Since $E$ is a
 $SL(2,\IC)$ rather than $GL(2,\IC)$ bundle, the resulting line bundle
 $L\to \Sigma_\beta$ satisfies the equivariance condition
 \be\label{eq:eqcondA} \iota^*L = L^\vee\otimes
 p_{\beta}^{*}K_{\Sigma}.  \ee One can show that there is a one-to-one
 correspondence between pairs $(E,\phi)$ in $\CN$ with fixed spectral
 cover $\wsigma$ and line bundles $L$ on $\wsigma$ satisfying
 condition \eqref{eq:eqcondA}.  This correspondence can be extended to
 singular, reducible or non-reduced spectral covers by allowing $L$ to
 be a torsion free rank one sheaf on $\wsigma$.

 This shows that the fiber $\CN_\beta = \bh^{-1}(\beta)$ is isomorphic
 to the subvariety of the Picard variety
 $\hbox{Pic}^{2g-2}(\wsigma_\beta)$ of line bundles on $\wsigma$ of
 degree $2g-2$. This subvariety is defined by the equation
 \eqref{eq:eqcondA}. Equivalently it can be identified with the fiber
 $\op{Nm}^{-1}([K_{\Sigma}])$ of the norm map $\op{Nm} :
 \op{Pic}^{2g-2}(\wsigma) \to \op{Pic}^{2g-2}(\Sigma)$ over the
 canonical point $[K_{\Sigma}] \in \op{Pic}^{2g-2}(\Sigma)$. According to
 \cite{RD:decomp}, the degree zero Picard $\hbox{Pic}^0(\wsigma)$
 decomposes up to isogeny into a direct product of Abelian varieties
 \[ 
 \hbox{Pic}^0(\wsigma)\simeq
 \hbox{Pic}^0(\wsigma)^{+}\times
 \hbox{Pic}^{0}(\wsigma)^{-} 
 \] 
 where $\hbox{Pic}^{0}(\wsigma)^{+}$ is the fixed locus of the inversion
 involution $L \mapsto L^{\vee}$ on $\hbox{Pic}^{0}(\wsigma)$ and
 $\hbox{Pic}^0(\wsigma_\beta)^{-}$  is the anti-invariant part
 of this involution. The abelian subvariety
 $\hbox{Pic}^0(\wsigma_\beta)^{-}$ is usually denoted by
 $\hbox{Prym}(\wsigma/\Sigma)$ and is called the Prym variety of the 
 spectral cover. The natural translation action of
 $\hbox{Pic}^{0}(\wsigma)$  on $\hbox{Pic}^{2g-2}(\wsigma)$ intertwines
 the inversion involution with the involution $\iota$ and realizes the
 Hitchin fiber $\CN_{\beta} \subset \op{Pic}^{2g-2}(\wsigma)$ as a
 principal homogeneous space over the Prym variety
 $\op{Prym}(\wsigma/\Sigma) \subset \op{Pic}^{0}(\wsigma)$. So
 $\CN_{\beta}$ is (non-canonically) isomorphic to
 $\op{Prym}(\wsigma/\Sigma)$. 

 Since we have an isomorphism $\op{Pic}^{0}(\wsigma) \to J(\wsigma)$
 determined by the Abel-Jacobi map,  
 it follows that the Jacobian 
 \[ 
 J(\wsigma) = H^{0}(\wsigma,\Omega^1_\wsigma)^\vee/H_1(\wsigma,\IZ) 
 \] 
 also decomposes up to isogeny into a direct sum of invariant and
 anti-invariant parts.  
 The Abel-Jacobi map maps the  
 Prym to the abelian subvariety of $J(\wsigma)^{-} \subset J(\wsigma)$,
 given by 
 \be \label{eq:prym}
 J(\wsigma)^{-}={(H^{0}(\wsigma,\Omega^1_\wsigma)^{-})}^\vee /
 H_1(\wsigma,\IZ)^{-} 
 \ee 
 where the superscript ``$-$'' denotes the anti-invariant part. Note that
 \[
 \begin{split}
 H^{0}(\wsigma,K_{\wsigma}) & = H^{0}(\wsigma,
 p^{*}K_{\Sigma}^{\otimes 2}) \\ & 
 = H^{0}(\Sigma,
 K_{\Sigma}^{\otimes 2}\otimes p_{*}{\mathcal
   O}_{\wsigma})  \\
 & =
 H^{0}(\Sigma,K_{\Sigma})\oplus H^{0}(\Sigma,K_{\Sigma}^{\otimes 2}).
 \end{split}
 \]
 Under this isomorphism, the space $H^0(\wsigma,\Omega^1_\wsigma)^{-} =
 H^{0}(\wsigma,K_{\wsigma})^{-}$ of anti-invariant
 holomorphic differentials on $\wsigma$ is identified with $H^0(\Sigma,
 K_{\Sigma}^{\otimes 2})$. Therefore 
 $\dim \CN_{\beta} = \dim \hbox{Prym}(\wsigma/\Sigma)$ equals the
 dimension $3g-3$ of the space of 
 quadratic differentials on $\Sigma$.

 So far we have constructed a family $\bh:\CN\to\CB$ of Abelian
 varieties over the open subset $\CB\subset B$ so that the dimension of
 the fibers equals the dimension of the base. To complete the data of
 an ACIHS we have to construct a holomorphic symplectic form $\sigma$
 on $\CN$ so that the fibers $\CN_\beta$ are Lagrangian cycles with
 respect to $\sigma$. This can be seen easily by noticing that the
 cotangent space to the moduli space of rank two stable bundles with
 trivial determinant at a point $E$ is given by the cohomology group
 $H^{0}(\Sigma,\op{End}_{0}(E)\otimes K_{\Sigma})$. In other words the
 total space of the cotangent bundle to the moduli of stable bundles
 (of rank two and with trivial determinant) is a Zariski open and dense
 set in $\CM$. Since $\CN \subset \CM$ is also Zariski open and dense
 and since cotangent bundles are naturally symplectic, we obtain a
 holomorphic symplectic form defined on an open dense set in
 $\CN$. Hitchin showed \cite{NH:selfdual} that this
 form extends to a holomorphic 
 symplectic form on all of $\CN$ and argued that all the fibers
 $\CN_{\beta} = 
 \bh^{-1}(\beta)$ are necessarily Lagrangian (see also
 \cite{NH:stable,DM:spectral}).   

 \

 \begin{rem} \label{rem:cubic} 
 The holomorphic symplectic structure $\sigma$ admits also an explicit
 interpretation in terms of the spectral data $(\wsigma,L)$. We recall this
 interpretation since it has a direct physical significance: it is
 related to the natural special K\"{a}hler geometry on the base
 $\mycal{B}$ of the  ACIHS $\bh : \CN \to \mycal{B}$. 
 Let us denote by $\CV$ the vector bundle over $\CB$ whose sections are
 vertical vector fields  
 on $\CN$ which are constant on each torus fiber, i.e. $\CV =
 \bh_{*}T_{\CN/\CB}$.  Note that the fiber $\CV_\beta$ is isomorphic to
 the  
 space ${(H^{0}(\wsigma,\Omega^1_\wsigma)^{-})}^\vee$, which is
 isomorphic in turn to $B^{\vee} = 
 H^0(\Sigma,K_{\Sigma}^{\otimes 2})^{\vee}$. 
 On the other hand 
 the tangent space $\CT_\beta\CB$ to the base at the point $\beta$ is
 isomorphic  
 to $B =H^0(\Sigma,K_{\Sigma}^{\otimes 2})$.
 Therefore we have the isomorphisms 
 \[ 
 \CT^{\vee}_\CB \simeq \CV \simeq B^{\vee}\otimes \CO_\CB.
 \]  
 The integrable structure can be characterized by the 
 cubic criterion of Donagi and Markman \cite{DM:spectral,DM:cubics}. 
 Let us choose a marking of the Abelian varieties $\CN_\beta$, i.e. a
 continuously varying symplectic  
 basis of $H_1(\CN_\beta, \IZ)$, $\beta\in \CB$. For example we can
 choose a symplectic basis of  
 anti-invariant cycles in $H_1(\wsigma_\beta,\IZ)^-$, for $\beta \in \CB$. 

 Then, locally on $\CB$ the family $\CN\to \CB$ determines a period map
 $\varrho:\CB \to 
 {\mathbb H}_{3g-3}$  
 where ${\mathbb H}_{3g-3}$ denotes the Siegel upper half space 
 \[ 
 {\mathbb H}_{3g-3} =\{(3g-3)\times (3g-3)\ \hbox{symmetric complex
   matrices {\it Z} with}\ \hbox{Im}(Z) >0\} 
 \] 
 Note that we can identify ${\mathbb H}_{3g-3}$ with a subspace of 
 $S^2B$ by choosing a 
 basis of holomorphic quadratic differentials on $\Sigma$. 
 Then the following conditions are equivalent
 \cite[Lemma~7.4]{DM:spectral} 

 \

 \begin{description}
 \item[(i)] There exists a holomorphic symplectic form $\sigma$ on
   $\CN$ so 
 that the fibers of $\bh:  
 \CN\to \CB$ are Lagrangian, and $\sigma$ induces the identity isomorphism 
 \[ 
 \hbox{Id}\in \Hom(\CT_{\CN/\CB}, \bh^*\CT^{\vee}_\CB)\simeq 
 \Hom(\bh^*\CV,\bh^*\CV). 
 \]
 \item[(ii)] The period map $\varrho:\CB \to S^2V$ can be locally written in
 $\CB$ as the Hessian   
 of a holomorphic function on $\CB$ (the holomorphic prepotential.)
 \item[(iii)] The differential of the period map $d\varrho\in
   \Hom(\CT_\CB, S^2\CV)  
 \simeq \CV\otimes S^2\CV$
 is a section of $S^3\CV$ (the cubic condition). 
 \end{description}

 \

 \noindent
 For the family $\CN\to \CB$ of $SL(2,\IC)$-Hitchin Pryms the cubic
 $d\varrho$ can be computed explicitly and is given by
 \be\label{eq:cubicA} d\varrho_{\beta} : S^{3}(\CT_\beta\CB) \to \IC,
 \qquad \gamma_{1}\cdot\gamma_{2}\cdot\gamma_{3} \to \hbox{Res}^2\left(
 \frac{\gamma_{1}\cdot\gamma_{2}\cdot\gamma_{3}}{\beta^2}\right).  \ee
 where
 $\hbox{Res}^2\left(
 \frac{\gamma_{1}\cdot\gamma_{2}\cdot\gamma_{3}}{\beta^2}\right)$
 is the quadratic residue of a quadratic differential.
 \end{rem}

 \

 \medskip

 \noindent
 For future reference we recast the Hodge theoretic interpretation
 \eqref{eq:prym} of the spectral Prym in terms of data on the base
 curve $\Sigma$. Recall that $\op{Prym}(\wsigma/\Sigma)$ was naturally
 identified with the kernel of the norm map $\op{Nm} :
 \op{Pic}^{0}(\wsigma) \to \op{Pic}^{0}(\Sigma)$ between the degree
 zero Picard varieties of $\wsigma$ and $\Sigma$. Topologically we have
 identifications $\op{Pic}^{0}(\wsigma) = H_{1}(\wsigma,{\mathbb
 Z})\otimes ({\mathbb R}/{\mathbb Z})$ and $\op{Pic}^{0}(\Sigma) =
 H_{1}(\Sigma,{\mathbb Z})\otimes ({\mathbb R}/{\mathbb Z})$. In fact
 these identifications can be thought of as isomorphisms of
 abelian varieties if we endow the right hand sides with the complex
 structures coming from the Hodge structures on the first homology of
 $\wsigma$ and $\Sigma$. This gives
 a natural topological identification
 \be \label{eq:trace}
 \begin{split}
 \op{Prym}(\wsigma/\Sigma) & = \ker\left[ \op{Pic}^{0}(\wsigma)
   \stackrel{\op{Nm}}{\to} \op{Pic}^{0}(\Sigma)\right] \\
 & = \ker\left[ H_{1}(\wsigma,{\mathbb Z}) {\to}
   H_{1}(\Sigma,{\mathbb Z})\right] \otimes_{\mathbb Z} ({\mathbb
   R}/{\mathbb Z}) \\
 & = H^{1}(\Sigma,\CK)\otimes_{{\mathbb Z}} ({\mathbb
   R}/{\mathbb Z}),
 \end{split}
 \ee
 where 
 \[
 \CK := \ker\left(p_{\beta*}{\mathbb Z}_{\wsigma} \stackrel{\op{Tr}}{\to}
     {\mathbb Z}_{\Sigma}\right) 
 \]
 is the kernel of the natural trace map. In the last step
 of \eqref{eq:trace}, we used the long exact sequence:
 \[
 0 \to H^0(\Sigma,{\mathbb Z}/n) \to H^1(\Sigma,\CK) \to
 H^1(\wsigma,{\mathbb Z}) \to H^1(\Sigma,{\mathbb Z})
 \]
 which implies that $H^1(\Sigma,\CK)$ agrees with
 $\ker(H_1(\wsigma,{\mathbb Z}) \to
 H_1(\wsigma,{\mathbb Z}))$ up to torsion, which disappears when
 tensoring with ${\mathbb R}/{\mathbb Z}$.

 Alternatively, the sheaf $\CK$
 can be described as follows.  Write $\bB \subset \Sigma$ and $\bR
 \subset \widetilde{\Sigma}$ for the branch and ramification divisors
 of $p_\beta$. Let $\Sigma^{o} := \Sigma - \bB$, $\widetilde{\Sigma}^{o} :=
 \Sigma - \bR$, and let $p_\beta^{o} := \widetilde{\Sigma}^{o} \to
 \Sigma^{o}$ be the restriction of the projection. If we now denote by
 $j : \Sigma^{o} \hookrightarrow \Sigma$, then $\CK$ can be viewed as
 $j_{*}$ of the local system on $\Sigma^{o}$ of anti-invariant
 ${\mathbb Z}$-valued functions on $\widetilde{\Sigma}^{o}$. More
 invariantly, we have 
 \begin{equation} \label{eq:K}
 \CK = (p_{\beta*}\Lambda_{\op{r}})^{W} =
 j_{*}((p^{o}_{\beta*}\Lambda_{\op{r}})^{W}), 
 \end{equation}
 where $\Lambda_{\op{r}}$ and $W$ denote the root lattice and Weyl
 group of of $SL(2,{\mathbb C})$ respectively, and we view the covering
 involution of $\wsigma \to \Sigma$ as the generator of $W$.

\

The analysis of the moduli space $\CM_{{\mathbb P}GL(2,\IC)}$ of
topologically trivial ${\mathbb P}GL(2,\IC)$ Higgs bundles is
similar. In fact, the moduli space $\CM$ determines $\CM_{{\mathbb
P}GL(2,\IC)}$. To see this we first note that with any Hitchin pair
$(E,\phi)$ consisting of a rank two vector bundle $E$ with trivial
determinant and a Higgs field $\phi : E \to E\otimes K_{\Sigma}$ gives
rise to a ${\mathbb P}GL(2,\IC)$ Hitchin pair
$(P_{E},\op{ad}_{\phi})$, where $P_{E}$ is the ${\mathbb
P}GL(2,\IC)$-bundle associated with the frame bundle of $E$ via the
adjoint representation of $SL(2,{\mathbb C})$. It is easy to check
that this procedure preserves semistability and so one gets a well
defined morphism  $\bbad : \CM \to \CM_{{\mathbb
P}GL(2,\IC)}$. Furthermore, since a principal ${\mathbb P}GL(2,\IC)$
bundle can be viewed as a ${\mathbb P}^{2}$ bundle, and since on a
curve all projective bundles are projectivizations of vector bundles,
we see that the morphism $\CM \to \CM_{{\mathbb P}GL(2,\IC)}$ is
surjective. A more careful analysis (see e.g. \cite{TH}) shows that
$\CM_{{\mathbb P}GL(2,\IC)}$ is in fact the quotient of $\CM$ by the
finite group $H^{1}(\Sigma,{\mathbb Z}/2) = \op{Pic}^{0}(\Sigma)[2]$
of $2$-torsion line bundles on $\Sigma$. Here an element $\xi \in
\op{Pic}^{0}(\Sigma)[2]$ acts as $(E,\phi) \mapsto (E\otimes \alpha,
\phi\otimes \op{id})$ and so this action preserves the fibers
$\CN_{\beta}$ of the Hitchin map $\bh$ and the symplectic form on
$\CM$. Thus $\CM$ and $\CM_{{\mathbb P}GL(2,\IC)}$ are ACIHS which are
fibered by Lagrangian tori over the same base space $B =
H^{0}(\Sigma,K_{\Sigma}^{\otimes 2})$ and related by a finite map that
respects the symplectic forms and the Lagrangian fibrations:
\[
\xymatrix{
\CM \ar[rr]^-{\bbad} \ar[dr]_-{\bh} & &\CM_{{\mathbb
P}GL(2,\IC)} \ar[dl]^-{\bh} \\ 
& B &
}
\]
In particular, the finite map $\bbad$ gives an explicit identification
of the fiber of the ${\mathbb P}GL(2,\IC)$ Hitchin map over $\beta \in
\CB$ with the quotient $\CN_{\beta}/\op{Pic}^{0}(\Sigma)[2]$. Thus the
fibers of the $SL(2,\IC)$ and ${\mathbb P}GL(2,\IC)$ Hitchin maps over
the same point $\beta \in \CB$ are isogenous abelian varieties.

 \

 This concludes our review of the $A_1$-Hitchin system. Next, we will
 explain the connection  
 between this ACIHS and the family of noncompact threefolds constructed
 in the previous section.  

 \subsection{Intermediate Jacobians and the Calabi-Yau integrable
   system}  \label{ss:IJ} 

 Let us start with some general considerations regarding Jacobian
 fibrations and integrable systems associated to families of Calabi-Yau
 threefolds \cite{DM:spectral,DM:cubics}.

 Suppose $\CB$ is a component of an enlarged moduli space of smooth
 projective Calabi-Yau threefolds supporting a universal family $\CX\to
 \CB$. Recall that the enlarged moduli space parameterizes Calabi-Yau
 threefolds together with a choice of a nontrivial global holomorphic
 three-form.  The family $\CX\to \CB$ determines a complex torus
 fibration $\CJ \to \CB$ whose fiber over a point $\beta \in
 \CB$ is the intermediate Jacobian 
 \be\label{eq:intjacA}
 J_{3}(X_\beta)=F^1H^3(X_\beta,{\mathbb C})^\vee/H_3(X_\beta, \IZ).  
 \ee 
 Here
 \[ 
 F^1H^3(X_\beta) = H^{3,0}(X_\beta,{\mathbb C}) \oplus H^{2,1}(X_\beta) 
 \] 
 denotes the first step in the Hodge filtration on the third cohomology of
 $X_\beta$.  The intermediate Jacobian has a natural
 non-degenerate indefinite polarization corresponding to the
 intersection pairing on $H_3(X_\beta, \IZ)$.
 Therefore $J_{3}(X_\beta)$ are complex analytic tori, but not Abelian
 varieties. It is also possible to describe the intermediate Jacobian
 without passing to quotients. Namely we have a natural identification:
 \[
 J_{3}(X_\beta) = H_3(X_\beta, \IZ)\otimes_{\mathbb Z} S^{1},
 \]
 where $S^{1}$ is the circle $S^{1} = {\mathbb R}/{\mathbb Z}$.  This
 point of view is particularly good for studying the algebraic
 properties of $J_{3}(X_\beta)$ as a torus but the complex structure on
 $J_{3}(X_\beta)$ is disguised in this interpretation.

 According to \cite{DM:spectral,DM:cubics} the resulting fibration $\CJ
 \to \CB$ underlies an analytically completely integrable Hamiltonian
 system. This is very similar to the ACIHS structure encountered in the
 previous section, except that we have to employ analytic spaces
 instead of algebraic varieties in the defining properties {\bf (i)-(iii)}
 listed on page \pageref{p:acihs}.  The existence of an analytic integrable
 structure follows again from the cubic criterion of
 \cite{DM:spectral,DM:cubics}. The required cubic form is in this case
 the normalized Yukawa coupling (see for example \cite{DM:guide}.)

 In the section \ref{ss:higher} we have constructed a family of noncompact
 Calabi-Yau threefolds $\CX\to \bL$ parameterized by $\bL=\bS\times B$
 where $\bS$ is the moduli space of rank two bundles on $\Sigma$ with
 canonical determinant, and $B = H^0(\Sigma, K_\Sigma^{\otimes 2})$.
 The threefold $X_l$, $l=(s,\beta)\in \bS\times B$ is smooth if and
 only if $\beta\in B$ has distinct simple zeroes, that is if and only
 if $\beta\in \CB$. Since the dependence on the point $s\in \bS$ is
 inessential, throughout this section we will consider the family
 obtained by restricting $\CX\to \bL$ to a subspace of the form
 $\{s\}\times B \subset \bS\times B$. From now on, until the end of
 this section,  we
 will drop the point $s$ from the labeling. Moreover, abusing notation
 we will denote by $\CX\to \CB$ the restriction of the family $\CX$
 to the open subset $\CB \subset \{s \}\times B$ parameterizing smooth
 threefolds.

 Our main goal is to establish a connection between the intermediate
 Jacobian fibration of the family $\CX\to \CB$ and the Hitchin
 integrable system.  In contrast with the case of compact threefolds
 described above, there is no general result concerning the existence
 of an integrable structure on the intermediate Jacobian fibration of a
 family of noncompact Calabi-Yau manifolds.  In general, these
 Jacobians are noncompact and may not even have the same dimension as
 the base of the fibration.

 However in our case the fibers of the family $\CX \to \CB$ are simple
 enough so that we can analyze their intermediate Jacobians in
 detail. Fix a point $\beta \in \CB$ and let $X_{\beta}$ be the
 corresponding smooth non-compact Calabi-Yau threefold. The
 intermediate Jacobians of $X_{\beta}$ are Hodge theoretic invariants
 of the complex structure of $X_{\beta}$. They are generalized tori (=
 quotients of a vector space by a discrete abelian subgroup) defined in
 terms of the mixed Hodge structure on the cohomology or the homology
 of $X_{\beta}$. More precisely, by the work of Deligne
 \cite{deligne-hodge2} we know that the abelian group $H =
 H^{k}(X_{\beta},{\mathbb Z})$ (respectively $H =
 H_{k}(X_{\beta},{\mathbb Z})$) is equipped with a mixed Hodge
 structure $(W_{\bullet},F^{\bullet})$, where $W_{\bullet}$ is an
 increasing {\em weight} filtration on the rational vector space
 $H_{{\mathbb Q}} := H\otimes
 {\mathbb Q}$ and $F^{\bullet}$ is a decreasing {\em Hodge} filtration on
 the complex vector space $H_{{\mathbb C}} :=H\otimes {\mathbb C}$.

 The weight and Hodge filtrations should be compatible in the sense
 that $F^{\bullet}$ induces a Hodge decomposition of weight $\ell$ on
 the $\ell$-th graded piece of $\op{gr}^{W}_{\ell}  = W_{\ell}/W_{\ell
   - 1}$ of the weight filtration. This means that
 $\op{gr}^{W}_{\ell}\otimes {\mathbb C} = \oplus_{p+q = \ell} H^{p,q}$,
 where 
 \[
 H^{p,q} = [(F^{p}W_{\ell} + W_{\ell -1})/W_{\ell -1}]\cap
 [(\overline{F}^{q}W_{\ell} + W_{\ell -1})/W_{\ell -1}].
 \]
 Equivalently the three filtrations $W_{\bullet}$, $F^{\bullet}$ and
 $\overline{F}^{\bullet}$ should satisfy
 \[
 \op{gr}_{F}^{p}\op{gr}_{\overline{F}}^{q}\op{gr}_{\ell}^{W}H = 0, \quad
 \text{ for all } p+q \neq \ell.
 \]
 Given a mixed Hodge structure $(H,W_{\bullet},F^{\bullet})$ we can
 consider the smallest interval $[a,b]$ such that $\op{gr}^{W}_{\ell} =
 0$ for $\ell \notin [a,b]$. The integer $b-a$ is called {\em length}
 of the mixed Hodge structure and $a$ and $b$ are the {\em lowest} and
 {\em highest} weight respectively. A mixed Hodge structure of length
 zero is pure of weight $a=b$.

 \

 \medskip

 \noindent
 With every mixed Hodge structure $(H,W_{\bullet},F^{\bullet})$
 one associates a sequence of intermediate Jacobians. If
 $(H,W_{\bullet},F^{\bullet})$ is a mixed Hodge structure and $p$ is
 any integer satisfying
 \[
 p > \frac{1}{2}(\text{highest weight of } (H,W_{\bullet},F^{\bullet})),
 \]
 then the level $p$ intermediate Jacobian of $(H,W_{\bullet},F^{\bullet})$ is
 \[
 H_{{\mathbb C}}/(F^{p}H_{{\mathbb C}} + H).
 \]
 The condition on $p$ here is imposed to ensure that $H_{{\mathbb Z}}$
 projects to a discrete subgroup in $H_{{\mathbb C}}/F^{p}H$, i.e. that
 the Jacobian is a generalized torus.

  Since $X_{\beta}$ is non-compact we will
 have to take extra care in distinguishing the intermediate Jacobians
 associated with the mixed Hodge structures on
 $H^{3}(X_{\beta},{\mathbb Z})$ and $H_{3}(X_{\beta},{\mathbb Z})$. We
 will denote these generalized tori by $J^{3}(X_{\beta})$ and
 $J_{3}(X_{\beta})$ respectively. Explicitly
 \begin{align}
 J^{3}(X_{\beta})  & = H^{3}(X_{\beta},{\mathbb
   C})/(F^{2}H^{3}(X_{\beta},{\mathbb C}) + H^{3}(X_{\beta},{\mathbb
   Z})), \label{eq:cohomology_jacobian}\\
 J_{3}(X_{\beta})  & = H_{3}(X_{\beta},{\mathbb
   C})/(F^{-1}H_{3}(X_{\beta},{\mathbb C}) + H_{3}(X_{\beta},{\mathbb
   Z})), \\
 & = H^{3}(X_{\beta},{\mathbb
   C})/(F^{2}H^{3}(X_{\beta},{\mathbb C}) + H_{3}(X_{\beta},{\mathbb
   Z})), \label{eq:homology_jacobian} 
 \end{align}
 where in the  formula \eqref{eq:homology_jacobian} the inclusion
 $H_{3}(X_{\beta},{\mathbb  
 Z})/(\text{torsion}) \hookrightarrow H^{3}(X_{\beta},{\mathbb C})$ is
 given by 
 the intersection pairing map on three dimensional cycles in
 $X_{\beta}$.  More precisely, by the universal coefficients theorem
 we can identify $H^{3}(X_{\beta},{\mathbb Z})/(\text{torsion})$ with
 the dual lattice $H_{3}(X_{\beta},{\mathbb Z})^{\vee} :=
 \op{Hom}_{{\mathbb Z}}(H_{3}(X_{\beta},{\mathbb Z}),{\mathbb
   Z})$. Combining this identification with the intersection pairing on
 the third homology of $X_{\beta}$ we get a well defined map
 \[
 \xymatrix@R-2pc{i : \!\!\!\!\!\! & H_{3}(X_{\beta},{\mathbb Z}) \ar[r] &
 H^{3}(X_{\beta},{\mathbb
   Z})/(\text{torsion}) \\ 
 & L \ar@{|->}[r] & (L,\bullet)
 }
 \]
 which is injective on the free part of $H_{3}(X_{\beta},{\mathbb
   Z})$. Combining $i$ with the natural inclusion $H^{3}(X_{\beta},{\mathbb
   Z})/(\text{torsion}) \subset H^{3}(X_{\beta},{\mathbb
   C})$ we obtain the map appearing in
   \eqref{eq:homology_jacobian}. Furthermore since $i$ is injective
   modulo torsion, it follows that the induced surjective map on intermediate
   Jacobians 
 \begin{equation}
 J_{3}(X_{\beta}) \to J^{3}(X_{\beta}) \label{eq:compare_jacobians}
 \end{equation}
 is a finite isogeny of generalized tori. Note that when $X_{\beta}$ is
 compact the unimodularity of the Poincare pairing implies that
 \eqref{eq:compare_jacobians} is an isomorphism and so we do not have
 to worry about the distinction between $J_{3}(X_{\beta})$ and
 $J^{3}(X_{\beta})$. In fact, we will see below that for our
 non-compact $X_{\beta}$, $\beta \in \CB$ the mixed Hodge structure on
 $H^{3}(X,{\mathbb Z})$ is actually pure and of weight $3$. This
 implies that 
 \[
 \begin{split}
 J_{3}(X_{\beta}) & = H_{3}(X_{\beta},{\mathbb Z})\otimes_{{\mathbb Z}}
 S^{1} \\
 J^{3}(X_{\beta}) & = H^{3}(X_{\beta},{\mathbb Z})\otimes_{{\mathbb Z}}
 S^{1}
 \end{split}
 \]
 and so the two Jacobians are compact complex tori. Furthermore the
 isogeny \eqref{eq:compare_jacobians} can be identified explicitly as 
 \[
 \xymatrix@R-1pc{
 J_{3}(X_{\beta}) \ar[r] \ar@{=}[d] & J^{3}(X_{\beta}) \ar@{=}[d] \\
 H_{3}(X_{\beta},{\mathbb Z})\otimes_{{\mathbb Z}} S^{1}
 \ar[r]_-{i\otimes \op{id}} & 
 H^{3}(X_{\beta},{\mathbb
   Z})\otimes_{{\mathbb Z}} S^{1}
 } 
 \]
 However we will also check that the map $i\otimes \op{id}$ is not an
 isomorphism and has a finite kernel that can be identified
 explicitly.

 To demonstrate the purity of the Hodge structure on $H^{3}(X,{\mathbb Z})$
  we look at the map \linebreak $\pi_{\beta} : X_{\beta} \to \Sigma$ onto the
  compact Riemann surface $\Sigma$. As we saw in the previous section,
  the fibers $X_{\beta,t} := \pi_{\beta}^{-1}(t)$ are smooth affine
  quadrics for $t$ not in the divisor of the quadratic differential
  $\beta \in H^{0}(\Sigma,K_{\Sigma}^{\otimes 2})$ and are irreducible
  quadratic cones for those $t$ for which $\beta(t) = 0$. On the other
  hand, every two dimensional smooth affine quadric $Q \subset {\mathbb
  C}^{3}$ is deformation equivalent\footnote{The deformation
  equivalence of $Q$ and $\op{tot}({\mathcal O}_{{\mathbb P}^{1}}(-2))$
  is realized explicitly by the resolved conifold $\widetilde{Z}$
  discussed in detail 
  in section~\ref{sec:DV}.} to
  the surface $\op{tot}({\mathcal O}_{{\mathbb P}^{1}}(-2))$. Thus by
  Ehresmann's fibration theorem $Q$ is homeomorphic to
  $\op{tot}({\mathcal O}_{{\mathbb P}^{1}}(-2))$  which in
  turn  is homotopy equivalent to ${\mathbb P}^{1}$.  Therefore
 \[
 \begin{split}
 H^{0}(Q,{\mathbb Z}) & = {\mathbb Z} \quad H^{2}(Q,{\mathbb Z}) =
 {\mathbb Z} \\
 H_{0}(Q,{\mathbb Z}) & = {\mathbb Z} \quad H_{2}(Q,{\mathbb Z}) =
 {\mathbb Z}
 \end{split}
 \]
 and the rest  of the cohomology and homology of $Q$ vanishes. Also,
 under the deformation equivalence $Q \sim \op{tot}({\mathcal
   O}_{{\mathbb P}^{1}}(-2))$ the generator $c$ of $H_{2}(Q,{\mathbb Z})$
 can be identified with  the zero section of ${\mathcal
   O}_{{\mathbb P}^{1}}(-2))$ and so the intersection form on
 $H_{2}(Q,{\mathbb Z})$ is given by $c\cdot c = -2$.  Thus the second
 homology $H_{2}(Q,{\mathbb Z})$ can be intrinsically identified with the
 root lattice $\Lambda_{\op{r}}$ of $SL(2,{\mathbb C})$. If we now use the
 universal coefficients theorem to identify $H^{2}(Q,{\mathbb Z})$ with 
 $H_{2}(Q,{\mathbb Z})^{\vee}$ we get a natural map 
 \[
 \xymatrix@R-2pc{
 H_{2}(Q,{\mathbb Z}) \ar[r] & H^{2}(Q,{\mathbb Z}) \\
 \alpha \ar[r] & \alpha\cdot \bullet
 }
 \]
 given by the intersection pairing on two cycles. Since $c \cdot c = -
 2$, it follows that the image of $H_{2}(Q,{\mathbb Z})$ is a subgroup
 of index two in $H^{2}(Q,{\mathbb Z})$. In other words we get a
 natural isomorphism of $H^{2}(Q,{\mathbb Z})$ with the weight lattice
 $\Lambda_{\op{w}}$ of $SL(2,{\mathbb C})$. Now fix a base point $t_{0}
 \in \Sigma^{o} = \Sigma - (\text{divisor of } \beta)$ and identify the
 fiber $\pi_{\beta}^{-1}(t_{0})$ with $Q$. Choose a collection
 $\{\gamma_{x} \}_{x \in \text{div}(\beta)}$ of non-intersecting paths
 in $\Sigma$ connecting $t_{0}$ with each zero $x$ of $\beta$. Since
 $\pi_{\beta} : X_{\beta} \to \Sigma$ is a Lefschetz fibration, it
 follows that the sphere $c \subset \pi_{\beta}^{-1}(t_{0}) \cong Q$
 vanishes along each $\gamma_{x}$ and that the local monodromy on
 $H_{2}(Q,{\mathbb Z})$ is given by the Picard-Lefschetz transformation
 $c \to c + (c,c)\cdot c = c - 2c = -c$. Thus the local monodromy action
 on $H^{2}(Q,{\mathbb Z}) = \Lambda_{\op{w}}$ is naturally equal to the
 action of the Weyl group $W$ on the weight lattice of $SL(2,{\mathbb
 C})$. Furthermore, it is not hard to compute the global monodromy 
 \begin{equation} \label{eq:monodromy}
 \op{mon} : \pi_{1}(\Sigma^{o},t_{0}) \to \{ \pm 1 \} =
 \op{Aut}(H^{2}(\pi_{\beta}^{-1}(t_{0}),{\mathbb Z})).
 \end{equation}
 Indeed, as explained at the end of section \ref{ss:higher}, the
 threefold $X_{\beta}$ is given by the equation
 \begin{equation} \label{eq:Xbeta}
 \op{det}_{W} - \pi_{W}^{*}\beta = 0
 \end {equation}
 in the total space $W$ of the rank three vector bundle $S^{2}V$.
 In particular, the global monodromy \eqref{eq:Xbeta} is the same as
 the monodromy of a double cover of $\Sigma$ which is branched
 precisely at the zeroes of $\beta$. If we further look at the
 compactification $\overline{X}_{\beta}$ we can identify the
 representation \eqref{eq:monodromy} with the monodromy on the two
 families of rulings in the fibers of $\overline{X}_{\beta|\Sigma^{o}}
 \to \Sigma^{o}$. However from the equation \eqref{eq:Xbeta} it is
 manifest that the covering parameterizing the two families of rulings
 is given by the equation $y^{2} - \beta$ in $T^{*}\Sigma$, i.e. is the
 spectral cover $p_{\beta} : \widetilde{\Sigma}_{\beta} \to
 \Sigma$. 

 This description of the local and global monodromies implies that we
 have a natural identification 
 \[
 R^{2}\pi_{\beta *}{\mathbb Z} = j_{*}((p_{\beta
   *}^{o}\Lambda_{\op{w}})^{W}) = (p_{\beta *}\Lambda_{\op{w}})^{W},
 \]
where $j : \Sigma^{o} \hookrightarrow \Sigma$ and $p_{\beta}^{o} :
\widetilde{\Sigma}^{o} \to \Sigma^{o}$ are the maps described at the
end of the previous section. 

 Finally from the Leray spectral sequence for the map $\pi_{\beta} : X_{\beta}
 \to \Sigma$ we immediately see that $H^{3}(\Sigma, {\mathbb Z}) =
 H^{1}(\Sigma,R^{2}\pi_{\beta *}{\mathbb Z})$ or equivalently
 \begin{equation} \label{eq:coh3}
 H^{3}(\Sigma, {\mathbb Z}) = H^{1}(\Sigma,(p_{\beta
   *}\Lambda_{\op{w}})^{W}). 
 \end{equation}
 Similarly, we have $R^{2}\pi_{\beta *}{\mathbb C} = j_{*}(R^{2}\pi_{\beta
   *}^{o}{\mathbb C})$ and $H^{3}(X_{\beta}, {\mathbb C}) =
 H^{1}(\Sigma, j_{*}(R^{2}\pi_{\beta *}^{o}{\mathbb C}))$. Since the
 Leray filtration is compatible with mixed Hodge structures and for the
 affine quadric $Q$ the cohomology $H^{2}(Q,{\mathbb C})$ is spanned by
 a single class of type $(1,1)$, it follows that the Hodge structure on
 $H^{3}(X_{\beta}, {\mathbb C})$ is pure of weight three. The last
 statement follows from the fact that the local system $R^{2}\pi_{\beta
   *}^{o}{\mathbb C}$ is a variation of pure Hodge structures of Tate
 type and weight two, and from the fact that for every complex local
 system ${\mathbb U}$ on $\Sigma^{o}$ we have
 $H^{1}(\Sigma,j_{*}{\mathbb U}) = \op{im}\left[
   H^{1}_{c}(\Sigma^{o},{\mathbb U}) \to H^{1}(\Sigma^{o},{\mathbb U})
   \right]$. In particular the only non-trivial pieces in the Hodge
 decomposition on $H^{3}(X_{\beta}, {\mathbb C})$ are of Hodge types
 $(2,1)$ and $(1,2)$. Twisting $H^{3}(X_{\beta}, {\mathbb C})$ by the
 Tate Hodge structure of weight $(-2)$ we get a Hodge decomposition on
 $H^{3}(X_{\beta}, {\mathbb C})$ which involves only $(1,0)$ and
 $(0,1)$ components. Thus $J^{3}(X_{\beta})$ and $J_{3}(X_{\beta})$ are
 both abelian varieties which are dual to each other. Finally, the two
 intermediate Jacobians come with a canonical isogeny $J_{3}(X_{\beta})
 \twoheadrightarrow J^{3}(X_{\beta})$ which combined with the duality
 gives natural polarizations on the abelian varieties
 $J^{3}(X_{\beta})$ and $J_{3}(X_{\beta})$.

 With all of this in place, we are now ready to compare the
 intermediate Jacobians $J^{3}(X_{\beta})$ and $J_{3}(X_{\beta})$ with
 the Hitchin fiber $\CN_{\beta} =
 \op{Prym}(\widetilde{\Sigma}_{\beta}/\Sigma_{\beta})$. According to
 equations \eqref{eq:trace}, \eqref{eq:K} and \eqref{eq:coh3} we have
 \[
 \begin{split}
 \CN_{\beta} & = H^{1}(\Sigma,(p_{\beta
   *}\Lambda_{\op{r}})^{W})\otimes_{\mathbb Z} S^{1}, \\
 J^{3}(X_{\beta}) & = H^{1}(\Sigma,(p_{\beta
   *}\Lambda_{\op{w}})^{W})\otimes_{\mathbb Z} S^{1},\\
 J_{3}(X_{\beta}) & = H^{1}(\Sigma,(p_{\beta
   *}\Lambda_{\op{w}})^{W})^{\vee}\otimes_{\mathbb Z} S^{1},
 \end{split}
 \]
 with complex structures all coming from the Hodge decomposition on the
 space \linebreak $H^{1}(\Sigma,(p_{\beta *} {\mathbb
 C}_{\widetilde{\Sigma}_{\beta}})^{W})$, where the $W$ action on the
 constant sheaf ${\mathbb C}_{\widetilde{\Sigma}_{\beta}} \to
 \widetilde{\Sigma}_{\beta}$ corresponds to the diagonal action of $W$
 on $\widetilde{\Sigma}_{\beta}\times {\mathbb C}$ which is the
 covering involution on $\widetilde{\Sigma}_{\beta}$ and the
 multiplication by $(-1)$ on ${\mathbb C}$. Therefore the comparison of
 $J^{3}(X_{\beta})$ and $J_{3}(X_{\beta})$ with $\CN_{\beta}$ as
 polarized abelian varieties amounts to a comparison of the torsion free parts
 of the abelian groups  $H^{1}(\Sigma,(p_{\beta
   *}\Lambda_{\op{w}})^{W})$ and $H^{1}(\Sigma,(p_{\beta
   *}\Lambda_{\op{r}})^{W})$.

 To make the comparison more explicit we choose a basis in
 $H^{1}(\widetilde{\Sigma}_{\beta},{\mathbb Z}) =
 H_{1}(\widetilde{\Sigma}_{\beta},{\mathbb Z})$ which is adapted to the
 double cover $p_{\beta} : \widetilde{\Sigma}_{\beta} \to \Sigma$. Up
 to homotopy we can bring all the branch points of $p_{\beta}$ inside a
 fixed disk $D \subset \Sigma$. Thinking of $D$ as a genus zero surface
 with a single boundary component, we can build the cover
 $\widetilde{\Sigma}_{\beta}$ topologically by first taking a double
 cover of $D$ corresponding to branch cuts between pairs of branch
 points and then attaching two copies of $\Sigma - D$ to the two
 boundary circles of this double cover as depicted on
 Figure~\ref{fig:cover} below. 

 \begin{figure}[!ht]
 \begin{center}
 \psfrag{S}[c][c][1][0]{{$\Sigma$}}
 \psfrag{SS}[c][c][1][0]{{$\widetilde{\Sigma}_{\beta}$}}
 \psfrag{p}[c][c][1][0]{{$p_{\beta}$}}
 \psfrag{a}[c][c][1][0]{{$\alpha_{i}$}}
 \psfrag{a1}[c][c][1][0]{{$\alpha_{i}'$}}
 \psfrag{a2}[c][c][1][0]{{$\alpha_{i}''$}}
 \psfrag{g}[c][c][1][0]{{$\gamma_{j}$}}
 \epsfig{file=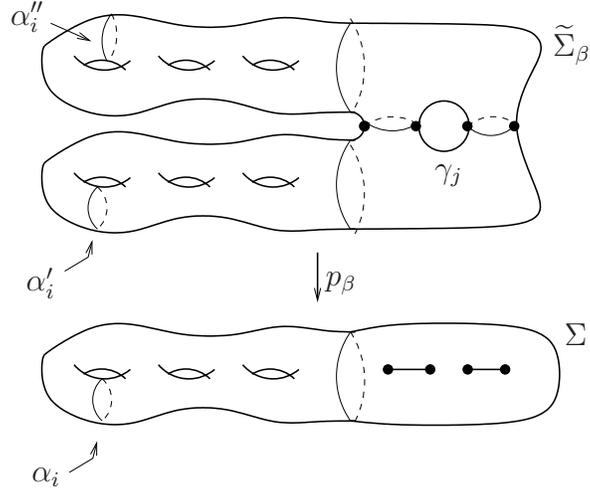,width=3in} 
 \end{center}
 \caption{The $p_{\beta}$-adapted basis in the homology of
   $\widetilde{\Sigma}_{\beta}$} 
 \label{fig:cover} 
 \end{figure}

 Now choose a basis $\{ \alpha_{i} \}_{i = 1}^{2g}$ in $H_{1}(\Sigma -
 D,{\mathbb Z}) = H_{1}(\Sigma,{\mathbb Z})$ consisting of loops
 contained in $\Sigma - D$. The inverse image of an $\alpha_{i}$ in
 $\widetilde{\Sigma}_{\beta}$ consists of two independent disjoint
 loops $\alpha_{i}'$ and $\alpha_{i}''$. Adding to $\alpha_{i}'$ and
 $\alpha_{i}''$ the extra loops $\gamma_{j}$ coming from the branch
 cuts on $D$ we get a basis
 \begin{equation} \label{eq:basis}
 \left\{ \alpha_{i}',
 \alpha_{i}'',\gamma_{j} \left| i = 1, \ldots, 2g, \ j = 1, \ldots, 4g
 - 6 \right.  \right\}
 \end{equation}
 of $H^{1}(\widetilde{\Sigma}_{\beta},{\mathbb Z})$ for which:

 \begin{itemize}
 \item Under the map $p_{\beta *} :
   H^{1}(\widetilde{\Sigma}_{\beta},{\mathbb 
   Z}) \to H^{1}(\Sigma,{\mathbb Z})$ we have 
 \[
 p_{\beta}(\alpha_{i}') =
   p_{\beta *}(\alpha_{i}'') = \alpha_{i}, \qquad
   \text{and} \qquad  p_{\beta * }(\gamma_{j}) = 0.
 \]
 \item The covering involution $\imath : \widetilde{\Sigma}_{\beta} \to
   \widetilde{\Sigma}_{\beta}$ for the cover $p_{\beta}$ transforms
   this basis as 
 \[
 \imath(\alpha_{i}') = \alpha_{i}'', \qquad \imath(\gamma_{j}) = - \gamma_{j}.
 \]
 \item The classes $\alpha_{i}'$,
 $\alpha_{i}''$, and $\gamma_{j}$ do not intersect with each other.
 \end{itemize}

 With the basis \eqref{eq:basis} at our disposal we can now proceed
 with the comparison of \linebreak $H^{1}(\Sigma,(p_{\beta
 *}\Lambda_{\op{w}})^{W})$ and $H^{1}(\Sigma,(p_{\beta
 *}\Lambda_{\op{r}})^{W})$. Choose a generator of $\Lambda_{\op{r}}$
 and identify $\Lambda_{\op{r}}$ with ${\mathbb Z}$ and
 $\Lambda_{\op{r}}\otimes {\mathbb Q}$ with ${\mathbb Q}$. Using the
 natural injection $\Lambda_{\op{r}} \to \Lambda_{\op{w}}$ given by the
 Cartan form on $\Lambda_{\op{r}}$ we can identify $\Lambda_{\op{w}}$
 with the sub-lattice $\frac{1}{2}{\mathbb Z} \subset {\mathbb Q}$. With
 this identification the generator of the Weyl group $W$ acts
 as multiplication by $(-1)$. Now by the calculation we did at the end
 of section~\ref{ss:pryms} we get that
 \[
 \begin{split}
 H^{1}(\Sigma,(p_{\beta
 *}\Lambda_{\op{r}})^{W}) & = \left(
 H^{1}(\widetilde{\Sigma}_{\beta},{\mathbb Z})\otimes
 \Lambda_{\op{r}}\right)^{W} =
 H^{1}(\widetilde{\Sigma}_{\beta},{\mathbb Z})^{-} = \left(\oplus_{i}
 {\mathbb Z}(\alpha_{i}' - \alpha_{i}'')\right) \oplus \left(
 \oplus_{j} \gamma_{j} \right)\\
 H^{1}(\Sigma,(p_{\beta
 *}\Lambda_{\op{w}})^{W}) & = \left(
 H^{1}(\widetilde{\Sigma}_{\beta},{\mathbb Z})\otimes
 \Lambda_{\op{w}}\right)^{W} =
 \frac{1}{2} H^{1}(\widetilde{\Sigma}_{\beta},{\mathbb Z})^{-} =
 \frac{1}{2}\left( 
 \left(\oplus_{i} 
 {\mathbb Z}(\alpha_{i}' - \alpha_{i}'')\right) \oplus \left(
 \oplus_{j} \gamma_{j} \right)\right).
 \end{split}
 \]
 Here  $H^{1}(\widetilde{\Sigma}_{\beta},{\mathbb
   Z})^{-}$ denotes the anti-invariants of the action of the covering
   involution $\imath$.

 From this identification it follows that the abelian varieties 
 \[
 \begin{split}
 \CN_{\beta} & = H^{1}(\Sigma,(p_{\beta *}\Lambda_{\op{r}})^{W})\otimes
 S^{1} \\
 J^{3}(X) & = H^{1}(\Sigma,(p_{\beta *}\Lambda_{\op{w}})^{W})\otimes
 S^{1}
 \end{split}
 \]
 are isomorphic but the natural map between them is given by
 a composition of an isomorphism and a multiplication-by-two map. 

 Similarly we get that the dual abelian varieties $\CN_{\beta}^{\vee}$
 and $J_{3}(X_{\beta})$ are also isomorphic. Explicitly we compute 
 \[
 \begin{split}
 H^{1}(\Sigma,(p_{\beta *}\Lambda_{\op{r}})^{W})^{\vee} & = \left(
 H^{1}(\widetilde{\Sigma}_{\beta},{\mathbb Z})\otimes
 \Lambda_{\op{w}}\right)_{W} = \left( 
 \left(\oplus_{i} 
 {\mathbb Z}\frac{\alpha_{i}' - \alpha_{i}''}{2}\right) \oplus \left(
 \oplus_{j} \gamma_{j} \right)\right) \\
 H^{1}(\Sigma,(p_{\beta *}\Lambda_{\op{w}})^{W})^{\vee} & = \left(
 H^{1}(\widetilde{\Sigma}_{\beta},{\mathbb Z})\otimes
 \Lambda_{\op{r}}\right)_{W} = 2\left( 
 \left(\oplus_{i} 
 {\mathbb Z}\frac{\alpha_{i}' - \alpha_{i}''}{2}\right) \oplus \left(
 \oplus_{j} \gamma_{j} \right)\right).
 \end{split}
 \]
Here for a $W$-module $\Gamma$ we write $\Gamma_{W}$ for the
group of $W$-coinvariants in $\Gamma$. That is, $\Gamma_{W}$ is the
quotient of $\Gamma$ by the additive subgroup in $\Gamma$ generated by
all elements of the form $\gamma - w\cdot \gamma$, $w \in W$, $\gamma
\in \Gamma$. 
 In particular we see that $\CN_{\beta}^{\vee} = J_{3}(X_{\beta})$ is
 also isomorphic to the quotient of $\CN_{\beta} = J^{3}(X_{\beta})$ by
 the group $\op{Pic}^{0}(\Sigma)[2]$ of $2$-torsion points on the
 Jacobian of $\Sigma$. On the other hand, as we saw in the previous
 section (see also \cite{TH}), the quotient
 $\CN_{\beta}/\op{Pic}^{0}(\Sigma)[2]$ is the Hitchin fiber for the
 Hitchin system of the Langlands dual group ${\mathbb P}GL(2,{\mathbb
   C})$. 

 \

 \noindent
 {\bf To summarize:} The family of homology intermediate Jacobians
 associated with the family of non-compact Calabi-Yau manifolds $\CX \to
 \CB$ is an ACIHS isomorphic to the Hitchin system for the group
 ${\mathbb P}GL(2,{\mathbb C})$. The family of cohomology intermediate
 Jacobians associated with $\CX \to
 \CB$ is an ACIHS isomorphic to the Hitchin system for the group
 $SL(2,{\mathbb C})$.

\

\bigskip

\noindent
In \cite{DDP} we will show how to generalize the above statement to
an isomorphism of the Calabi-Yau integrable system of the Calabi-Yau
varieties described in Remark~\ref{rem:balazs} and the Hitchin
integrable system for the corresponding $ADE$ group.

 \section{Large $N$ quantization for linear transitions}
 \label{sec:quantization} 

 In this section we give a physical proof of genus zero large N duality
 for the linear transitions constructed in section four. The open
 string side of the duality is constructed by wrapping topological {\bf
 B}-branes on the exceptional curves in a threefold $\wX_\wm$
 corresponding to a generic point in $\wbM$. The resulting topological
 open-closed string theory is expected to be related to closed
 topological string theory on a smoothing $X_l$, $l\in \bL$ of the
 nodal singularities.  We will give a physical proof for this
 equivalence at genus zero by showing that the large $N$ dynamics of
 topological {\bf B}-branes in the planar limit is governed by the
 Hitchin integrable system constructed in section five.

 Following the strategy of the previous sections, we will fix a point
 $s\in\bS$ and work only along the normal to $\bS$ slice
 $\bL_s=H^0(\Sigma, K_{\Sigma}^{\otimes 2})$ of the open stratum $\bL$
 of the moduli space.  We are interested in topological open-closed
 string theory on a resolution $\wX_\alpha$ of a nodal threefold
 $X_{\alpha^2}$. Assuming $\alpha$ to be generic, let us denote by
 $C_1,\ldots, C_{2g-2}$ the exceptional curves on $\wX_\alpha$.  We
 construct an open-closed topological string theory by wrapping $N_i$
 ${\bf B}$-type branes on the curve $C_i$, $i=1,\ldots, 2g-2$ so that the
 net D-brane number $\sum_{i=1}^{2g-2} N_i$ is zero. This means that on
 each curve $C_i$ we may have either branes or anti-branes depending
 on the sign of $N_i$. We will denote by $N$ the total number of
 branes in the system, which equals the total number of anti-branes
 i.e.
 \[
 N = \sum_{N_i>0} N_i = -\sum_{N_i<0} N_i.
 \]

 For a more precise mathematical definition of the boundary {\bf
 B}-model, recall that topological {\bf B}-branes on a Calabi-Yau
 space should be thought of as derived objects
 \cite{MRD:category,AL:category,ES:derived} (see also
 \cite{CL:complex,CL:unitarity,DED:enhanced}.)  The off-shell dynamics
 of {\bf B}-branes is captured by a topological string field theory
 whose action is a graded version of holomorphic Chern-Simons theory
 on $\wX_\alpha$. For a physical proof of large $N$ duality, we
 need to solve this theory at least in the large $N$
 planar limit. We
 will employ a strategy inspired from \cite{DV:matrix}. Since we are
 working with linearized deformations, we can write the holomorphic
 Chern-Simons theory on $\wX_\alpha$ as a perturbation of holomorphic
 Chern-Simons on $\wX$.  The latter can be reformulated in terms of a
 holomorphic gauge theory on the curve $\Sigma$ by dimensional
 reduction. The effect of complex structure deformations of $\wX$ can
 then be taken into account by a perturbation of the gauge theory on
 $\Sigma$ which is more tractable than holomorphic Chern-Simons theory
 on $\wX_\alpha$.

 There is however an important subtlety in this approach. The
 holomorphic Chern-Simons theory only captures the open string
 background in a fixed closed string background. The dynamics of the
 full-open closed topological string theory should be described in
 terms of holomorphic Chern-Simons theory coupled to Kodaira-Spencer
 theory. Although we do not have a rigorous justification, we will
 assume that the Kodaira-Spencer theory decouples from the holomorphic
 Chern-Simons theory in the genus zero sector of the theory. Therefore
 if we are only interested in the large $N$ planar limit of the theory
 we can quantize open strings in a fixed closed string background.
 This assumption will be a posteriori justified by the results modulo
 a subtle caveat related to the integration measure for the open
 string theory which will be discussed in section 6.3.

  \subsection{Holomorphic Chern-Simons theory and twisted Higgs
   complexes} \label{ss:holoCS} 

 Following the general outline, let us start with holomorphic
 Chern-Simons theory on $\wX$.  Recall that $\wX$ contains a ruled
 surface $S$ obtained by resolving the curve $\Sigma$ of $A_1$
 singularities of $X$.  We consider a D-brane configuration consisting
 of $N$ branes and $N$ anti-branes wrapping fibers of $S$. For
 simplicity let us consider the generic case in which the branes and
 antibranes wrap distinct fibers of $S$. More precisely we specify two
 divisors
 \[
 D_+ = \sum_{a =1}^N p_a, \qquad 
 D_-= \sum_{a=1}^N q_a
 \] 
 on $\Sigma$ so that the branes are supported on the fibers $S_{p_a}$
 and the antibranes are supported on the fibers $S_{q_a}$. Generically
 we will have $p_a\neq p_b$, $q_a\neq q_b$ for any $a,b=1\ldots N$,
 $a\neq b$, and $p_a\neq q_b$ for any $a,b=1,\ldots, N$.

 The corresponding boundary topological {\bf B}-model is given by the 
 complex $\CQ=\CQ^+ \oplus \CQ^-[-1]$
 where
 \[ 
 \CQ^+ = \oplus_{a=1}^N \CO_{S_{p_a}},\qquad 
 \CQ^- = \oplus_{a=1}^N \CO_{S_{q_a}}.
 \]
 The boundary chiral ring is isomorphic to the Ext algebra 
 \[
 \oplus_{k=0}^3 \rm{Ext}^k_{\wX}(\CQ,\CQ)
 \]
 In order to write down a physical action for off-shell fluctuations
 around this open string background, it is more convenient to work
 with a locally free resolution $\CE$ of $\CQ$.  Since $\CE$ and $\CQ$
 are quasi-isomorphic complexes, $\CE$ defines an equivalent boundary
 ${\bf B}$-model. Then the space of off-shell open string states is
 given by
 \[
 \CH_{\wX} = \bigoplus_{k=0}^3\bigoplus_{m,n} \Omega^{0,k}_{\wX}(E_m^\vee
 \otimes E_n). 
 \]
 Note that there is a integral ghost number grading on this vector
 space defined by $k+(n-m)$.  The physical states are elements of
 ghost number $k+(n-m)=1$.  The string field theory action for the
 physical states is a graded version of holomorphic Chern-Simons
 theory \cite{DED:enhanced} This action is not very tractable for
 concrete practical applications.

 However, since $\wX$ is isomorphic to the total space of a line
 bundle over the ruled surface $S$ we can find a better starting point
 for large $N$ quantization invoking Koszul duality. Very briefly, in
 this situation Koszul duality establishes an equivalence between
 coherent sheaves on $\wX$ finite over $S$ and Higgs sheaves on $S$.
 For convenience, recall that a Higgs sheaf on $S$ is a pair $(\CQ,
 \Phi)$ where $\CQ$ is a coherent sheaf on $S$ and $\Phi : \CQ \to \CQ
 \otimes K_{S}$ is a morphism from $\CQ$ to $\CQ\otimes K_{S}$.
 In general $\Phi$ should satisfy an integrability condition which is
 empty in our particular case.

 From a physical point of view, a Higgs sheaf on $S$ can be interpreted as a topological 
 {\bf B}-brane wrapping the surface $S\subset \wX$ as follows. For simplicity suppose that 
 $\CQ$ is locally free and denote by $Q$ the underlying vector bundle. Then the data $(S,Q)$ 
 determines a topological boundary {\bf B}-model with a nilpotent BRST symmetry. Such models 
 have been analyzed in great detail in \cite{KS:ext}. Their results will be very useful in 
 the following. 
 In particular, according to \cite{KS:ext}, the spectrum of {\bf B}-model boundary physical 
 states is realized as the limit of a local to global spectral sequence with second term 
 \[
 E_2^{p,q} = H^{0,p}(End(Q) \otimes \Lambda^q(N_{S/X}))
 \]
 which converges to $\rm{Ext}^k_{\wX}(\CQ,\CQ)$, $k=p+q$. 
 Koszul duality implies that 
 this spectral sequence collapses at the second term, and it has a canonical 
 split filtration so that 
 \[
 {\rm Ext}^k_{\wX}(\CQ,\CQ) = \oplus_{p+q=k}H^{0,p}(End(Q)\otimes \Lambda^q(N_{S/X})).
 \]
Since $N_{S/\wX} \cong K_{S}$, this shows that instead of working with
 bundles on $\wX$, it suffices to consider bundles on the compact
 surface $S$ as long as we take into account the Higgs field data.

 It is also helpful to discuss this data from the point of view of
 holomorphic Chern-Simons theory. Adopting a differential geometric
 point of view, we can think of $Q$ as a $C^\infty$ bundle on $S$
 equipped with a connection $A$ satisfying the integrability condition
 $F^{0,2}_A=0$. Then the covariant Dolbeault operator $\dbar_A$
 determines a holomorphic structure on $Q$. The off-shell open string
 states of the boundary {\bf B}-model are elements of the infinite
 dimensional vector space
 \begin{equation} \label{eq:AS}
 \CH_{S} = \oplus_{p=0}^2 \oplus_{q=0}^1 
 \Omega^{0,p}_{\wX}\left(End(Q) \otimes \Lambda^q(N_{S/X})\right)
 \end{equation}
 In order to construct a string field action for the off-shell
 fluctuations around this background we define a DG-algebra structure
 on $\CH$ as follows.  We first construct the $\IZ$-graded
 superalgebra 
\be\label{eq:superA} \Omega_S= \left(\oplus_{p=0}^2
 \Omega_S^{0,p}\right)\otimes_{\Omega^{0,0}_S}
 \left(\oplus_{q=0}^1\Omega^{0,0}_S(\Lambda^q(N_{S/\wX}))\right) 
\ee
 where the two factors are the exterior algebras of the
 antiholomorphic cotangent bundle of $S$ and respectively the
 holomorphic normal bundle to $S$ in $\wX$.  The grading ${\rm deg} :
 \Omega_S \to \IZ$ is defined by 
\be\label{eq:grading} {\rm deg}
 (\omega) = p+q, \qquad {\rm for} \ \omega \in
 \Omega_S^{0,p}(\Lambda^q(N_{S/\wX})).  
\ee Since
 $\Lambda^1(N_{S/\wX})\simeq \Omega^{2,0}_S$, we have an isomorphism of
 graded vector spaces 
\be\label{eq:superB} \Omega_S \simeq
 \oplus_{p=0}^2\oplus_{q=0}^1 \Omega^{2q,p}_S.  
\ee 
Using this
 isomorphism, the superalgebra structure on $\Omega_S$ can be
 explicitly written in the form 
\be\label{eq:superC} \omega \omega' =
 (-1)^{p'q} \omega\wedge \omega' 
\ee where $\omega\in
 \Omega^{2q,p}_S$, $\omega'\in \Omega^{2q',p'}_S$.  Then we give the
 space $\CH_{S}$ defined in \eqref{eq:AS}
 a tensor product superalgebra structure of the form
 \be\label{eq:superalgA} 
\CH_{S} = \Omega_S \otimes_{\Omega^{0,0}_S}
 \Omega^{0,0}_S(End(Q)) 
\ee 
where the last factor can be regarded as a
 superalgebra with trivial odd component.  The grading
 \eqref{eq:grading} extends trivially to $\CH_{S}$.

 Next, note that the covariant Dolbeault operator with respect to the
 background connection $A$ defines a degree one differential operator
 $\dbar^{(0)}:\CH_{S}\to \CH_{S}$ satisfying the Leibnitz rule
 \[
 {\dbar^{(0)}}(\omega\omega') = (\dbar^{(0)}\omega) \omega'
 +(-1)^{deg(\omega)} \omega(\dbar^{(0)}\omega') 
 \] 
 for any $\omega, \omega'\in \CH_{S}$. 
 Therefore we obtain an associative  DG-algebra structure on $\CH_{S}$. 

 In order to write down a holomorphic Chern-Simons action for
 off-shell fluctuations  
 we also need a trace ${\rm tr}: \CH \ra \IC$, which in this case is given by 
 \[
 {\rm tr} = \int_S \str 
 \]
 where $\str : \CH \ra \Omega_S$ is the supertrace. 
 The string field action is defined for ghost number one fields 
 $\phi\in \CH_{S}$, ${\rm deg}(\phi) = 1$ which parameterize arbitrary
 deformations  
 $\dbar^{(0)} \to \dbar^{(0)} + \phi$ of the $DG$ structure on $\CH_{S}$. 
 More precisely, $\phi$ can be written as a sum of homogeneous elements
 \be\label{eq:compA}
 \begin{aligned}
 \phi = \phi^{0,1} + \phi^{2,0},\cr 
 \end{aligned} 
 \ee 
where $\phi^{0,1} \in \Omega_{S}^{0,1}(End(Q))$ is an arbitrary
 deformation of the background Dolbeault operator $\dbar^{(0)}$.
 on $Q$ and $\phi^{2,0}\in \Omega_{S}^{0,0}(End(Q)\otimes N_{S/\wX})$ is a Higgs field 
on $S$. 
These are the expected off-shell $C^\infty$ deformations of a
 topological ${\bf B}$-brane supported on $S$. Applying the general
 reasoning of \cite{EW:CS} to the present case, it follows that the
 string field action reduces to a holomorphic Chern-Simons action on
 $S$ of the form 
\be\label{eq:holCSA} {\mathfrak S}_{CS} = \int_S \str
 \left(\frac{1}{2}\phi\dbar^{(0)}\phi + \frac{1}{3} \phi^3\right).
 \ee 
Substituting equation \eqref{eq:compA} in this expression we
 obtain 
\be\label{eq:holCSB}
 \begin{aligned} 
 {\mathfrak S}_{CS} & = \int_S {\rm Tr}\left(\phi^{2,0}\wedge
 (\dbar^{(0)}\phi^{0,1} + \phi^{0,1}\wedge \phi^{0,1})\right)\cr
 & = \int_S {\rm Tr}\left(\phi^{2,0}\wedge F^{0,2}\right)\cr
 \end{aligned} 
 \ee
 where $F^{0,2}$ is the $(0,2)$ 
 component of the curvature of the deformed connection 
 $A+\phi^{0,1}$. 
 Note that the action ${\mathfrak S}_{CS}$ is left
 invariant
by gauge transformations of the string field $\phi$ of the form 
 \[
 \delta \phi = \dbar^{(0)} \lambda +[\phi,\lambda] 
 \]
 where $\lambda \in \CH$ is an arbitrary ghost number zero field. 

 The solutions to the equations of motion for $\phi$ modulo gauge
 transformations parameterize deformations of the boundary topological
 {\bf B}-model. Applying the variational principle to the action
 \eqref{eq:holCSA} yields the Maurer-Cartan equation
 \[ 
 \dbar^{(0)} \phi + \phi \phi =0.
 \] 
 In components, we obtain
 \[
 \dbar \phi^{2,0} =0,\qquad F^{0,2} =0
 \]
 hence the solutions are in one to one correspondence to Higgs bundle
 structures on a fixed underlying $C^\infty$ bundle $Q$.  Gauge
 equivalent solutions correspond to isomorphic Higgs bundles,
 therefore we can conclude that deformations of the boundary {\bf
 B}-model are in one to one correspondence to isomorphism classes of
 Higgs bundles.

 In our case, we have to extend the above construction to an open
 string background specified by a complex $\CQ$ of coherent sheaves on
 $S$.  We first pick a locally free resolution $\CE$ of $\CQ$ on
 $S$. Note that $\CQ$ and $\CE$ are quasi-isomorphic as complexes of
 coherent sheaves on $\wX$, therefore they define equivalent boundary
 ${\bf B}$-models.  Then we construct a holomorphic Chern-Simons
 action on $S$ for off-shell fluctuations around this open string
 background following the same steps. The main difference is that we
 will have to take into account the $\IZ$-grading of the complex
 $\CE$, as explained in a similar context in \cite{DED:enhanced}. The
 discussion is fairly general, so in the following we can take $\CE$
 to be an arbitrary complex of locally free sheaves.

 In differential-geometric language, the open string background is specified by 
 a finite sequence of smooth complex bundles and maps 
 \be\label{eq:complexA}
 \cdots \to E_{n-1} \stackrel{e_{n,n-1}}{\ra} E_n \stackrel{e_{n+1, n}}{\ra} 
 E_{n+1} \to \cdots 
 \ee 
 The bundles are equipped with background connections $A_n$ subject to the 
 integrability condition 
 $F^{0,2}_{A_n}=0$ for all $n\in \IZ$. Therefore the covariant Dolbeault operators 
 $\dbar_n^{(0)}=\dbar_{A_n}$ 
 define holomorphic structures on the bundles $E_n$. 
 The maps $e_{n+1,n}$, $n\in \IZ$ are required 
 to be holomorphic with respect to the resulting complex structures and satisfy the condition 
 \[
 e_{n+1,n} e_{n,n-1} =0
 \]
 for all $n\in \IZ$. 

 In this context, the space of off-shell open string states is given by 
 \[
 \CH_{S} = \oplus_{p=0}^2 \oplus_{q=0}^1 \oplus_{m,n\in \IZ}
 \Omega^{0,p}_S({\rm Hom}(E_m,E_n)\otimes  
 \Lambda^q(N_{S/X}))
 \]
 The ghost number grading is defined by ${\rm deg} : \CH_{S} \to \IZ$, 
 \[
 {\rm deg}(\phi)= p+q + (n-m) 
 \] 
 for $\phi \in \Omega^{0,p}_S({\rm Hom}(E_m,E_n)\otimes
 \Lambda^q(N_{S/\wX}))$.  In order to define the correct superalgebra
 structure on $\CH_{S}$, we have to regard the $\IZ$-graded vector bundle
 $\{E_n\}$ as a $\IZ$-graded supervector bundle $\{{\check E}_n\}$
 \cite{DED:enhanced}, where
 \[
 {\check E}_n = \left\{ \begin{array}{cc}
 (E_n, 0), \qquad & {\rm for}\ n \ {\rm even} \cr
 (0, E_n),\qquad & {\rm for} \ n\ {\rm odd.}\cr
 \end{array}\right.
 \]
 Then we have a superalgebra $\Omega^{0,0}_S(End({\check E}))$ where 
 ${\check E}$ is the supervector 
 bundle ${\check E} =(E^+,E^-)$ where
 \[
 E^+ = \oplus_{n\in \IZ} E_{2n},\qquad 
 E^- = \oplus_{n\in \IZ} E_{2n+1}.
 \] 
 The superalgebra structure on $\CH_{S}$ is then defined by writing $\CH_{S}$ as a tensor product of 
 superalgebras 
 \be\label{eq:supertensor}
 \CH_{S} = \Omega_S \otimes_{\Omega_S^{0,0}}\Omega_S^{0,0}(End({\check E})).
 \ee 
 Note that the Dolbeault operators $\{\dbar_n^{(0)}\}$ define a degree one 
 differential operator $\dbar^{(0)} : \CH_{S} \ra \CH_{S}$ 
 satisfying the Leibnitz rule with respect to superalgebra multiplication. 
 Moreover the maps $e_{n+1,n}:E_n\to E_{n+1}$ define a degree one element 
 $e\in \CH_{S}$. Then we define the BRST operator $D^{(0)}:\CH_{S} \to
 \CH_{S}$ in the  
 background specified by the complex \eqref{eq:complexA} to be 
 \[
 D^{(0)}= \dbar^{(0)} + e.
 \] 
 Given a field $\phi \in \CH_{S}$, we have 
 \[ 
 D^{(0)}\phi = \dbar^{(0)} \phi + [e,\phi]
 \]
 where $[\,,\,]$ is the supercommutator 
 \[
 [\phi,\phi'] = \phi\phi'-(-1)^{{\rm deg}(\phi){\rm deg}(\phi')}\phi'\phi.
 \]
 Now we can write down the graded holomorphic Chern-Simons action for
 ghost number one  
 open string fields 
 $\phi \in \CH_{S}$ 
 \be\label{eq:holCSBB} 
 {\mathfrak S}_{CS} = \int_S \str \left(\frac{1}{2} \phi D^{(0)} \phi +
 \frac{1}{3} \phi^3\right). 
 \ee
 This action is left invariant by infinitesimal gauge transformations
 of the form  
 \be\label{eq:gaugeB} 
 \delta \phi = (D^{(0)}+\phi)\lambda 
 \ee
 where $\lambda$ is an arbitrary ghost number zero element of $\CH_{S}$.
 Note that the $D=D^{(0)}+\phi$ is an arbitrary off-shell deformation
 of the BRST operator.  
 The equations of motion derived from the holomorphic Chern-Simons action 
 can be written in compact form 
 \be\label{eq:eqmotionB} 
 D^{(0)}\phi + \phi^2 =0. 
 \ee
 which is equivalent with the integrability condition 
 $$ 
 D^2=0
 $$ 
 for the deformed BRST operator. 
 The solutions to these equations modulo gauge transformations
 parameterize deformations of  
 the open string background specified by the complex
 \eqref{eq:complexA}. In the ungraded case,  
 we identified these deformations with Higgs bundle structures on a
 fixed $C^\infty$ bundle  
 up to isomorphism. The equations \eqref{eq:eqmotionB} yield a
 generalization of the Higgs  
 bundle conditions which is better understood in the framework of
 D-brane categories,  
 which we explain next. 

 So far we have been studying fluctuations 
 around a fixed open string background. In principle one can consider
 more general situations  
 in which we have several topological D-branes wrapping the surface
 $S$. In that case  
 the algebraic structure of the resulting open string theory is
 encoded in a triangulated D-brane category  
 \cite{MRD:category,AL:category,ES:derived,CL:complex,CL:unitarity}.
 This category is a physical variant of the Bondal-Kapranov construction 
\cite{BK:category} introduced in the context of cubic string field theory in 
\cite{CL:complex,CL:unitarity}.

 In the present context, we start with a DG-category 
 $\CC$ given as follows. The objects of $\CC$ are holomorphic vector
 bundles $E\to S$. The space  
 of morphisms between two objects $E,E'$ is given by the complex 
 \[
 \begin{aligned} 
 {\rm Hom}_\CC(E,E') & = 
 \oplus_{p=0}^2\oplus_{q=0}^1 \Omega^{0,p}_S(Hom(E,E')\otimes
 \Lambda^q(N_{S/\wX}))\cr  
 & \simeq \oplus_{p=0}^2\oplus_{q=0}^1 \Omega^{2q,p}_S(Hom(E,E'))\cr
 \end{aligned} 
 \] 
 where the grading is given by $p+q$ and the differential is given by
 the covariant  
 Dolbeault operator $\dbar^{(0)}$. Here $E,E'$ are regarded again as
 $C^\infty$ bundles equipped with  
 $(0,1)$-connections.  

 Next we construct a new DG-category ${\widetilde \CC}$ by taking the
 shift completion of $\CC$.  
 The objects of $\wCC$ are pairs $(E,n)$, where $E$ is an object of
 $C$ and $n\in \IZ$. The space  
 of morphisms between two objects $(E,n)$, $(E',n')$ is the shifted complex 
 \[
 {\rm Hom}_{\wCC}\left((E,n),(E',n')\right) = {\rm Hom}_\CC(E,E')[n-n'].
 \]
 From a physical point of view, the integer $n$ represents the D-brane 
 grading introduced in \cite{MRD:category}. For future reference, we
 should also keep in mind that  
 composition of morphisms in ${\widetilde C}$ is given by 
 \be\label{eq:shiftcompB} 
 (gf)_{\wCC} = (-1)^{(k+n-n')(n'-n'')}(gf)_{\CC},
\ee
 for any $f\in
 \op{Hom}^k_{\wCC}((E,n),(E',n'))$ and any $g \in
 \op{Hom}^l_{\wCC}((E',n'), (E'', n'')$.

 The D-brane category is the triangulated category $\tr(\wCC)$ of
 twisted complexes  
 over $\wCC$ defined in \cite{BK:category}. 
 Twisted complexes over $\wCC$ are finite collections 
 of objects $\{(E_i,n_i)\}$ and degree one morphisms  
 $\Phi_{ji}\in {\rm Hom}_{\wCC}^1((E_i,n_i), (E_j,n_j))$ satisfying
 the Maurer-Cartan  
 equation 
 \be\label{eq:pretr}
 {\dbar^{(0)}} \Phi_{ji} + \sum_{k} \Phi_{jk}\Phi_{ki} = 0.
 \ee
 Twisted complexes form a DG-category denoted by $\pt(\wCC)$ in
 \cite{BK:category}.  
 The morphisms between two such objects $(E_i,n_i,\Phi_{ji})$ and
 $(E'_i,n'_i, \Phi'_{ji})$  
 are DG-complexes
 \be\label{eq:pretrA}
 {\rm
 Hom}_{\pt(\wCC)}^k\left((E_i,n_i,\Phi_{ji}),(E'_j,n'_j.\Phi'_{lj})\right)  
 = \oplus_{i,j} {\rm Hom}_{\wCC}^{k}((E_i,n_i),(E'_j,n_j'))
 \ee 
 The action of the differential on morphisms $\eta\in {\rm
 Hom}_{\wCC}^{l}((E_i,n_i),(E'_j,n_j'))$  
 is defined by
 \be\label{eq:pretrB}
 d\eta = \dbar^{(0)} \eta + \sum_{k} \Phi'_{kj}\eta -
 (-1)^{k}\eta\Phi_{ik}. 
 \ee
 Then one can obtain a triangulated category \cite{BK:category} with
 same objects as  
 $\pt(\wCC)$ by taking the space of morphisms between two objects to
 be the degree  
 zero cohomology of the complex \eqref{eq:pretrA},\eqref{eq:pretrB}. 
 For our purposes it is more convenient to work with the $\IZ$ graded
 category obtained from  
 $\pt(\wCC)$ by taking the space of morphisms between two objects to
 be the full cohomology of  
 the complex \eqref{eq:pretrA},\eqref{eq:pretrB}. We will denote the
 resulting enriched  
 triangulated category by $\tr(\wCC)$. 

 In order to see the connection between twisted complexes and graded
 holomorphic Chern-Simons  
 theory, note that the solutions to the equation of motion
 \eqref{eq:eqmotionB} are in fact 
 diagonal twisted complexes characterized by $n_i=i$ for all $i$. 

More precisely, if $\phi\in \CH^1$ is a solution to \eqref{eq:eqmotionB}, 
one can easily  check that $\Phi=\phi +e$ satisfies the equation 
\be\label{eq:eqmotionBB}
 \dbar^{(0)}\Phi + \Phi^2 =0.
 \ee
By construction the collection of fields  
\[ 
\Phi_{nm} \in \Omega_S^{2q,p}(Hom(E_m,E_n))
\] 
with $p+q=1+m-n$ can be regarded as morphisms 
\[ 
\Phi_{nm} \in \hbox{Hom}^1_{\wCC}(E_m,E_m) \simeq 
\hbox{Hom}^{1+m-n}_{\CC}(E_m,E_n).
\] 
between the bundles $E_m, E_n$ in the category $\wCC$. 
Moreover, the equation of motion \eqref{eq:eqmotionBB} is identical to 
the Maurer-Cartan equation \eqref{eq:pretr}. In order to check this equivalence, 
notice that the sign rule \eqref{eq:shiftcompB} for composition of morphisms in $\wCC$ 
is compatible with multiplication in the superalgebra \eqref{eq:supertensor}. 
Therefore the collection $\{E_n, \Phi_{nm}\}$ is a twisted complex with $n_i=i$. 

Keeping this correspondence in mind, from now on we will refer to the objects of 
 $\tr(\wCC)$ as twisted Higgs complexes. 

 In order to apply this construction to D-branes wrapping fibers of the ruling 
 $S\to \Sigma$, recall that we have canonical locally free resolutions 
 of divisors on $S$ 
 \[
 \begin{aligned} 
 & 0\ra \CO_S\left(-\sum_{a=1}^N S_{p_a} \right)\stackrel{e_1}{\ra}\CO_S\ra 
 \oplus_{a=1}^N \CO_{S_{p_a}}\ra 0 \cr 
 & 0 \ra \CO_S\left(-\sum_{a=1}^N S_{q_a}\right)\stackrel{e_2}{\ra} \CO_S\ra 
 \oplus_{a=1}^N \CO_{S_{q_a}}\ra 0.\cr
 \end{aligned} 
 \]
 Then we can construct a complex $\CE$ of locally free sheaves
which is quasi-isomorphic to $\CQ$ and has the form 
 \be\label{eq:complexB}
\xymatrix@1{
 0\ar[r] &  \CO_S\left(-\sum_{a=1}^N S_{q_a} \right) 
\ar[r]^-{\binom{e_1}{0}} & \CO_S\oplus 
 \CO_S\left(-\sum_{a=1}^N S_{p_a}\right) \ar[r]^-{(0 \;
   e_2)} &  \CO_S \ar[r] & 0.} 
 \ee
 The graded holomorphic Chern-Simons action we are searching for given by 
 \eqref{eq:holCSBB} in which the locally free complex $\CE$ is taken to be 
 \eqref{eq:complexB}.

 According to the general approach explained in section \ref{sec:DV},
 we need to understand the moduli space of solutions to the equations
 of of motion of the Chern-Simons action modulo gauge transformations.
 In this case the problem can be further simplified by taking
 dimensional reduction of the action along the fibers of the ruling
 $q:S \to \Sigma$. More precisely, we pick up a K{\"a}hler metric on
 $S$ so that the volume of the fibers is very small compared to the
 volume of the base. Then we can reduce the Chern-Simons
 action along the fibers obtaining a two-dimensional field
 theory. Strictly speaking this procedure is employed in the physics
 literature only when $S$ is a direct product $S= \Sigma \times
 {\mathbb P}^1$. In fact this restriction is too severe. Dimensional
 reduction can be applied equally well in all cases when the
 projective bundle $S= {\mathbb P}(V)$, where $V\to \Sigma$ is a rank
 $2$ holomorphic bundle, having locally constant transition
 functions \cite{BGK}. This will be the case if $V \to \Sigma$ is a polystable
 holomorphic bundle.  For simplicity, we will assume in the following
 that $S$ is a direct product, that is $V$ is a trivial rank $2$
 bundle. If $V$ is non-trivial, but polystable, the result remains
 unchanged.

 Let $\CE=q^*\CF$ be the pull-back of a complex of holomorphic vector
 bundles $\CF$  
 \be\label{eq:pullbackA}
 \cdots \to F_{n-1} \stackrel{f_{n,n-1}}{\ra} F_n
 \stackrel{f_{n+1,n}}{\ra}  F_{n+1} \to \cdots 
 \ee
 As above the bundles $F_n$ are regarded as $C^\infty$ bundles
 equipped with $(0,2)$ connections  
 $B_n$ which determine covariant Dolbeault operators $\dbar_n$. In
 order to perform the dimensional  
 reduction we write the off-shell field $\phi=\sum
 {\phi}^{2q,p}_{n,m}\in \CH$ as  
 \be\label{eq:dimredA}
 \phi^{2q,p}_{n,m} = \psi_{n,m}^{2q,p} \eta^{0,0} +
 \chi_{n,m}^{2q-1,p} \eta^{1,0} +  
 \chi_{n,m}^{2q,p-1}\eta^{0,1} + \psi_{n,m}^{2q-1,p-1} \eta^{1,1}
 \ee
 where $\psi_{n,m}^{k,l}, \chi^{k,l}_{n,m}\in \Omega^{k,l}_\Sigma(Hom(F_m,F_n))$,
 $\eta^{r,s}\in \Omega^{r,s}_{\IP^1}$. In the right hand side of
 \eqref{eq:dimredA},  
 all forms should be pulled back to $S$. Next, we take 
 $\eta^{r,s}$ to be solutions to the linearized equations of motion 
 modulo gauge transformations. Since the background complex is pulled back form $\Sigma$, 
 the linearized equations of motion read 
 \[ 
 \dbar \eta^{r,s} =0 
 \]
 for any $r,s=0,1$. Therefore the space of solutions to the linearized
 equations of motion  
 modulo gauge transformations is parameterized by
 $H^{r,s}(\IP^1)$. Let us choose generators  
 $1\in H^{0,0}(\IP^1)$, $[\eta]\in H^{1,1}(\IP^1)$ of the nontrivial
 cohomology groups.  
 Then the dimensional reduction ansatz \eqref{eq:dimredA} reduces to 
 \be\label{eq:dimredB} 
 \phi^{2q,p}_{n,m} = \psi^{2q,p}_{n,m} + \eta \psi^{2q-1,p-1}_{n,m}.
 \ee 
 Therefore we obtain the following space of off-shell fields on $\Sigma$ 
 \be\label{eq:dimredC}  
 \CB = \oplus_{q=0}^1\oplus_{p=0}^1 \oplus_{m,n\in
 \IZ}\Omega^{q,p}_\Sigma\left(Hom(F_m,F_n) \right) 
 \ee 
 where the ghost number grading is given by
 $\hbox{deg}(\psi^{q,p}_{n,m})= 2q+p+(n-m)$. We can give  
 $\CB$ a $\IZ$-graded superalgebra structure by adopting the
 construction used below equation  
 \eqref{eq:complexA}. 
 Let $\Omega_\Sigma$ be the $\IZ$-graded superalgebra obtained by
 reducing $\Omega_S$ along the  
 fibers of the ruling. We have 
 \[
 \Omega_\Sigma = \oplus_{q=0}^1 \oplus_{p=0}^1 \Omega_\Sigma^{q,p} 
 \] 
 where the multiplication is defined by 
 \[
 \omega \omega' = (-1)^{p'q} \omega\wedge \omega'.
 \]
 Let ${\check F}=(F^+,F^-)$ be the supervector bundle obtained by
 rolling the $\IZ$-graded vector space  
 $\oplus_{n\in \IZ} F_n$. Then we take 
 \be\label{eq:dimredD} 
 \CB = \Omega_\Sigma \otimes_{\Omega^{0,0}_\Sigma}
 \Omega^{0,0}_\Sigma(End({\check F})). 
 \ee 
 The background connections $B_n$ together with the maps $f_{n+1,n}:
 F_n \to F_{n+1}$  
 define a differential operator $D^{(0)}=\dbar^{(0)} + [f,\,]: \CB \to
 \CB$ satisfying the Leibnitz rule.  
 By dimensional reduction, the Chern-Simons action \eqref{eq:holCSBB}
 yields the following  
 action on $\Sigma$ 
 \be\label{eq:holCSC} 
 {\mathfrak S}=\int_\Sigma \str \left( \half \psi D^{(0)}\psi + \frac{1}{3}
 \psi^3\right). 
 \ee 
 This is again left invariant by infinitesimal gauge transformations of the form 
 \[ 
 \delta \psi = D\lambda 
 \] 
 where $\lambda$ is an arbitrary ghost number zero element and $D = D^{(0)}+\psi$. 
 The equations of motion of this action are again of the form 
 \be\label{eq:eqmotionC} 
 D^{(0)}\psi + \psi^2=0.
 \ee

 In order to facilitate the construction of the moduli space of solutions to the 
 equations \eqref{eq:eqmotionC} modulo gauge transformations, it is very helpful to 
 rephrase this construction in terms of D-brane categories. One can construct 
 a triangulated category of twisted complexes on $\Sigma$ by performing dimensional reduction 
 on twisted complexes on $S$. More precisely, let us start with the DG category $\CD$ of 
 holomorphic vector bundles $F\to \Sigma$ so that the space of morphisms between two objects 
 $F,F'$ is given by the complex 
 \be\label{eq:dimredE} 
 \hbox{Hom}_{\CD}(F,F') = \oplus_{q=0}^1\oplus_{p=0}^1 \Omega^{q,p}(Hom(F,F')).
 \ee
 The grading is defined by $p+2q$ and the differential is given by the covariant Dolbeault 
 operator as in the previous case. The D-brane category in question is the enriched 
 triangulated category $\tr(\wCD)$ of 
 twisted complexes on $\Sigma$ associated to the shift completion $\wCD$. It is a straightforward 
 exercise to check that twisted complexes on $\Sigma$ can be obtained by dimensional reduction of 
 twisted complexes on $S$. 

 This concludes our discussion of holomorphic D-branes on $\wX$ from a classical point of view. 
 In order to reach our goal we have to understand the quantum dynamics of holomorphic branes 
 on $\wX$, at least in the large $N$ limit. 

 \subsection{Quantization and moduli space} \label{ss:moduli}

 As explained in section \ref{sec:DV}, the quantization of the
 holomorphic Chern-Simons theory involves a formal integral of a
 holomorphic measure on a middle dimensional real cycle in the space of
 fields. Exploiting the topological symmetry of this theory, this
 functional integral localizes to a finite dimensional integral on a
 middle dimensional cycle in the moduli space $\CM$ of solutions to the
 classical field equations of motion.

 We would like to carry out this construction for the holomorphic
 Chern-Simons action associated to the complex
 \eqref{eq:complexB}. Since this complex is pulled back from $\Sigma$,
 the classical moduli space can be determined using dimensional
 reduction and truncation to zero modes. Therefore it suffices to
 consider the cubic action \eqref{eq:holCSC} for a complex of the form
 \be\label{eq:quantA} 0\to \CO_{\Sigma}(-D_+) \stackrel{{f_1\choose
 0}}{\ra} \CO_\Sigma \oplus \CO_\Sigma(-D_-) \stackrel{(f_2\, 0)}{\ra}
 \CO_\Sigma \to 0 \ee where $D_+ = \sum_{a=1}^N p_a$, $D_-=\sum_{a=1}^N
 q_a$ are disjoint divisors on $\Sigma$ and
 $$
 \begin{aligned} 
 & 0\to \CO_{\Sigma}(-D_+){\buildrel f_1\over \ra}\CO_\Sigma \ra \CO_{D_+}\to 0\cr
 & 0\to \CO_{\Sigma}(-D_-){\buildrel f_2 \over \ra}\CO_\Sigma \ra \CO_{D_-}\to 0\cr
 \end{aligned}
 $$
 are canonical maps. 
 On common grounds the functional integral \eqref{eq:fctintA} reduces to an integral over a middle 
 dimensional real cycle on the moduli space $\CM$ of critical points of the action 
 modulo gauge transformations. The holomorphic measure of the moduli space integral 
 should be determined in principle by integrating out the massive modes, provided that the 
 original measure $D\psi$ is at least formally well defined.
 A direct approach to this problem is beyond the purpose of the present paper so we will 
 employ a different technique. We first determine the moduli space and then find an expression 
 for the measure using holomorphy and physical constraints. In the process we will discover 
 a new aspect of this problem involving spin structures on $\Sigma$. 

 In order to find the moduli space it is convenient to write 
 the action \eqref{eq:holCSC} in terms of the field 
$\Psi = f + \psi-\psi^{0,1}$ and the 
 Dolbeault operator $\dbar = \dbar^{(0)} + \psi^{0,1}$. Then we have 
 \be\label{eq:quantB} 
 S_{CS} = \int_S \tr_s\left(\Psi^{1,0}_{01} \dbar \Psi^{0,0}_{10} + \Psi^{1,0}_{12} \dbar 
 \Psi^{0,0}_{21} + \Psi^{1,1}_{02}\Psi^{0,0}_{21}\Psi^{0,0}_{10}\right).
 \ee
 The equations of motion \eqref{eq:eqmotionC} become 
 \be\label{eq:eqmotionD}
 \begin{aligned}
 & {\dbar}_{10}\Psi^{0,0}_{10} = 0 & & \Psi^{0,0}_{21} \Psi^{0,0}_{10} =0\\
 & \dbar_{21} \Psi_{21}^{0,0} =0 & & \Psi^{1,0}_{01} \Psi^{0,0}_{10} =0\\
 & \dbar_{01}\Psi^{1,0}_{01} + \Psi^{1,1}_{02}\Psi^{0,0}_{21} =0 & 
 & \Psi^{0,0}_{10}\Psi^{1,0}_{01} - \Psi^{1,0}_{12}\Psi^{0,0}_{21} =0 \\
 & \dbar_{21}\Psi_{21}^{1,0} + \Psi^{0,0}_{10}\Psi^{1,1}_{02} =0& & 
 \Psi^{0,0}_{21} \Psi^{1,0}_{12} = 0\\
 \end{aligned}
 \ee

 As explained above, our goal is to determine the moduli space of solutions to the equations 
 of motion modulo gauge transformations. A better formulation of the problem can be achieved 
 in the framework of D-brane categories developed in the previous subsection. We have to find the moduli 
 space of three term twisted complexes of the form $\CT=(F_i,\Psi_{ji})$, $i,j=0,1,2$ where 
 \[
 F_0 = \CO_\Sigma(-D_+),\qquad F_1=\CO_\Sigma \oplus \CO_\Sigma(-D_-),\qquad F_2 = \CO_\Sigma 
 \] 
 as $C^\infty$-bundles. By convention we fix the degrees of $F_0,F_1,F_2$ to be $-1,0,1$. 
 Two such twisted complexes are said to be equivalent if there exist morphisms 
 $\eta \in H^0_{\tr(\wCD)}(\CT,\CT')$, $\rho\in H^0_{\tr(\wCD)}(\CT',\CT)$ such that 
 $\rho\circ \eta= {\mathbb I}_{\CT}$ and $\eta\circ  \rho = {\mathbb I}_{\CT'}$.

 In order to formulate a well defined moduli problem, we have to translate this data into
 algebraic-geometric language. We start with the three term twisted complexes $\CT$. 
 The fields $\Psi^{01,}_{m,m}$, $m=0,1,2$ correspond to integrable complex structures on the 
 bundles $F_m$, $m=0,1,2$. We will denote by $\CF_m$, $m=0,1,2$ the associated 
 locally free sheaves. $\Psi^{0,0}_{m+1,m}: \CF_{m}\to \CF_{m+1}$ determine morphisms 
 $\Psi_{10}:\CF_0\to \CF_1$, $\Psi_{21}:\CF_1\to \CF_2$ 
 of locally free sheaves which compose to zero. Therefore we obtain a complex of locally free 
 sheaves 
 \be\label{eq:complexC} 
 0\to \CF_0 \stackrel{\Psi_{10}}{\ra } \CF_1
 \stackrel{\Psi_{21}}{\ra } \CF_2 \to 0.
 \ee
 The remaining fields $\Psi^{1,0}_{m,m+1}$, $m=0,1,2$, $\Psi_{02}^{1,1}$ can be interpreted as 
 follows. $\Psi_{02}^{1,1}$ determines an extension 
 \be\label{eq:extensionA}
 0\to \CF_0\otimes \CO(K_\Sigma) \to \CS \to \CF_2 \to 0.
 \ee
 and $\Psi_{01}^{1,0}$, $\Psi_{12}^{1,0}$ determine splittings 
 \be\label{eq:splittingA}
 \xymatrix{0 \ar[r] & \CF_0\otimes \CO(K_\Sigma) \ar[r]\ar[d]_{\simeq} & 
 \Psi_{21}^*\CS \ar[r]\ar[d] & \CF_1 \ar@/_1pc/[l]_{\Psi_{01}} \ar[r] \ar[d]^{\Psi_{21}} 
 \ar[dl]^{{\widetilde \Psi}_{21}} & 0 \\
 0 \ar[r] & \CF_0\otimes \CO(K_\Sigma) \ar[r] & \CS \ar[r] & \CF_2 \ar[r] & 0 \\}
 \ee
 \be\label{eq:splittingB} 
 \xymatrix{0 \ar[r] & \CF_1\otimes \CO(K_\Sigma) \ar[r] & 
 \Psi_{10\ast}\CS \ar[r]\ar@/_1pc/[l]_{\Pi_{12}} & \CF_2 \ar@/_1pc/[l]_{\Psi_{12}} \ar[r] 
 & 0 \\
 0 \ar[r] & \CF_0\otimes \CO(K_\Sigma) \ar[r] \ar[u]^{\Psi_{10} \otimes \II_{\CO(K_\Sigma)}} 
 & \CS \ar[u]
 \ar[r]\ar[ul]^{{\widetilde \Psi}_{10}} & \CF_2 \ar[u]_{\simeq} \ar[r] & 0 \\}
 \ee
 where $\Psi_{12}^*\CS$, $\Psi_{10*}\CS$ denote the pullback and respectively pushforward extensions.
 For further reference note that the splitting $\Psi_{01}$ induces a map ${\widetilde \Psi}_{21}:\CF_1
 \ra \CS$ lifting $\Psi_{21}:\CF_1\ra \CF_2$. The splitting $\Psi_{12}$ induces a canonical 
 projection $\Pi_{12}: \Psi_{10\ast}\CS\to \CF_1\otimes \CO(K_\Sigma)$ which induces in turn a map 
 ${\widetilde \Psi}_{10}: \CS \to \CF_1 \otimes \CO(K_\Sigma)$ extending
 $\Psi_{10}\otimes \II_{\CO(K_\Sigma)}:\CF_0\otimes \CO(K_\Sigma)\to \CF_1\otimes \CO(K_\Sigma)$.

 In order to interpret the remaining constraints
 \be\label{eq:remaining}
 \begin{aligned}
 & \Psi^{1,0}_{01} \Psi^{0,0}_{10} =0\cr
 & \Psi^{0,0}_{10}\Psi^{1,0}_{01} - \Psi^{1,0}_{12}\Psi^{0,0}_{21} =0\cr
 & \Psi^{0,0}_{21} \Psi^{1,0}_{12} = 0\cr
 \end{aligned}
 \ee
 note for example that $\Psi_{10}:\CF_0\ra \CF_1$ induces a splitting 
 \[
 \xymatrix{0 \ar[r] & \CF_0\otimes \CO(K_\Sigma) \ar[r]\ar[d]_{\simeq} & 
 \Psi_{10}^*\Psi_{21}^*\CS \ar[r]\ar[d] & \CF_0\ar@/_1pc/[l]_{\Psi_{10}^*\Psi_{01}} \ar[r] 
 \ar[d]^{\Psi_{10}}& 0 \\
 0 \ar[r] & \CF_0\otimes \CO(K_\Sigma) \ar[r] & 
 \Psi_{21}^*\CS \ar[r] & \CF_1 \ar@/_1pc/[l]_{\Psi_{01}} \ar[r] 
 & 0. \\}
 \]
 The first equation in \eqref{eq:remaining} is equivalent to the condition
 \be\label{eq:splittingC} 
 \Psi_{10}^*\Psi_{21}^*\CS = \CF_0\oplus \CF_0\otimes \CO(K_\Sigma)\ \ \hbox{and}\ \ 
  \Psi_{10}^*\Psi_{01} = \left(\II_{\CF_0},0\right).
 \ee
 Similarly, $\Psi_{21}:\CF_1\ra \CF_2$ induces a splitting 
 $$ 
 \xymatrix{0 \ar[r] & \CF_2\otimes \CO(K_\Sigma) \ar[r] & 
 \left(\Psi_{21}\otimes \II_{\CO(K_\Sigma)}\right)_*\Psi_{10*}\CS \ar[r] & 
 \CF_2\ar@/_2pc/[l]_{\left(\Psi_{21}\otimes \II_{\CO(K_\Sigma)}\right)_*\Psi_{12}} \ar[r] 
 & 0 \\
 0 \ar[r] & \CF_1\otimes \CO(K_\Sigma) \ar[r]\ar[u]_{\Psi_{21}\otimes \II_{\CO(K_\Sigma)}} & 
 \Psi_{10\ast}\CS \ar[r]\ar[u] & \CF_2 \ar@/_1pc/[l]_{\Psi_{12}} \ar[r] \ar[u]^{\simeq}
 & 0.\\}
 $$
 The last equation is equivalent to 
 \be\label{eq:splittingD}
 \left(\Psi_{21}\otimes \II_{\CO(K_\Sigma)}\right)_*\Psi_{10*} \CS 
 = \CF_2 \oplus \CF_2\otimes \CO(K_\Sigma)\ \hbox{and}\
 \left(\Psi_{21}\otimes \II_{\CO(K_\Sigma)}\right)_*\Psi_{12} = \left(\II_{\CF_2},0\right).
 \ee
 Finally, in order to find the algebraic interpretation of the second equation in \eqref{eq:remaining} 
 note that we have two induced splittings 
 \[
 \xymatrix{0 \ar[r] & \CF_1\otimes \CO(K_\Sigma) \ar[r]& 
 \left(\Psi_{10}\otimes \II_{\CO(K_\Sigma)}\right)_*\Psi_{21}^*\CS \ar[r] & 
 \CF_1\ar@/_2pc/[l]_{\left(\Psi_{10}\otimes \II_{\CO(K_\Sigma)}\right)_*\Psi_{01}} \ar[r] & 0 \\
 0 \ar[r] & \CF_0\otimes \CO(K_\Sigma) \ar[r]\ar[u]_{\Psi_{10}\otimes \II_{\CO(K_\Sigma)}} & 
 \Psi_{21}^*\CS \ar[r]\ar[u] & \CF_1 \ar@/_1pc/[l]_{\Psi_{01}} \ar[r] \ar[u]^{\simeq}
 & 0. \\}
 \]
 \[
 \xymatrix{0 \ar[r] & \CF_1\otimes \CO(K_\Sigma) \ar[r]\ar[d]_{\simeq} & 
 \Psi_{21}^*\Psi_{10*}\CS \ar[r]\ar[d] & 
 \CF_1\ar@/_1pc/[l]_{\Psi_{21}^*\Psi_{12}} \ar[r] \ar[d]^{\Psi_{21}}
 & 0 \\
 0 \ar[r] & \CF_1\otimes \CO(K_\Sigma) \ar[r]& 
 \Psi_{10\ast}\CS \ar[r] & \CF_2 \ar@/_1pc/[l]_{\Psi_{12}} \ar[r]
 & 0.\\}
 \]
 Note also that the extensions $\left(\Psi_{10}\otimes \II_{\CO(K_\Sigma)}\right)_*\Psi_{21}^*\CS$ and 
 $\Psi_{21}^*\Psi_{10*}\CS$ are canonically isomorphic. Then the middle equation in 
 \eqref{eq:remaining} is equivalent to the condition that the two splittings agree 
 \be\label{eq:splittingE}
 \left(\Psi_{10}\otimes\II_{\CO(K_\Sigma)}\right)_\ast\Psi_{01} = \Psi_{21}^* \Psi_{12}. 
 \ee
 To summarize, we have shown that the data given by a three term twisted complex 
 $\CT=(\CF_i, \Psi_{ji})$, 
 $i,j=0,1,2$ on $\Sigma$ is equivalent to the following algebraic data 

 $i)$ a three term complex $\CF$ of locally free sheaves \eqref{eq:complexC},

 $ii)$ an extension of the form \eqref{eq:extensionA}, and 

 $iii)$ splittings \eqref{eq:splittingA}, \eqref{eq:splittingB} satisfying conditions 
 \eqref{eq:splittingC}, \eqref{eq:splittingD} and \eqref{eq:splittingE}.

 Next we have to specify equivalence relations between two sets of algebraic data. 
 This will be achieved 
 by finding an algebraic description for the morphism space between three term twisted complexes 
 in $\tr(\wCD)$. Let $(\CF_i, \Psi_{02}, \Psi_{01}, \Psi_{12})$ and 
 $(\CF'_i, \Psi'_{02},\Psi'_{01}, \Psi'_{12})$ $i=0,1,2$  be two sets of algebraic data satisfying
 conditions $(i)-(iii)$ above. 

 Given the data $(\CF_i,\Psi_{ji})$ as above, we construct the five term complex $\CK$ 
 \be\label{eq:complexE} 
 \CF_0 \stackrel{\Psi_{10}}{\ra} \CF_1 \stackrel{{\widetilde \Psi}_{21}}{\ra } 
 \CS \stackrel{{\widetilde \Psi}_{10}}{\ra}\CF_1\otimes \CO(K_\Sigma) \stackrel{\Psi_{21}\otimes 
 \II_{\CO(K_\Sigma)}}{\ra} \CF_2\otimes \CO(K_\Sigma). 
 \ee
 in which the first term has degree $-2$. 
 We also have a similar complex $\CK'$ for the data $(\CF'_i,\Psi_{ji}')$
 \be\label{eq:complexF} 
 \CF'_0 \stackrel{\Psi'_{10}}{\ra} \CF'_1 \stackrel{{\widetilde \Psi}'_{21}}{\ra } 
 \CS' \stackrel{{\widetilde \Psi}'_{10}}{\ra}\CF'_1\otimes \CO(K_\Sigma) 
 \stackrel{\Psi'_{21}\otimes 
 \II_{\CO(K_\Sigma)}}{\ra} \CF'_2\otimes \CO(K_\Sigma). 
 \ee
 Note that we have short exact sequences of complexes 
 \be\label{eq:exseqA}
 \begin{aligned} 
 & 0\to \CF\otimes \CO(K_\Sigma)[-2] \to \CK \to \CF \to 0 \cr
 & 0\to \CF'\otimes \CO(K_\Sigma)[-2] \to \CK' \to \CF' \to 0 \cr
 \end{aligned} 
 \ee
 Applying the functor ${Hom}(\ ,\CF')\otimes \CO(K_\Sigma)[-2]$ to the first complex in 
 \eqref{eq:exseqA}, we obtain the short exact sequence of complexes 
 \be\label{eq:exseqC} 
 0\to {Hom}(\CF,\CF')\otimes \CO(K_\Sigma)[-2] \to 
 {Hom}(\CK\otimes \CO(-K_\Sigma)[2],\CF') \to {Hom}(\CF,\CF') \to 0.
 \ee 
 Applying the functor ${Hom}(\CF,\ )$ to the second complex in \eqref{eq:exseqA}
 we obtain the short exact sequence of complexes 
 \be\label{eq:exseqD} 
 0\to {Hom}(\CF,\CF')\otimes \CO(K_\Sigma)[-2] \to 
 {Hom}(\CF,\CK') \to {Hom}(\CF,\CF') \to 0.
 \ee
 Therefore we have produced two extensions of the complex ${Hom}(\CF,\CF')$ by 
 ${Hom}(\CF,\CF')\otimes \CO(K_\Sigma)[-2]$. Next we construct a new complex $C(\CK,\CK')$ 
 by taking the difference of the two extensions. 
 This results in a five term complex whose terms are explicitly computed in appendix A. 
 We claim the $\IZ$ graded space of morphisms between the twisted complexes $\CT, \CT'$ in 
 $\tr(\wCD)$ is isomorphic to the hypercohomology of $C(\CK,\CK')$ on $\Sigma$
 \be\label{eq:algmorphisms}
 H^{\bullet}_{\tr(\wCD)}(\CT,\CT') \simeq  {\mathbb H}^{\bullet}(C(\CK,\CK')).
 \ee 
 The proof of this assertion reduces to a rather lengthly homological algebra computation 
 which is performed in appendix A.

 Using the algebraic formulation we can set up a well defined moduli problem for a 
 three term complex $\CF$ of the form \eqref{eq:quantA} in the category of twisted complexes. 
 In principle, one can construct a moduli stack associated to this moduli problem, but we will 
 perform such a construction here. For a semiclassical analysis it suffices to identify 
the irreducible component of the moduli space which contains equivalence classes of 
complexes of the form $\CF$ up to birational equivalence. This is an easier task, 
which can be accomplished as follows. 

 First note that we can construct a family of equivalence classes of complexes $\CF$ 
 in the category of twisted complexes by varying the divisors $D_+, D_-$ on $\Sigma$. 
This gives rise to a map 
$\phi: \hbox{Sym}^N(\Sigma)^2 \to \CM$ from $\hbox{Sym}^N(\Sigma)^2$ 
to the moduli space sending a point $(D_+,D_-)\in (\Sigma)^2$ to a 
 three term complex of the form \eqref{eq:quantA}. 
 
 Now let us compute the space of infinitesimal first order deformations 
 $H^1_{\tr(\wCD)}(\CF,\CF)$ of the complex $\CF$ in the category of twisted complexes. 
 According to equation \eqref{eq:algmorphisms}, we have 
 \[
 H^1_{\tr(\wCD)}(\CF,\CF) \simeq {\mathbb H}^1(C(\CK,\CK)) 
 \]
 where $\CK,\CK'$ are the five term complexes defined in \eqref{eq:complexE},\eqref{eq:complexF}.
 Since $\Psi_{02}=0$ in this case, the extension \eqref{eq:extensionA} is canonically split. 
 Then one can check that 
 \[
 \CK = \CF \oplus \CF \otimes \CO(K_\Sigma)[-2] 
 \] 
 and 
 \[ 
 C(\CK, \CK) = End(\CF) \oplus End(\CF) \otimes \CO(K_\Sigma)[-2].
 \]
 Therefore we have to evaluate the hypercohomology group 
 \[
 {\mathbb H}^1(End(\CF) \oplus End(\CF) \otimes \CO(K_\Sigma)[-2]) =
 \hbox{Hom}_{D^b(\Sigma)}(\CF,\CF[1]) \oplus 
 \hbox{Hom}_{D^b(\Sigma)}(\CF,\CF\otimes \CO(K_\Sigma)[-1]).
 \]
 The complex $\CF$ is quasi-isomorphic to its cohomology $\CO_{D_+}\oplus \CO_{D_-}[-1]$, hence we 
 have 
 \be\label{eq:derhomA}
 \begin{aligned} 
 & \hbox{Hom}_{D^b(\Sigma)}(\CF,\CF[1]) \oplus 
 \hbox{Hom}_{D^b(\Sigma)}(\CF,\CF\otimes \CO(K_\Sigma)[-1])=\cr
 & \qquad \qquad \qquad
 \hbox{Hom}_{D^b(\Sigma)}\left(\CO_{D_+}\oplus \CO_{D_-}[-1],\CO_{D_+}[1]\oplus \CO_{D_-}\right)
 \oplus \cr
 &  \qquad \qquad \qquad
 \hbox{Hom}_{D^b(\Sigma)}\left(\CO_{D_+}\oplus \CO_{D_-}[-1],\CO_{D_+}(K_\Sigma)[-1]\oplus 
 \CO_{D_-}(K_\Sigma)[-2]\right)=\cr
 & \qquad \qquad \qquad \hbox{Hom}_{D^b(\Sigma)}(\CO_{D_+},\CO_{D_+}[1])\oplus 
 \hbox{Hom}_{D^b(\Sigma)}(\CO_{D_-},\CO_{D_-}[1]).\cr
 \end{aligned}
 \ee
 Using Serre duality, one can show that all other terms in the right hand side of the above 
 equation vanish provided that  $D_+, D_-$ have disjoint supports. Therefore, 
 if $\hbox{Supp}(D_+)\cap \hbox{Supp}(D_-) = \emptyset$, we find that 
 \[ 
 {\mathbb H}^1(C(\CK,\CK))=\hbox{Ext}^1(\CO_{D_+},\CO_{D_+}) \oplus \hbox{Ext}^1(\CO_{D_-},\CO_{D_-}),
 \] 
 which is the space of infinitesimal deformations of the pair $(D_+, D_-)$ i.e. the 
 tangent space $T_{(D_+,D_-)}\hbox{Sym}^N(\Sigma)^2$.

Let $\Delta_{\pm}$ denote the big diagonals in the two factors of the direct product 
$\hbox{Sym}^N(\Sigma) \times \hbox{Sym}^N(\Sigma)$; the points in 
$\hbox{Sym}^N(\Sigma)\setminus \Delta_\pm$ parameterize effective divisors 
$D_{\pm}$ with distinct simple points. We will also denote by 
 \[
 \Delta \subset \hbox{Sym}^N(\Sigma)^2, \quad \Delta= \{(D_+,D_-)\in 
 \hbox{Sym}^N(\Sigma)^2| \hbox{Supp}(D_+) \cap 
 \hbox{Supp}(D_-) \neq \emptyset \}.
 \]
the big diagonal of the direct product.

The above considerations show that $\phi$ is an isomorphism between the open subset 
\[ 
\left(\hbox{Sym}^N(\Sigma) \times \hbox{Sym}^N(\Sigma)\right) \setminus \left(\Delta_+\times 
 \hbox{Sym}^N(\Sigma) \cup \hbox{Sym}^N(\Sigma) \times \Delta_- \cup \Delta\right)
\]
and an open subset of the moduli space $\CM$. Note that $\phi$ is also well defined 
and induces an isomorphism of virtual tangent spaces along the divisor
\[
\Delta_+ \times\hbox{Sym}^N(\Sigma) \cup \hbox{Sym}^N(\Sigma) \times \Delta_-.
\]
However, $\phi$ does not extend to this divisor as in isomorphism of stacks, since the moduli 
space $\CM$ is a higher stack. 
In principle, it could extend as an isomorphism of stacks in the presence of a suitable 
stability condition on twisted complexes which would make $\CM$ an stack. 
We will not attempt to formulate such a condition here, but we will assume that with right choice 
of a stability condition,
we can identify an open subset of the D-brane moduli space with the 
open subset 
\[
\CM_0= [\Sigma^{N}/S_{N}] \times [\Sigma^{N}/S_{N}] \setminus \Delta,
\]
where $[\Sigma^{N}/S_{N}]$ denotes the (stacky) orbifold quotient of
the cartesian product $\Sigma^{N}$ by the action of the symmetric
group $S_{N}$. 

 \subsection{The measure} \label{ss:measure}

 The next step in the quantization process requires a holomorphic
 measure on the moduli space. In the following we will concentrate on
 the open subset $\CM_0$ of the moduli space constructed in the
 previous subsection.  Note that $\CM_0$ has a finite cover of the
 form 
\be\label{eq:finitecover} 
\CM_0' = \left( \Sigma^{N} \times \Sigma^{N}\right)
 \setminus \Delta 
\ee 
where $\Delta$ is defined again as the divisor
 of $\Sigma^{N}\times \Sigma^N$ where a point $p_a$ coincides with a
 point $q_b$ for some $a,b=1,\ldots, N$.  In fact $\CM_0$ is a
 quotient of $\CM_0'$ by the obvious action of $S_N\times S_N$
 which from a physical point of view should be thought of as a
 residual discrete gauge action.

Following the common practice in gauge theories, we will write the
measure on a finite cover of the moduli space of the form and divide
the resulting functional integral by \linebreak $|S_N \times S_N| = (N!)^2$.

The restriction of the measure to $\CM_0'$ should be a holomorphic
 $2N$ differential form. Since $\CM_0$ is isomorphic to the
 complement of the diagonal $\Delta$ in $\Sigma^{2N}$, any such form
 can be extended to a meromorphic $2N$ form $\Omega$ on
 $\Sigma^{2N}$.
 
In principle one should be able to derive a formula for $\Omega$
starting from the path integral formulation of holomorphic
Chern-Simons theory. However, this approach may be quite cumbersome in
practice, so it is more economical to find a formula for $\Omega$
based on holomorphy and physical constraints. The main idea is that
physical constraints specify the polar structure of $\Omega$ along the
divisor 
\be\label{eq:polardiv} 
\left( \Delta_{+} \times\Sigma^N\right) \cup \left(\Sigma^N
\times \Delta_{-}\right)  \cup \Delta.  
\ee 
Although these conditions do not
determine $\Omega$ uniquely, with some additional physical insight we
can write down the measure uniquely up to a scale.

The physical constraints on $\Omega$ are imposed by the universal
character of the local effective interactions among branes at very
short distances.  This means that when the branes are very close to
each other the dominant interaction terms are identical to their local
counterparts discussed in section two.  Therefore we should have
Coulomb repulsion between two branes or two anti-branes approaching
each other and also Coulomb attraction between a brane anti-brane pair
\cite{DV:matrix, robbert:santabarbara}.  This means that $\Omega$
should have a zero of order two along each divisor in $\CM_0'$ given
by $p_a=p_b$ or $q_a=q_b$ for any $a,b=1,\ldots, N$, $a\neq b$ and a
pole of order two along divisors given by $p_a=q_b$ for all
$a,b=1,\ldots, N$.  It is straightforward to check that any such
differential must have extra zeroes on $\CM_0$. These zeroes do not have a
direct physical interpretation.

 Although a natural meromorphic form $\Omega$ with these properties
 does not exist, we can construct such a preferred  form once we choose
 a spin structure on $\Sigma$.  More precisely, let us pick a
 theta-characteristic $\epsilon$, and take $\Omega$ to be the square of
 the free fermion correlator \be\label{eq:fermcorrA} \langle
 \psi(p_1)\ldots \psi(p_N) {\overline \psi}(q_1)\ldots {\overline
 \psi}(q_N)\rangle.  \ee where $\psi$ is a complex spinor on $\Sigma$
 with respect to the spin structure $\epsilon$.  This is a well defined
 meromorphic top form on $\Sigma^N \times \Sigma^N$ 
exhibiting the physical behavior explained in the last paragraph. 

Now, one may legitimately ask why this peculiar construction is the
correct moduli space measure for the holomorphic Chern-Simons
theory. A short answer to this question is that although the physical
constraints do not fix the measure uniquely, they do fix the relevant
part of the measure in the large $N$ planar limit. That is the
structure of the semiclassical large $N$ vacua of the theory is
insensitive to changing the measure by adding a holomorphic top form
without zeroes or poles along the divisor \eqref{eq:polardiv}. This
will be manifest in the calculations performed in the next
section. Therefore this choice of the measure is as good as any other
choice exhibiting identical polar structure along the divisor
\eqref{eq:polardiv}.

However this is not a conceptually satisfying answer since the full
quantum theory involves much more than the large $N$ planar limit. To
gain a different perspective on this problem, recall that so far we
have ignored the coupling between holomorphic Chern-Simons theory and
Kodaira-Spencer theory. Indeed, although this coupling does not play
an important role in the large $N$ planar limit, it is certainly
expected to play an important role in the full quantum theory. Without
getting into too much detail at this point, let us mention that in the
present set-up Kodaira-Spencer theory can be described in terms of a
chiral boson on the Riemann surface $\Sigma$ by analogy with the
examples considered in \cite{AKMV:vertex,ADKMV:int}.  It is well known
that in order to define the theory of a chiral boson at quantum level,
one has to choose a spin structure on $\Sigma$.  Moreover, the branes
should be regarded as fermion operators in this theory related to the
Kodaira-Spencer field by bosonization \cite{AKMV:vertex,ADKMV:int}.
This explains the choice of a spin structure. The construction of the
measure in terms of fermionic correlators can be justified starting
from the identification between the Kodaira-Spencer field and the
collective field of the D-branes in the large $N$ limit
\cite{DV:geometry,ADKMV:int}.  We will fully develop these ideas
elsewhere.

It is also worth noting that the choice of a spin structure has a
natural geometric interpretation in the topological closed string
theory discussed in section five.  Recall that the genus zero
structure of the theory was shown to be encoded in an $A_1$ Hitchin
integrable system $\varpi:{\mathscr N} \to {\mathscr B}$. The fiber
${\mathscr N}_\beta$ are torsors over the Prym variety
$\hbox{Prym}(\widetilde{\Sigma}_{\beta}/\Sigma)$. The choice of a spin
structure on $\Sigma$ determines a section of the fibration
$\varpi:{\mathscr N} \to {\mathscr B}$, hence also an isomorphism
between each fiber ${\mathscr N}_\beta$ and the Prym. It would be very
interesting to understand the connection between this section and
Kodaira-Spencer theory on the open-closed side of the transition.

For computational purposes it is helpful to write the fermionic
correlator \eqref{eq:fermcorrA} in terms of $\vartheta$-functions and
prime forms.  To keep the technical complications to a minimum, we
will choose $\epsilon$ so that the corresponding Dirac operator has no
zero modes: $h^0(\Sigma, \epsilon) = h^1(\Sigma, \epsilon)=0$. In
particular $\epsilon$ has to be an even spin structure.

Let us briefly recall the construction of the prime form associated to
the Riemann surface $\Sigma$. We denote by ${\widetilde \Sigma}$ the
universal cover of $\Sigma$ and by $\tp, {\tq}, \ldots $ points on the
universal cover projecting to $p,q,\ldots$ on $\Sigma$.  The prime
form \cite{fay:theta,mumford:theta} is a $(-\half, -\half)$
differential on ${\widetilde \Sigma} \times {\widetilde \Sigma}$ so
that $E(\tp,\tq)$
 \[ 
 E(\tp, \tq) = 0 \Longleftrightarrow \ p=q,
 \] 
 and the order of vanishing along $p=q$ is one. One can construct such
 a form by picking up a nonsingular odd theta characteristic $\delta$
 with $h^0(\Sigma, \delta) =1$.  Then we have \be\label{eq:primeformA}
 E(\tp, \tq) = \frac{\vartheta(\tp-\tq+\delta)}{h(\tp) h(\tq)} \ee
 where $\vartheta$ is Riemann's theta function and $h$ is the unique
 (up to multiplication by a nonzero constant) section of $\delta$.
 Note that the function $E$ does not depend on the choice of $\delta$
 but it depends on the choice of a homology basis for $\Sigma$
 \cite[pg 17]{fay:theta}. Somewhat more invariantly we can view $E$ as
 a section in a line bundle on the surface $\Sigma\times \Sigma$. Let
 $\theta \to \op{Pic}^{g-1}(\Sigma)$ be the canonical theta line
 bundle on the degree $(g-1)$ Jacobian. Let $\op{AJ}_{\delta} :
 \Sigma\times \Sigma \to \op{Pic}^{g-1}(\Sigma)$ be the Abel-Jacobi
 map $\op{AJ}_{\delta}(p,q) = p-q+\delta$. Now the theta function
 $\vartheta(\tp-\tq+\delta)$ descends to a section of the line bundle
 $\op{AJ}_{\delta}^{*}\theta$ and so $E$ descends to a section (which
 will be denoted again by $E$) in the
 line bundle $(\op{AJ}_{\delta}^{*}\theta)\otimes p_{1}^{*} \delta
 \otimes p_{2}^{*}\delta$. It is easy to check that this section is
 holomorphic. Furthermore a straightforward application of the see-saw
 principle shows that $(\op{AJ}_{\delta}^{*}\theta)\otimes p_{1}^{*} \delta
 \otimes p_{2}^{*}\delta = \mathcal{O}_{\Sigma\times \Sigma}({\sf{Diag}})$
 and so up to scale $E$ is the unique holomorphic section in the line bundle
 $\mathcal{O}_{\Sigma\times \Sigma}({\sf{Diag}})$ corresponding to the
 diagonal  divisor ${\sf{Diag}} := \{ (x,y) \in \Sigma\times \Sigma |
 x= y\}$. Since the self-intersection ${\sf{Diag}}^{2} =     2 - 2g$
 of the divisor ${\sf{Diag}}$ is negative, it follows that
 $H^{0}({\Sigma\times \Sigma},\mathcal{O}({\sf{Diag}}))$ is one
 dimensional and so 
 $E$ necessarily vanishes along the diagonal.

 Now consider the following top degree meromorphic differential on
 ${\widetilde  \Sigma}^{2N}$    
 \be\label{eq:measureA} 
 \Omega(\tp_a, \tq_a) = 
 \frac{\vartheta(\sum_{a=1}^N \tp_a - \sum_{a=1}^N \tq_a +
   \epsilon)^2}{\vartheta(\epsilon)^2} 
 \frac{{\mathop{\prod_{a,b=1}^N}_{a\neq
       b}}E(\tp_a,\tp_b)E(\tq_a,\tq_b)}{\prod_{a,b=1}^N   
 E(\tp_a,\tq_b)^2}.
 \ee
 Using the modular transformation properties of $E(\tp, \tq)$ and the
 $\vartheta$-function,  
 one can check that $\Omega$ descends to a differential (denoted by
 the same letter) 
 on $\Sigma^{2N}$. Again we can interpret $\Omega$ as a 
 section in a line bundle on $\Sigma^{2N}$. If we write
 ${\sf{a}}_{\epsilon}  :
 \Sigma^{2N} \to \op{Pic}^{g-1}(\Sigma)$ for the Abel-Jacobi map given
 by ${\sf{a}}_{\epsilon}((\{ p_{a} \}_{a=1}^{N}, \{ q_{a}
 \}_{a=1}^{N})) := \epsilon + \sum_{a=1}^{N} (p_{a} - q_{a})$, then
 $\Omega$ is by definition a meromorphic section in the line bundle 
${\sf{a}}_{\epsilon}^{*}\theta\otimes
 \mathcal{O}_{\Sigma^{2N}}(2\Delta_{+} + 2\Delta_{-} - 2\Delta)$ with
 a double pole along the divisor $\Delta$.
Again  a quick computation with the see-saw
 principle identifies  ${\sf{a}}_{\epsilon}^{*}\theta\otimes
 \mathcal{O}_{\Sigma^{2N}}(2\Delta_{+} + 2\Delta_{-} - 2\Delta)$ 
with the line bundle
 $K_{\Sigma^{2N}}$. In other words, $\Omega$  is a meromorphic $2N$
 form on $\Sigma^{2N}$ which has a double pole precisely along the
 ``diagonal'' divisor $\Delta \subset \Sigma^{N}\times \Sigma^{N}$.
 According to \cite{VV:chiral}, the formula \eqref{eq:measureA}
 represents the 
 square of the fermionic correlator \eqref{eq:fermcorrA}. In the 
 following we will adopt \eqref{eq:measureA} as the measure for the
 moduli space integral. 

 For a complete definition of the quantum theory, we have to specify
 an integration contour in $\Sigma^{2N}$. A priori, there is no
 canonical choice of contour, and at the moment it is unclear what
 physical constraints should be imposed on such a contour. In all
 holomorphic Chern-Simons theories studied so far
 \cite{DV:matrix,EW:twistors}, the moduli space has a natural
 antiholomorphic involution $\tau : \CM \to \CM$ so that $\tau^*
 \Omega = {\overline \Omega}$.  In these case the contour ${\mathfrak G}$ is
 chosen to be the fixed point set of this involution, which is a
 special Lagrangian cycle calibrated by the measure $\Omega$.  In
 particular, the restriction of $\Omega$ to ${\mathfrak G}$ is a real
 differential form up to multiplication by a nonzero complex number.

This reality condition is important for a semiclassical analysis for
the following simple reason. The semiclassical vacua are determined as
the critical points of a complex function on $\Gamma$ obtained by
dividing $\Omega$ by some reference classical measure $\Omega_0$,
which is real. Generically, a complex valued function on the cycle
${\mathfrak G}$ does not have any critical points. Therefore, if we
choose ${\mathfrak G}$ to be an arbitrary cycle, we will not be able to
develop a semiclassical expansion of the theory. While this is not a
consistency requirement of a nonperturbative quantum field theory, we
will adopt the reality condition as a selection criterion for
${\mathfrak G}$. The issue of the dependence of the action on the
choice of a contour has been raised in the physics literature
before. In particular a reality condition appears in
\cite{DV:matrix}, and the subtleties of extending this condition to the
context of holomorphic matrix models are analyzed in detail in
\cite{BM:special}.

In our case, for a generic $\Sigma$, the moduli space does not admit
antiholomorphic involutions, therefore there is no canonical choice
for ${\mathfrak G}$.  Following the arguments of the previous
paragraph, we will choose ${\mathfrak G}$ to be any special Lagrangian
cycle with respect to the measure $\Omega$.  If $\Sigma$ admits an
antiholomorphic involution $\tau:\Sigma \to \Sigma$, we can construct
such a cycle as ${\mathfrak G} = \Gamma^{2N}$ where $\Gamma$ is a
component of the fixed point set of $\tau$ on $\Sigma$. In the absence
of a real structure, we will simply assume that we can find a closed
contour $\Gamma$ on $\Sigma$ so that $\Gamma^{2N}$ is special
Lagrangian with respect to $\Omega$.

 \subsection{Deformations and classical superpotential}  \label{ss:superp}

 So far we have formulated holomorphic Chern-Simons theory on
 ${\widetilde X}$ in terms of a contour integral of a meromorphic top
 form on the classical moduli space $\CM_0$. This is only a first step
 in our program since we are interested in ${\bf B}$-branes on a
 deformed threefold ${\widetilde X}_\alpha$.

 Recall that the threefold ${\widetilde X}_\alpha$ constructed in section four 
 has an affine bundle structure over the ruled 
 surface $S$. If $\alpha$ is generic, that is 
 \[
\hbox{div}(\alpha)=v_1+\ldots + v_{2g-2}
\] 
 with $v_1,\ldots, v_{2g-2}$ distinct points on $\Sigma$, then only
 the $2g-2$ fibers of the ruling $q:S\to \Sigma$ sitting over the
 points $\{ v_{i} \}$ will lift to holomorphic isolated $(-1,-1)$
 curves $C_1,\ldots, C_{2g-2}$ curves in 
 $\wX_\alpha$.

 The classical vacua of this theory consist of configurations of 
 $N_k^+$, $k=1,\ldots,r$ branes wrapped on $r$ curves $C_{s_k}$,
 $k=1,\ldots, r$,  
 and $N^-_k$, $k=r+1,\ldots,2g-2$ 
 antibranes wrapped on the remaining curves $C_{s_k}$, $k=r+1,\ldots, 2g-2$ 
 for some $1\leq r \leq 2g-3$. The multiplicities are subject
 to the constraint 
 \be\label{eq:mult} 
 \sum_{k=1}^r N_k^+ =\sum_{k=r+1}^{2g-2} N_k^-=N.
 \ee 
 In terms of the holomorphic gauge theory on $\Sigma$, these D-brane
 configurations  
 correspond to points of the form 
 \be\label{eq:clvacB} 
 D_+=\sum_{k=0}^r N_k^+ v_{s_k}, \qquad D_- = \sum_{k=r+1}^{2g-2}
 N_k^-v_{s_k} 
 \ee 
 of the undeformed moduli space. Without loss of generality 
 we can set $s_k =k$, $k=1,\ldots 2g-2$ in the following. 

Following the strategy of Dijkgraaf-Vafa transitions, holomorphic
Chern-Simons theory on $\wX_\alpha$ can be defined as a superpotential
deformation of the $\alpha=0$ theory. The superpotential in question
should be a (possibly multivalued) holomorphic function on the finite
cover $\CM_0'$ of the moduli space whose critical points are in
one-to-one correspondence with classical D-brane configurations.

As explained in section~\ref{sec:DV}, such a function has a natural
geometric origin.  Indeed, one can construct a transverse holomorphic
family $\CC$ of two-cycles on ${\widetilde X}_\alpha$ parameterized by
$\Sigma$ including the holomorphic curves $C_1,\ldots, C_{2g-2}$ at
the points $v_1,\ldots, v_{2g-2}$. The generic fiber $C_p$ of this
family over a point $p\in \Sigma$ is a smooth non holomorphic
two-cycle on the affine quadric ${(\wX_\alpha)}_p$, similarly to the
local situation described in section 2.

This family determines a Donaldson-Thomas superpotential on the moduli
space via the Abel-Jacobi map by analogy with the considerations of
section~\ref{sec:DV}.  The only difference is that in the present case, the
Donaldson-Thomas superpotential is multivalued since $\Sigma$ contains
nontrivial homology one-cycles.  Therefore in order to obtain a single
valued expression we have to work on the universal cover ${\widetilde
\Sigma}^{2N}$ of the direct product $\Sigma^{2N}$.  The superpotential
is then given by
\be\label{eq:superdef}
  W_\alpha(\tp_a, \tq_a) = \sum_{a=1}^N \int_{\tq_a}^{\tp_a} \alpha.
 \ee
 It is a simple exercise to check that the critical points of 
 $W_\alpha$ are classical vacua of the form \eqref{eq:clvacB}. 

 Summarizing this discussion, the deformed holomorphic Chern-Simons
 action will be defined by a measure \be\label{eq:deformA}
 \Omega_\alpha(\tp_a,\tq_a) = \Omega(\tp_a,\tq_a) e^{-\frac{1}{g_s}
 W_\alpha(\tp_a,\tq_a)} \ee on ${\widetilde \Sigma}^{2N}$.  To
 complete the construction, we have to specify a middle dimensional
 cycle ${\mathfrak G}$ on ${\widetilde \Sigma}^{2N}$. According to the
 discussion of the previous subsection, ${\mathfrak G}$ should be a
 special Lagrangian cycle with respect to the deformed measure
 \eqref{eq:deformA}.  In the following we will assume that the cycle
 $\mathfrak{G}$ can be chosen of the form ${\mathfrak G} =
 \Gamma^{2N}$ where $\Gamma$ is a closed one-cycle on $\Sigma$ passing
 through the zeroes of $\alpha$. This assumption is not
 unreasonable. For instance if $\Sigma$ is a real curve equipped
 with an antiholomorphic involution $\tau:\Sigma \to \Sigma$, and
 $\alpha$ satisfies an appropriate reality condition, $\Gamma$ can be
 chosen to be a component of the fixed point set of $\tau$.

 \subsection{Semiclassical vacua at large $N$ and Hitchin systems}
 \label{ss:semiclassical} 

 In this subsection we determine the semiclassical vacua of the
 deformed holomorphic Chern-Simons theory in the large $N$ limit. The
 main result is that the semiclassical vacua are in one-to-one
 correspondence to $A_1$ Hitchin spectral covers of $\Sigma$. This
 establishes a direct connection with the algebraic integrable systems
 found in the previous section.

 In order to derive the semiclassical equations of motion we will
 rewrite the measure $\Omega_\alpha$ in a more explicit form. Let $U$
 denote the complement of the divisor $h = 0$ in 
 ${\widetilde \Sigma}$. We define a local coordinate function $z:U\ra
 \IC$ so that  
 \[ 
 dz(\tp) = h(\tp)^2
 \]
 as in \cite[pg. 3.207-3.211]{mumford:theta}. 
 The restriction of $\Omega$ to $U^{2N}$ can then be written as
 \be\label{eq:locmeasureA} 
 \Omega\big |_{U} = \frac{{\vartheta(\sum_{a=1}^N \tp_a - \sum_{a=1}^N
 \tq_a+\epsilon)^2}}{\vartheta(\epsilon)^2}\cdot
 \frac{\mathop{\prod_{a,b=1}^N}_{a\neq 
 b}\vartheta(\tp_a-\tp_b+\delta)
 \vartheta(\tq_a-\tq_b+\delta)}{\prod_{a,b=1}^N
 \vartheta(\tp_a-\tq_b+\delta)^2}  
 \prod_{a=1}^N dz(\tp_a)dz(\tq_a)
 \ee

 The deformed measure $\Omega_\alpha$ gives rise to an effective semiclassical 
 superpotential 
 \be\label{eq:sclsupA}
 \begin{aligned}
 {1\over g_s} W^{\op{eff}}_\alpha(\tp_a,\tq_a) = &  
 {1\over g_s} W_\alpha(\tp_a,\tq_a) - \mathop{\sum_{a,b=1}^N}_{a\neq b}(
 \log \vartheta(\tp_a-\tp_b + \delta) + \log \vartheta(\tq_a-\tq_b+\delta)) \cr
 & + \sum_{a,b=1}^N \log \vartheta(\tp_a-\tq_b+\delta)
 - 2\log\vartheta\left(\sum_{a=1}^N \tp_a -\sum_{a=1}^N \tq_a
 +\epsilon\right).\cr 
 \end{aligned} 
 \ee
 Applying the variational principle to $W^{\op{eff}}_\alpha(\tp_a,\tq_a)$
 we obtain the following  
 semiclassical equations of motion 
 \be\label{eq:scleqmotion} 
 \begin{split}
 -{1\over g_s} \alpha(\tp_a) + 2 \mathop{\sum_{b=1}^N}_{b\neq a}
 d_{\tp_a}  
 \log \vartheta(\tp_a-\tp_b+\delta) - & 2 \mathop{\sum_{b=1}^N}_{b\neq a}  
 d_{\tp_a} 
 \log\vartheta(\tp_a-\tq_b+\delta) \\
 & +2 d_{\tp_a}\log\vartheta\left(\sum_{b=1}^N\tp_b 
 -\sum_{b=1}^N\tq_b +\epsilon\right)=0 \\
 {1\over g_s} \alpha(\tq_a) + 2 \mathop{\sum_{b=1}^N}_{b\neq a} d_{\tq_a} 
 \log \vartheta(\tq_a-\tq_b+\delta) - & 2 \mathop{\sum_{b=1}^N}_{b\neq a} 
 d_{\tq_a} \log \vartheta(\tp_b-\tq_a+\delta)\\ 
 & +2 d_{\tq_a}\log\vartheta\left(\sum_{b=1}^N\tp_b 
 -\sum_{b=1}^N\tq_b +\epsilon\right)=0.
 \end{split} 
 \ee
 In order to study the large $N$ limit of these equations, let us
 introduce the meromorphic  
 differential 
 \be\label{eq:resolventA} 
 \omega (\tp) = {1\over N} d_\tp \sum_{a=1}^N \log {E(\tp,\tp_a)\over
 E(\tp,\tq_a)}. 
 \ee
 It is easy to check that $\omega$ is invariant under the action of
 the fundamental group,  
 therefore it descends to a meromorphic one-form on $\Sigma$. 
 This is an Abelian differential of the third kind with simple poles
 with residue   
${1\over N}$ at 
 $p_a$, $a=1,\ldots, N$ and simple poles with residue $-{1\over N}$ at
 $q_a$, $a=1,\ldots, N$.  
 In the present context $\omega$ plays the same 
 role as the resolvent in the large $N$ solution of matrix models. 
 Over $U$ we can write
 \be\label{eq:resolventB} 
 \omega(\tp) = {1\over N} \sum_{a=1}^N d_\tp \log
 {\vartheta(\tp-\tp_a+\delta) \over  
 \vartheta(\tp-\tq_a+\delta)}.
 \ee
 Our goal is to show that in the large $N$ limit the equations of
 motion \eqref{eq:scleqmotion}  
 give rise to an algebraic equation of the form 
 \[ 
 \omega^2 -{1\over \mu} \alpha \omega = \hbox{holomorphic deformation} 
 \]
 analogous to the loop equation of matrix models. Here $\mu = Ng_s$ is
the 't Hooft coupling constant which is kept finite in the large $N$
limit.  Note that the theta function $\vartheta(\tp-\tp_a+\delta)$ has
a first order zero at $\tp=\tp_a$ \cite[pg. 311]{mumford:theta}.
Therefore locally we can write 
\be\label{eq:localtheta}
\vartheta(\tp-\tp_a+\delta) = (z(\tp)-z(\tp_a)){\widetilde
\vartheta}(\tp,\tp_a) 
\ee where 
${\widetilde \vartheta}(\tp,\tp_a)$ is
a holomorphic function over $U$ non vanishing at $\tp=\tp_a$,
$a=1,\ldots, N$.  Then $\omega$ can be further rewritten in the form
\be\label{eq:resolventC} \omega(\tp) = {1\over N}
\sum_{a=1}^N\left({dz(\tp)\over z(\tp)-z(\tp_a)} -{dz(\tp)\over
z(\tp)-z(\tq_a)} + d_{\tp}\log {\widetilde \vartheta}(\tp,\tp_a)
-d_\tp \log {\widetilde \vartheta}(\tp,\tq_a)\right).  \ee The
equations of motion \eqref{eq:scleqmotion} yield
\be\label{eq:scleqmotionB}
 \begin{aligned} 
 & -{1\over g_s}\sum_{a=1}^N {1\over z(\tp)-z(\tp_a)} {\alpha(\tp_a)\over dz(\tp_a)} 
 + 2 \mathop{\sum_{a,b=1}^N}_{b\neq a} {1\over z(\tp)-z(\tp_a)} {d_{\tp_a} 
 \log \vartheta(\tp_a-\tp_b+\delta)\over dz(\tp_a)}\cr
 & -2 \mathop{\sum_{a,b=1}^N}_{b\neq a} {1\over z(\tp)-z(\tp_a)}
 {d_{\tp_a}\log\vartheta(\tp_a-\tq_b+\delta)\over dz(\tp_a)} \cr &
  +2 \sum_{a=1}^N {1\over z(\tp)-z(\tp_a)}{d_{\tp_a}\log\vartheta(\sum_{b=1}^N\tp_b 
 -\sum_{b=1}^N\tq_b +\epsilon)\over dz(\tp_a)}=0 \cr
 & \ \ \ {1\over g_s} \sum_{a=1}^N {1\over z(\tp)-z(\tq_a)}{\alpha(\tq_a)\over dz(\tq_a)} + 
 2 \mathop{\sum_{a,b=1}^N}_{b\neq a} {1\over z(\tp)-z(\tq_a)} {d_{\tq_a} 
 \log \vartheta(\tq_a-\tq_b+\delta)\over dz(\tq_a)} \cr
 & -2 \mathop{\sum_{a,b=1}^N}_{b\neq a} {1\over z(\tp)-z(\tq_a)}
 { d_{\tq_a} \log \vartheta(\tp_b-\tq_a+\delta)\over dz(\tq_a)}\cr & 
  +2 \sum_{a=1}^N {1\over z(\tp)-z(\tq_a)}{ d_{\tq_a}\log\vartheta(\sum_{b=1}^N\tp_b 
 -\sum_{b=1}^N\tq_b +\epsilon)\over dz(\tq_a)}=0.\cr
 \end{aligned} 
 \ee
 Using \eqref{eq:resolventC}, \eqref{eq:scleqmotionB}, a somewhat tedious computation yields 
 \be\label{eq:largeNE} 
 \begin{aligned} 
 & \left({\omega(\tp)\over dz(\tp)}\right)^2 - {1\over Ng_s}{\omega(\tp)\over dz(\tp)}
 {\alpha(\tp)\over dz(\tp)} = \cr
 & -{1\over N^2g_s} \sum_{a=1}^N\left[{1\over z(\tp)-z(\tp_a)}
 \left({\alpha(\tp)\over dz(\tp)}-{\alpha(\tp_a)\over 
 dz(\tp_a)}\right) - {1\over z(\tp)-z(\tq_a)}\left({\alpha(\tp)\over dz(\tp)}
 -{\alpha(\tq_a)\over dz(\tq_a)}\right)\right]\cr
 & -{1\over N^2g_s}\sum_{a=1}^N\left({d_{\tp}\log 
 {\widetilde \vartheta}(\tp,\tp_a) -d_\tp \log  {\widetilde \vartheta}(\tp,\tq_a)\over 
 dz(\tp)} {\alpha(\tp)\over dz(\tp)}\right)\cr
 & + {1\over N^2} \sum_{a=1}^N\left[ \left({d_\tp \log \vartheta(\tp-\tp_a+\delta)\over dz(\tp)}\right)^2
 + \left({d_\tp \log \vartheta(\tp-\tq_a+\delta)\over dz(\tp)}\right)^2\right]\cr
 & + {2\over N^2} \mathop{\sum_{a,b=1}^N}_{a\neq b} {1\over z(\tp)-z(\tp_a)}
 \left({d_\tp \log {\widetilde \vartheta}(\tp,\tp_b) \over dz(\tp)}- 
 {d_{\tp_a} \log {\widetilde \vartheta}(\tp_a,\tp_b) \over dz(\tp_a)}\right)\cr
 & + {2\over N^2} \mathop{\sum_{a,b=1}^N}_{a\neq b} {1\over z(\tp)-z(\tq_a)}
 \left({d_\tp \log {\widetilde \vartheta}(\tp,\tq_b) \over dz(\tp)}- 
 {d_{\tq_a} \log {\widetilde \vartheta}(\tq_a,\tq_b) \over dz(\tq_a)}\right)\cr
\end{aligned}
\ee 
\[
\begin{aligned}
& -{2\over N^2} \sum_{a,b=1}^N {1\over z(\tp)-z(\tp_a)}
 \left({d_\tp \log {\widetilde \vartheta}(\tp,\tq_b) \over dz(\tp)}- 
 {d_{\tp_a} \log {\widetilde \vartheta}(\tp_a,\tq_b) \over dz(\tq_a)}\right)\cr
 & -{2\over N^2} \sum_{a,b=1}^N {1\over z(\tp)-z(\tq_a)}
 \left({d_\tp \log {\widetilde \vartheta}(\tp,\tp_b) \over dz(\tp)}- 
 {d_{\tq_a} \log {\widetilde \vartheta}(\tq_a,\tp_b) \over dz(\tq_a)}\right)\cr
 & +{1\over N^2} \mathop{\sum_{a,b=1}^N}_{a\neq b} \left(
 {d_\tp \log {\widetilde \vartheta}(\tp,\tp_a) \over dz(\tp)}
 {d_\tp \log {\widetilde \vartheta}(\tp,\tp_b) \over dz(\tp)}+ 
 {d_\tp \log {\widetilde \vartheta}(\tp,\tq_a) \over dz(\tp)}
 {d_\tp \log {\widetilde \vartheta}(\tp,\tq_b) \over dz(\tp)}\right)\cr
 & -{2\over N^2} \sum_{a,b=1}^N {d_\tp \log {\widetilde \vartheta}(\tp,\tp_a) \over dz(\tp)}
 {d_\tp \log {\widetilde \vartheta}(\tp,\tq_b) \over dz(\tp)}\cr
 & -{2\over N^2} \sum_{a=1}^N {1\over z(\tp)-z(\tp_a)}{d_{\tp_a}\log\vartheta(\sum_{b=1}^N\tp_b 
 -\sum_{b=1}^N\tq_b +\epsilon)\over dz(\tp_a)}\cr
 & -{2\over N^2} \sum_{a=1}^N 
 {1\over z(\tp)-z(\tq_a)}{d_{\tq_a}\log\vartheta(\sum_{b=1}^N\tp_b 
 -\sum_{b=1}^N\tq_b +\epsilon)\over dz(\tq_a)}.\cr
 \end{aligned}
 \]
 Now we take the limit $N\to \infty$, $g_s\to 0$ keeping the 't Hooft coupling 
 $\mu = Ng_s$ fixed. The behavior of the terms in the equation
 \eqref{eq:largeNE} is  
determined by their scaling with $N$. Terms of the form 
\[ 
{1\over N} \sum_{a=1}^N \ldots \qquad \hbox{and} \qquad 
{1\over N^2} \sum_{a,b=1}^N \ldots 
\] 
are expected to have finite limit when $N\to \infty$, whereas terms  
of the form 
\[ 
{1\over N^2} \sum_{a=1}^N \cdots 
\] 
tend to zero because they are suppressed by an extra power of $N$.
Using these rules, we can check that the right hand side of equation
\eqref{eq:largeNE} is a holomorphic function on $U$ in the large
$N$ limit.  The polar part of that expression -- that is the terms in
the second and last two lines -- scales as $1/N$, therefore it
vanishes in the large $N$ limit. The remaining terms are nonzero, but
the poles cancel even at finite $N$.

The limit of the left hand side of equation \eqref{eq:largeNE} is a
quadratic expression of the form  
\be\label{eq:quadr} 
\omega_\infty^2 - {1\over \mu } \omega_\infty \alpha 
\ee
where $\omega_\infty$ is the large $N$ limit of the meromorphic form $\omega$ which characterizes the 
distribution of branes on $\Sigma$. We will assume that the limit $\omega_\infty$ exists and it is a 
well defined mathematical object on $\Sigma$. 
By construction, the expression \eqref{eq:quadr} can only have poles 
at the locations of the branes on $\Sigma$. However, we have shown in
the previous paragraph that these  
poles are absent from the left hand side of equation \eqref{eq:largeNE}.
Therefore we can conclude that the large $N$ limit of equation
\eqref{eq:largeNE} is of the form  
\be\label{eq:largeNEE} 
\omega_\infty -{1\over \mu} \alpha \omega_\infty = \beta 
\ee
where $\beta$ is a global holomorphic quadratic differential on $\Sigma$. 

Then $\omega_\infty$ must be a multivalued holomorphic differential on
$\Sigma$ with branch points located at the zeroes of $\beta$. Note
that by construction $\omega_\infty$ characterizes the distribution of
branes on $\Sigma$ in the large $N$ limit. For finite $N$, $\omega$
has poles at the locations of the branes, therefore we would naively
expect $\omega_\infty$ to have infinitely many poles on $\Sigma$. The
only way such a mathematical object can be well defined is if the
collection of poles of $\omega$ becomes a collection of branch cuts
$\Gamma_1,\ldots, \Gamma_{2g-2}$ in the large $N$ limit, and
$\omega_\infty$ is a multivalued differential. This is consistent with
the large $N$ limit \eqref{eq:largeNEE} of the equation of motion and
it also shows that the branch points must be located on the contour
$\Gamma$ and the branch cuts must be some line segments contained in
$\Gamma$.  The filling fraction associated to each branch cut is
determined by the period \be\label{eq:fillingA} \int_{\gamma_s}
\Omega_\infty \ee where $\gamma_s$ is a contour on $\Sigma$
surrounding the branch cut $\Gamma_s$, $s=1,\ldots 2g-2$.

To summarize: we have found that the large $N$ semiclassical vacuum
configurations are in one-to-one correspondence with quadratic
holomorphic differentials $\beta$. The distribution of branes in a
semiclassical vacuum determined by $\beta$ is encoded in the
multivalued differential $\omega_\infty$ which solves
\eqref{eq:largeNEE}.

Now the connection with Hitchin spectral covers becomes
manifest. Equation \eqref{eq:largeNEE} is the defining equation of a
spectral cover ${\widetilde \Sigma}_{\beta}$ and $\omega_\infty$ is
the canonical holomorphic differential on ${\widetilde \Sigma}_\beta$.
Each contour $\gamma_s$ on $\Sigma$ lifts to an anti-invariant closed
one-cycle ${\widetilde \gamma}_s$ on $\Sigma_\beta$, and the filling
fractions \eqref{eq:fillingA} are given by the periods of
$\omega_\infty$ on the cycles ${\widetilde \gamma}_s$.

Taking into account the relation between Hitchin Pryms and homology
intermediate Jacobians proved in section \ref{ss:IJ}, the filling fractions
\eqref{eq:fillingA} can be related to periods of the holomorphic
three-form $\Omega_{X_\beta}$ on the threefold $X_\beta$ constructed
in section \ref{sec:linear}.

In conclusion, we have shown that the large $N$ limit of the
holomorphic Chern-Simons theory is governed by a Hitchin integrable
system, which is in turn isomorphic to the Calabi-Yau integrable
system for the universal family of Calabi-Yau threefolds over the   of
the moduli space $\bL$. This is a physical proof of
large $N$ duality at genus zero.

\bigskip
\bigskip

\appendix

 \Appendix{Morphisms of Twisted Complexes} \label{app-twisted}

 \bigskip

 \noindent
 In this appendix we prove equation \eqref{eq:algmorphisms} in
 section~\ref{ss:moduli}.  For convenience, recall that we are given
 two sets of algebraic data $(\CF_i,\Psi_{ji})$, $(\CF'_i,
 \Psi'_{ji})$, $i,j=0,1,2$ satisfying conditions $(i)$-$(iii)$ below
 equation \eqref{eq:splittingE}. In particular we have two three term
 complexes $(\CF, \CF')$ and we construct the extensions
 \be\label{eq:extA}
 \begin{aligned} 
 & 0\to \CF\otimes \CO(K_\Sigma)[-2] \to \CK \to \CF \to 0 \cr
 & 0\to \CF'\otimes \CO(K_\Sigma)[-2] \to \CK' \to \CF' \to 0\cr
 \end{aligned} 
 \ee
 which yield the short exact sequences of complexes
 \be\label{eq:extB}
 \begin{aligned} 
 & 0\to {Hom}(\CF,\CF')\otimes \CO(K_\Sigma)[-2] \to 
 {Hom}(\CK\otimes \CO(-K_\Sigma)[2],\CF') \to {Hom}(\CF,\CF') \to 0.\cr
 & 0\to {Hom}(\CF,\CF')\otimes \CO(K_\Sigma)[-2] \to 
 {Hom}(\CF,\CK') \to {Hom}(\CF,\CF') \to 0.\cr
 \end{aligned} 
 \ee
 More explicitly, using the notation of section 5.2, we have 
 \be\label{eq:extBA}
 \begin{aligned} 
 & \CF_0 {\buildrel \Psi_{10}\over \ra} \CF_1 {\buildrel {\widetilde \Psi}_{21} \over \ra } 
 \CS {\buildrel {\widetilde \Psi}_{10} \over \ra}\CF_1\otimes \CO(K_\Sigma) 
 {\buildrel \Psi_{21}\otimes 
 \II_{\CO(K_\Sigma)}\over \ra} \CF_2\otimes \CO(K_\Sigma).\cr
 & \CF'_0 {\buildrel \Psi'_{10}\over \ra} \CF'_1 {\buildrel {\widetilde \Psi}'_{21} \over \ra } 
 \CS' {\buildrel {\widetilde \Psi}'_{10} \over \ra}\CF'_1\otimes \CO(K_\Sigma) 
 {\buildrel \Psi'_{21}\otimes 
 \II_{\CO(K_\Sigma)}\over \ra} \CF'_2\otimes \CO(K_\Sigma). \cr
 \end{aligned} 
 \ee
 where the sheaves $\CS, \CS'$ are given by extensions 
 \be\label{eq:extBB} 
 \begin{aligned} 
 & 0\to \CF_0(K_\Sigma){\buildrel i\over \ra} \CS {\buildrel \rho \over \ra} \CF_2 \to 0\cr
 & 0\to \CF'_0(K_\Sigma){\buildrel i'\over \ra} \CS' {\buildrel \rho' \over \ra} \CF'_2 \to 0.\cr
 \end{aligned} 
 \ee 
 Our problem is to construct the difference $C(\CK,\CK')$ of the two extensions 
 \eqref{eq:extB} and compute its hypercohomology. 

 Suppose we have two exact sequences of complexes of coherent sheaves on 
 a smooth projective variety  
 \be\label{eq:extC} 
 \begin{aligned} 
 & 0 \to A \to C \to B\to 0 \cr
 & 0 \to A \to C' \to B\to 0.\cr
 \end{aligned} 
 \ee
 The difference $C'-C$ can be constructed in two steps. First take the pull-back extension  
 \be\label{eq:extD} 
 \xymatrix{ 
 0 \ar[r]& A\oplus A  \ar[r]& C\oplus C'  \ar[r]& B\oplus B \ar[r] & 0  \\
 0 \ar[r] & A\ar[u]^{\iota^-} \ar[r]\ar[u] & {C\oplus C'\over \iota^-(A)}\ar[u] \ar[r] & 
 B\oplus B\ar[u]^{\simeq} \ar[r] & 0\\}
 \ee
 where $\iota^-: A\to A\oplus A$ is the anti diagonal embedding. Then take a second 
 pullback 
 \be\label{eq:extE} 
 \xymatrix{ 
 0 \ar[r] & A \ar[r] & {C\oplus C'\over \iota^-(A)} \ar[r] & B\oplus B \ar[r] & 0\\
 0 \ar[r] & A \ar[r]\ar[u]^{\simeq} & C'-C \ar[r]\ar[u] & B\ar[u]^{\iota^+}\ar[r] & 0\\} 
 \ee 
 where $\iota^+:B \to B\oplus B$ is the diagonal embedding. 

 We have to carry out this construction for the extensions \eqref{eq:extB}. 
In order to keep the 
 formulas short, we will use the notation $H_{nm} \equiv Hom(\CF_m,\CF'_n)$.
We will first write down explicitly the complexes 
$Hom(\CF,\CF'), Hom(\CF,\CF')\otimes \CO(K_\Sigma)[-2]$
and then write down the complex $C(\CK,\CK')$ as an extension of complexes, 
obtaining the following diagrams.
\newpage  
 \be\label{eq:homA}
 \begin{aligned}
 &\qquad \qquad\quad Hom(\CF,\CF'): & &\qquad\quad Hom(\CF,\CF')\otimes \CO(K_\Sigma)[-2]:\cr
 &\xymatrix{ H_{02} \ar[d] \\ H_{12} \oplus H_{01} \ar[d] \\ 
 Hom(\CF_2,\CS')\oplus H_{11} \oplus H_{00} \ar[d]\\
 H_{12}(K_\Sigma) \oplus 
 Hom(\CF_1,\CS') \oplus H_{10} \ar[d] \\ H_{22}(K_\Sigma) \oplus 
 H_{11}(K_\Sigma) \oplus Hom(\CF_0,\CS') \ar[d] \\ 
 H_{21}(K_\Sigma) \oplus H_{10}(K_\Sigma) \ar[d] \\
 H_{20}(K_\Sigma)\\}& 
 & \xymatrix{ H_{02} \ar[d] \\H_{12} \oplus H_{01} \ar[d] \\ 
 H_{22} \oplus H_{11} \oplus Hom(\CS,\CF_0')(K_\Sigma) \ar[d]\\ 
 H_{21}\oplus Hom(\CS,\CF_1')(K_\Sigma) \oplus H_{01}(K_\Sigma)\ar[d] \\ 
 Hom(\CS,\CF_2')(K_\Sigma) \oplus H_{11}(K_\Sigma) \oplus H_{00}(K_\Sigma) \ar[d]\\
 H_{21}(K_\Sigma)\oplus H_{10}(K_\Sigma) \ar[d]\\ 
 H_{20}(K_\Sigma) \\} \cr
 \end{aligned} 
 \ee
  \be\label{eq:homC} 
 \xymatrix{ 0 \ar[r] \ar[d] & \ckk^{-2}\ar[r] \ar[d]^{c_{-1,-2}} & H_{02}\ar[d]\\ 
 0 \ar[r]\ar[d] & \ckk^{-1}\ar[r] \ar[d]^{c_{0,-1}} & H_{12}\oplus H_{01} \ar[d]\\
 H_{02}(K_\Sigma)\ar[d]\ar[r] & \ckk^{0} \ar[d]^{c_{1,0}}\ar[r] 
 & H_{22}\oplus H_{11}\oplus H_{00}\ar[d]\\
 H_{12}(K_\Sigma) \oplus H_{01}(K_\Sigma)\ar[d]\ar[r] & \ckk^1 \ar[r] \ar[d]^{c_{21}} 
 & H_{21}\oplus H_{10}\ar[d]\\
 H_{22}(K_\Sigma) \oplus H_{11}(K_\Sigma) \oplus H_{00}(K_\Sigma) \ar[d]\ar[r]
 & \ckk^2\ar[r]\ar[d]^{c_{32}} & H_{20}\ar[d]\\
 H_{21}(K_\Sigma) \oplus H_{10}(K_\Sigma) \ar[d]\ar[r]&\ckk^3\ar[d]^{c_{43}}\ar[r] & 0 \ar[d]\\
 H_{20}(K_\Sigma)\ar[r] & \ckk^4 \ar[r] &0 \\}
 \ee
 This determines the terms of degrees $-2,-1,3,4$. The remaining terms can be determined following
 the steps \eqref{eq:extD}-\eqref{eq:extE} described above. In degree $0$, the first step yields 
 \be\label{eq:diffA}
 {C\oplus C' \over \iota^-(A)} = {Hom(\CF_2,\CS') \oplus Hom(S,\CF_0')(K_\Sigma) 
 \over \iota^-(H_{02}(K_\Sigma))} \oplus H_{22} \oplus H_{11}^{\oplus 2} \oplus H_{00}
 \ee 
 and the final result is 
 \be\label{eq:diffB} 
 C(\CK',\CK)^0 = {Hom(\CF_2,\CS') \oplus Hom(S,\CF_0')(K_\Sigma) 
 \over \iota^-(H_{02}(K_\Sigma))} \oplus H_{11}.
 \ee 
 In degree one, we obtain at the first step 
 \be\label{eq:diffC} 
 \begin{aligned} 
 {C\oplus C' \over \iota^-(A)} & ={ H_{12}(K_\Sigma) \oplus Hom(\CF_1,\CS') \oplus H_{10}\oplus 
 H_{21}\oplus Hom(\CS,\CF_1')(K_\Sigma) \oplus H_{01}(K_\Sigma)\over \iota^-(H_{12}(K_\Sigma)
 \oplus H_{01}(K_\Sigma))}\cr
 & ={ H_{12}(K_\Sigma) \oplus Hom(\CF_1,\CS')\oplus 
 Hom(\CS,\CF_1')(K_\Sigma) \oplus H_{01}(K_\Sigma)\over \iota^-(H_{12}(K_\Sigma)
 \oplus H_{01}(K_\Sigma))}\oplus H_{10}\oplus H_{21}\cr
 & \simeq Hom(\CF_1,\CS')\oplus 
 Hom(\CS,\CF_1')(K_\Sigma)\oplus H_{10}\oplus H_{21}.\cr
 \end{aligned}
 \ee
 After the second step we obtain
 \be\label{eq:diffD} 
 \ckk^1 \simeq  Hom(\CF_1,\CS')\oplus Hom(\CS,\CF_1')(K_\Sigma).
 \ee 
 The term of degree two can be similarly determined to be 
 \be\label{eq:diffE} 
 \ckk^2 \simeq Hom(\CF_0,\CS') \times_{H_{20}} Hom(\CS,\CF_2')(K_\Sigma) \oplus H_{11}(K_\Sigma).
 \ee 
 Now let us determine the differentials. The first and the last differentials are standard, 
 hence we will focus on the remaining ones. We have 
 \be\label{eq:diffF}
 \begin{aligned} 
 c_{0,-1}: H_{12}\oplus H_{01} & \to  {Hom(\CF_2,\CS') \oplus Hom(\CS,\CF_0')(K_\Sigma) 
 \over \iota^-(H_{02}(K_\Sigma))} \oplus H_{11}\cr
 (s_{12},s_{01}) & \to \left(\left[{\widetilde \Psi}'_{21}s_{12}, 
  +(s_{01}\otimes {\mathbb I}_{K_\Sigma})
 {\widetilde \Psi}_{10}\right], s_{12}\Psi_{21} + \Psi'_{10} s_{01}\right) \cr
 \end{aligned}
 \ee
 where we use the notation $[\ ,\ ]$ for equivalence classes in the quotient 
 $(Hom(\CF_2,\CS') \oplus Hom(\CS,\CF_0')(K_\Sigma)) /\iota^-(H_{02}(K_\Sigma))$. 
 Next, 
 \be\label{eq:diffG} 
 \begin{aligned}
 c_{10}: & {Hom(\CF_2,\CS') \oplus Hom(\CS,\CF_0')(K_\Sigma) 
 \over \iota^-(H_{02}(K_\Sigma))} \oplus H_{11} \to
 Hom(\CF_1,\CS')\oplus Hom(\CS,\CF_1')(K_\Sigma)\cr
 & ([u,v],w)  \to \left(-u\Psi_{21}-
 i'v{\widetilde \Psi}_{21} + {\widetilde \Psi}'_{21}w, {\widetilde \Psi}'_{10}u\rho +(\Psi'_{10}
 \otimes {\mathbb I}_{K_\Sigma})v -w {\widetilde \Psi}_{10}\right)\cr
 \end{aligned} 
 \ee
 using the notation of equation \eqref{eq:extBB}. 
 Let us check that the map $c_{10}$ is well defined on equivalence classes $[u,v]$. It suffices 
 to show that $c_{10}([i's, -s\rho],0) =0$ for any element $s\in H_{02}(K_\Sigma)$. We have 
 \[ 
 \begin{aligned} 
 c_{10}([i's, s\rho],0) & = \left(i's\Psi_{21} -i's\rho{\widetilde \Psi}_{21}, 
 -{\widetilde \Psi}'_{10}i's\rho + (\Psi'_{10}\otimes {\mathbb I}_{K_\Sigma})s\rho\right)\cr
 & = \left(i's\Psi_{21} -i's\Psi_{21}, -(\Psi'_{10}\otimes {\mathbb I}_{K_\Sigma})s\rho + 
 (\Psi'_{10}\otimes {\mathbb I}_{K_\Sigma})s\rho\right) \cr & =0\cr
 \end{aligned} 
 \] 
 where we have used the relations 
 \[ \rho{\widetilde \Psi}_{21}=\Psi_{21},\qquad 
 {\widetilde \Psi}'_{10}i' = \Psi'_{10}\otimes {\mathbb I}_{K_\Sigma}
 \]
 following from diagrams \eqref{eq:splittingA},\eqref{eq:splittingB}. 
 One can similarly check that $c_{10}c_{0,-1}=0$ using the relations 
 \[
 {\widetilde \Psi}_{10}\Psi_{21} = (\Psi_{21}\otimes {\mathbb I_{K_\Sigma}}){\widetilde \Psi}_{10}=0,
 \qquad 
 {\widetilde \Psi}_{10}'{\widetilde \Psi}_{21}'={\widetilde \Psi}'_{21}\Psi'_{10} =0 
 \]
 following from \eqref{eq:extBA}. Proceeding in a similar manner, we find 
 \be\label{eq:diffH}
 \begin{aligned}
 c_{21}: & Hom(\CF_1,\CS')\oplus Hom(\CS,\CF_1')(K_\Sigma)  \to 
 Hom(\CF_0,\CS') \times_{H_{20}} Hom(\CS,\CF_2')(K_\Sigma) \oplus H_{11}(K_\Sigma)\cr
 & (x,y) \to \left(x\Psi_{10}+({\widetilde \Psi}'_{21}\otimes {\mathbb I}_{K_\Sigma})yi, 
 (\rho'\otimes {\mathbb I}_{K_\Sigma}) x{\widetilde \Psi}_{10} + 
 (\Psi'_{21}\otimes {\mathbb I}_{K_\Sigma})y, {\widetilde \Psi}'_{10}x + y {\widetilde \Psi}_{21}
 \right)\cr
 \end{aligned} 
 \ee
 and 
 \be\label{eq:diffI} 
 \begin{aligned}
 c_{32}:Hom(\CF_0,\CS') & \times_{H_{20}} Hom(\CS,\CF_2')(K_\Sigma)\oplus H_{11}(K_\Sigma)
 \to H_{21}(K_\Sigma) 
 \oplus H_{10}(K_\Sigma) \cr 
 & (r,t,z) \to \left(t{\widetilde \Psi}_{21}-{\Psi}'_{21}z, {\widetilde \Psi}'_{10}r-
 z\Psi_{10}\right)\cr
 \end{aligned}
 \ee
 It is a straightforward exercise to check that $c_{21}$ is well defined and $c_{21}c_{10}= 
 c_{32}c_{21}=0$. 

 In the remaining part of this section, we will write down the hypercohomology double complex of the 
 sheaf complex $\ckk$ and prove formula \eqref{eq:algmorphisms}. We will regard all locally free sheaves 
 $\CF_i, \CF'_i$, $i=0,1,2$ as $C^\infty$ vector bundles $F_i, F'_i$, $i=0,1,2$ equipped with Dolbeault 
 operators and use Dolbeault resolutions. The hypercohomology double complex is if the form 
 \[
 {\cal H}^{p,q} = \Omega^{0,p}(\ckk^q), \qquad D = {\dbar} + (-1)^p c
 \] 
 where $c$ is the differential of $\ckk$ and it gives rise to a single complex 
 \be\label{eq:hyperA} 
 {\cal H}^n = \oplus_{p+q=n} \Omega^{0,p}(\ckk^q),  \qquad D = {\dbar} + (-1)^p c.
 \ee
 For concreteness, we will write down explicit formulas for the subcomplex 
 \[ 
 {\cal H}^{-1} {\buildrel D_{0,-1}\over \ra} {\cal H}^0 {\buildrel D_{10}\over \ra} {\cal H}^1
 \]
 and show that $H^0({\mathcal H},D)$ is isomorphic to the space of degree zero 
 morphisms between three term twisted complexes as defined in section 5.1. 
 One can prove the same result for degree one morphisms by a very similar computation. 

 Let us write down the terms of the complex \eqref{eq:hyperA}. In degree $-1$ we have 
 \be\label{eq:hyperB}
 \begin{aligned} 
 & \Omega^{0,0}(\ckk^{-1}) \oplus \Omega^{0,1}(\ckk^{-2}) \simeq 
 \Omega^{0,0}(F_2,F'_1) \oplus \Omega^{0,0}(F_1,F_0') \oplus \Omega^{0,1}(F_2,F_0')\cr
 \end{aligned} 
 \ee
 where we have used the shorthand notation $\Omega^{0,p}(Hom(F_m,F'_n))\equiv \Omega^{0,p}(F_m,F'_n)$. 
 The Dolbeault operator is the direct sum of the Dolbeault operators for the individual terms in 
 \eqref{eq:hyperB}. 
 In order to write down the degree zero term, note that the extensions \eqref{eq:extBB} are split 
 as exact sequences of $C^\infty$ bundles. Therefore we have 
 \be\label{eq:hyperC} 
 \begin{aligned} 
 \Omega^{0,0}(\ckk^0)\oplus \Omega^{0,1}(\ckk^{-1}) \simeq &
 {\Omega^{0,0}(F_2,F_0'(K_\Sigma))\oplus \Omega^{0,0}(F_2,F_0'(K_\Sigma))\over 
 \iota^-\Omega^{0,0}(F_2,F_0'(K_\Sigma))}\ \oplus \cr & \Omega^{0,0}(F_2,F_2') \oplus 
 \Omega^{0,0}(F_0,F'_0) \oplus \Omega^{0,0}(F_1, F_1')\ \oplus \cr
 &\Omega^{0,1}(F_2,F'_1)\oplus \Omega^{0,1}(F_1,F_0') \cr
 &\simeq \Omega^{0,0}(F_2,F_0'(K_\Sigma))\oplus \Omega^{0,0}(F_2,F_2')\oplus \Omega^{0,0}(F_0,F'_0)\ 
 \oplus \cr 
 & \Omega^{0,0}(F_1, F_1')\oplus \Omega^{0,1}(F_2,F'_1)\oplus \Omega^{0,1}(F_1,F_0'). \cr
 \end{aligned}
 \ee
 The Dolbeault operator 
 \[
 \dbar : \Omega^{0,0}(\ckk^0)\to \Omega^{0,1}(\ckk^0) 
 \]
 is given by 
 \be\label{eq:hyperD} 
 \dbar\left[\begin{array}{c} 
 \lambda^{1,0}_{02} \cr \lambda^{0,0}_{22} \cr \lambda^{0,0}_{00}\cr \lambda^{0,0}_{11} \cr 
 \end{array} \right]=
 \left[\begin{array}{c} 
 \dbar \lambda^{1,0}_{02} + \Psi'^{1,1}_{02}\lambda^{0,0}_{22} -\lambda^{0,0}_{00}\Psi^{1,1}_{02} \cr
 \dbar \lambda^{0,0}_{22} \cr \dbar \lambda^{0,0}_{00} \cr \dbar \lambda^{0,0}_{11} \cr
 \end{array} \right]
 \ee 
 where $\lambda^{p,q}_{n,m}$ denotes an arbitrary element of $\Omega^{p,q}_{n,m}$. 
 The degree one term is 
 \be\label{eq:hyperE} 
 \begin{aligned} 
  \Omega^{0,0}& (\ckk^1) \oplus \Omega^{0,1}(\ckk^0) \simeq \cr
 & \Omega^{0,0}(F_1,F_0'(K_\Sigma)) \oplus \Omega^{0,0}(F_1,F_2')\oplus \Omega^{0,0}(F_0,F_1') 
 \oplus \Omega^{0,0}(F_2,F_1'(K_\Sigma)) \ \oplus \cr
 & \Omega^{0,1}(F_2,F_0'(K_\Sigma))\oplus \Omega^{0,1}(F_2,F_2')\oplus \Omega^{0,1}(F_0,F'_0)
 \oplus \Omega^{0,1}(F_1, F_1')\cr
 \end{aligned}
 \ee 
 Now let us compute the differentials. We have 
 \be\label{eq:hyperF} 
 \begin{aligned} 
 D_{0,-1}:\Omega^{0,0}(\ckk^{-1}) \oplus \Omega^{0,1}(\ckk^{-2}) & \to 
 \Omega^{0,0}(\ckk^{0}) \oplus \Omega^{0,1}(\ckk^{-1})\cr
 D_{0,-1}\left[\begin{array}{c} \lambda^{0,0}_{12}\cr \lambda^{0,0}_{01}\cr \lambda^{0,1}_{02} \cr
 \end{array}\right] & = \left[\begin{array}{c} \lambda^{0,0}_{01}\Psi^{1,0}_{12} + 
 {\Psi'}^{1,0}_{01}\lambda^{0.0}_{12} \cr {\Psi'}^{0,0}_{21}\lambda^{0,0}_{12} \cr
 \lambda^{0,0}_{01} \Psi^{0,0}_{10} \cr {\Psi'}^{0,0}_{10}\lambda^{0,0}_{01} + 
 \lambda^{0,0}_{12} \Psi^{0,0}_{21} \cr 
 \dbar\lambda^{0,0}_{12} - \Psi'^{0,0}_{10}\lambda^{0,1}_{02} \cr
 \dbar \lambda^{0,0}_{01} + \lambda^{0,1}_{02} \Psi^{0,0}_{21} \cr
 \end{array} \right] \cr
 \end{aligned}
 \ee
 In order to evaluate 
 \[ 
 D_{10}: \Omega^{0,0}(\ckk^{0}) \oplus \Omega^{0,1}(\ckk^{-1}) \to 
 \Omega^{0,0}(\ckk^{1}) \oplus \Omega^{0,1}(\ckk^{0}),
 \]
 we need some auxiliary results. The relevant Dolbeault operator has been written down in \eqref{eq:hyperD}. 
 We also have to determine the action of differentials $c_{0,-1}$, $c_{10}$ on Dolbeault elements 
 $\lambda^{p,q}_{nm}$. 
 Given the isomorphism \eqref{eq:hyperC}, the differential 
 \[
 c_{10}: \Omega^{0,0}(\ckk^{0}) \to \Omega^{0,0}(\ckk^1) 
 \] 
 can be determined by making the substitutions 
 \[ 
 u= \half \lambda^{1,0}_{02} + \lambda^{0,0}_{22},\quad 
 v= \half \lambda^{1,0}_{02} + \lambda^{0,0}_{00},\quad 
 w= \lambda^{0,0}_{11} 
 \]
 in \eqref{eq:diffG}. Using the fact that the extensions \eqref{eq:extBB} are split, 
 we have 
 \be\label{eq:hyperG} 
 \begin{aligned} 
  -u\Psi_{21} -i'v{\widetilde \Psi}_{21}+ {\widetilde \Psi}'_{21} w & = 
 -\left(\half \lambda^{1,0}_{02} + \lambda^{0,0}_{22} \right)\Psi^{0,0}_{21} 
 -\half \lambda^{1,0}_{02}\Psi^{0,0}_{21}-\lambda^{0,0}_{00}\Psi^{1,0}_{01} +\Psi'^{0,0}_{21} 
 \lambda^{0,0}_{11}\cr & \quad  +\Psi'^{1,0}_{01}\lambda^{0,0}_{11} \cr 
 & = -\lambda^{1,0}_{02}\Psi^{0,0}_{21} -\lambda^{0,0}_{00}\Psi^{1,0}_{01} +\Psi'^{1,0}_{01}\lambda^{0,0}_{11} 
 -\lambda^{0,0}_{22}\Psi^{0,0}_{21} +\Psi'^{0,0}_{21} \lambda^{0,0}_{11} \cr
 {\widetilde \Psi}'_{10} u\rho + (\Psi'_{10}\otimes {\mathbb I}_{K_\Sigma})v - w{\widetilde \Psi}_{10} 
 & = \half {\Psi'}\lambda^{1,0}_{02} +\Psi'^{1,0}_{12}\lambda^{0,0}_{22} +\Psi'^{0,0}_{10}
 \left(\half \lambda^{1,0}_{02} + \lambda^{0,0}_{00}\right) -\lambda_{11}^{0,0}\Psi^{0,0}_{10}\cr & \quad 
 -\lambda^{0,0}_{11} \Psi^{1,0}_{12} \cr
 & = \Psi'^{0,0}_{10} \lambda^{1,0}_{02} + \Psi'^{1,0}_{12}\lambda^{0,0}_{22} 
 -\lambda^{0,0}_{11} \Psi^{1,0}_{12} +\Psi'^{0,0}_{10}\lambda^{0,0}_{00} -\lambda^{0,0}_{11} \Psi^{0,0}_{10}
 \cr
 \end{aligned}
 \ee
 Then we obtain 
 \be\label{eq:hyperH} 
 D_{10}\left[\begin{array}{c} 
 \lambda^{1,0}_{02} \cr \lambda^{0,0}_{22}\cr \lambda^{0,0}_{00} \cr \lambda^{0,0}_{11} \cr 
 \lambda^{0,1}_{12} \cr \lambda^{0,1}_{01} \cr \end{array} \right]= 
 \left[\begin{array}{c} 
 \Psi'^{1,0}_{01}\lambda^{0,0}_{11}-\lambda^{1,0}_{02}\Psi^{0,0}_{21}-\lambda^{0,0}_{00}\Psi^{1,0}_{01}\cr
 \Psi'^{0,0}_{21} \lambda^{0,0}_{11}-\lambda^{0,0}_{22}\Psi^{0,0}_{21}\cr
 \Psi'^{0,0}_{10}\lambda^{0,0}_{00} -\lambda^{0,0}_{11} \Psi^{0,0}_{10}\cr
 \Psi'^{0,0}_{10} \lambda^{1,0}_{02} + \Psi'^{1,0}_{12}\lambda^{0,0}_{22} 
 -\lambda^{0,0}_{11} \Psi^{1,0}_{12}\cr
 \dbar \lambda^{1,0}_{02} - \Psi'^{1,0}_{01}\lambda^{0,1}_{12} - \lb^{0,1}_{01} \Psi^{1,0}_{12} 
 + \Psi'^{1,1}_{02}\lb^{0,0}_{22} -\lb^{0,0}_{00}\Psi^{1,1}_{02} \cr 
 \dbar \lb^{0,0}_{22} - \Psi'^{0,0}_{21}\lb^{0,1}_{12} \cr
 \dbar \lb^{0,0}_{00} - \lb^{0,1}_{01} \Psi^{0,0}_{10} \cr
 \dbar \lb^{0,0}_{11} - \Psi'^{0,0}_{10}\lb^{0,1}_{01} - \lb^{0,1}_{12}\Psi^{0,0}_{21} \cr
 \end{array} \right]
 \ee
 The hypercohomology group ${\mathbb H}^0(\ckk)$ is isomorphic to the quotient 
 $\hbox{Ker}(D_{10})/\hbox{Im}(D_{0,-1})$. We would like to compare this space with the space 
 of degree zero morphisms $H^0_{\tr(\wCD)}(\CT, \CT')$. The later is the $0$-th cohomology of 
 the complex $H_{\hbox{Pre-Tr}(\wCD)}(\CT,\CT')$ defined in section 5.1, equations \eqref{eq:pretrA}, 
 \eqref{eq:pretrB}. Given equations \eqref{eq:hyperB}, \eqref{eq:hyperC} and \eqref{eq:hyperE}, 
 it is straightforward to check that 
 \be\label{eq:hyperI} 
 H^{l}_{\hbox{Pre-Tr}(\wCD)}(\CT,\CT')\simeq {\mathcal H}^{l}
 \ee 
 for $l=-1,0,1$. One can also show that the differentials are identical by specializing formulas 
 \eqref{eq:pretrB} to the case at hand taking into account \eqref{eq:shiftcompB}. Then, 
 for $\lambda^{q,p}_{n',n} \in H^{l}_{\wCD}((F_n,n-1),(F'_{n'},n'-1))$, $n,n'=0,1,2$, $p,q=0,1$, 
 $l=2q+p+n'-n$, we obtain 
 \be\label{eq:hyperJ} 
 d \lambda^{q,p}_{n',n} = 
 \dbar \lambda^{q,p}_{n,m} + \sum_{m=0}^2\sum_{r,s=0}^1\left[
 (-1)^{p(n'-m)}\Psi'^{s,r}_{m,n'}\lambda^{q,p}_{n',n} 
 -(-1)^{(r+1)(n-n')+p} \lambda^{q,p}_{n',n} \Psi^{s,r}_{n,m}\right]
 \ee
 where by convention $\Psi'^{s,r}_{m,n'} =0$ unless $2s+r+m-n'=1$ and $\Psi^{s,r}_{n,m}=0$ unless 
 $2s+r+n-m=1$. It is a simple exercise to check that the differentials \eqref{eq:hyperJ} 
 agree with \eqref{eq:hyperF} and \eqref{eq:hyperH}. This proves formula \eqref{eq:algmorphisms} 
 in degree zero. The proof in degree one is very similar.


\end{document}